\documentclass[10pt]{article}
\usepackage{graphicx} 
\usepackage[a4paper,margin=2.2cm]{geometry}
\usepackage[ruled,vlined,linesnumbered]{algorithm2e}
\usepackage[T1]{fontenc}
\usepackage[utf8]{inputenc}
\usepackage{lmodern}
\usepackage{longtable}
\usepackage{microtype}
\usepackage{amsmath,amssymb,mathtools}
\usepackage{booktabs,tabularx,array,multirow}
\usepackage{enumitem}
\usepackage{appendix}
\usepackage{booktabs}
\usepackage{array}
\usepackage{xcolor}
\usepackage{tikz}
\usepackage{enumitem}
\usepackage{amsthm}
\usetikzlibrary{arrows.meta,
positioning,
fit,
shapes.geometric, 
backgrounds, 
calc}
\usepackage{listings}
\usepackage{hyperref}
\usepackage[nameinlink,noabbrev]{cleveref}
\usepackage{caption}
\usepackage{subcaption}
\usepackage{url}
\usepackage{balance}
\usepackage{tabularx}
\usepackage{booktabs}
\usepackage{array}
\usepackage{makecell}
\usepackage{ragged2e}
\newcolumntype{C}[1]{>{\Centering\arraybackslash}p{#1}}
\newcolumntype{Y}{>{\RaggedRight\arraybackslash}X}

\definecolor{acmblue}{RGB}{37,80,130}
\definecolor{acmgray}{RGB}{245,247,249}
\definecolor{passgreen}{RGB}{34,139,34}
\definecolor{deviationorange}{RGB}{210,105,30}
\definecolor{skipgray}{RGB}{105,105,105}
\definecolor{failred}{RGB}{178,34,34}
\definecolor{alertred}{RGB}{178,34,34}
\definecolor{preservedgreen}{RGB}{34,139,34}
\definecolor{approximatedorange}{RGB}{210,105,30}
\definecolor{unsupportedgray}{RGB}{105,105,105}
\hypersetup{
  colorlinks=true,
  linkcolor=acmblue,
  citecolor=acmblue,
  urlcolor=acmblue,
  pdftitle={Agentic Configuration Management},
  pdfauthor={AUTHOR NAME}
}

\lstdefinestyle{acmcode}{
  basicstyle=\ttfamily\small,
  backgroundcolor=\color{acmgray},
  frame=single,
  rulecolor=\color{black!15},
  breaklines=true,
  columns=fullflexible,
  showstringspaces=false
}

\title{\textbf{Agentic Configuration Management (ACM):\\A Reference Configuration Model for Governed Agentic Systems}}

\author{
  Audrey Quessadda-Vial\\
  PwC\\
  \texttt{audrey.quessada-vial@pwc.com}
}

\date{V1 --- August 2026}

\begin{document}

\maketitle

\begin{abstract}

Agentic systems are rapidly evolving from isolated LLM applications to complex software composed of interacting agents, tools, prompts, skills, composite subsystems, models, and execution workflows. While existing LLMOps and AgentOps platforms provide valuable support for orchestration and observability, they do not offer a framework-independent configuration and governance model capable of describing, validating, and governing the system as a coherent whole. As a result, configuration management remains fragmented across frameworks, making reproducibility, impact analysis, and long-term maintenance increasingly difficult.

This paper introduces \textbf{Agentic Configuration Management (ACM)}, a framework-independent governance and configuration reference model for heterogeneous agentic systems. ACM extends established Software Configuration Management principles through immutable configuration items, explicit baselines, deterministic governance semantics, lifecycle management, dependency-aware impact propagation, and runtime provenance. Rather than replacing existing orchestration frameworks, ACM provides a common semantic representation through which heterogeneous configurations can be projected, governed, and evaluated consistently.

We present a complete Python reference implementation together with semantic projection adapters for LangGraph, CrewAI, and the OpenAI Agents SDK. The implementation combines deterministic validation, governance evaluation, replayable runtime reconstruction, and preservation-based semantic projection into a unified operational pipeline. Experimental evaluation includes qualitative governance scenarios and quantitative impact analysis to assess semantic preservation, deterministic governance behavior, and framework-independent operational consistency across complementary introspection regimes.

The governance semantics are formally characterized through monotonic impact propagation over a finite lattice, establishing monotonicity, convergence, termination, and uniqueness of the least fixed point above the initial impact valuation under the assumptions of the proposed governance model.

The results demonstrate that, for the evaluated configurations, heterogeneous agentic representations can be projected into governance-equivalent ACM representations while preserving the governance-relevant information required for lifecycle management, assurance evaluation, dependency analysis, and deterministic impact propagation. Across three representative frameworks spanning complementary introspection regimes, the evaluation provides empirical evidence that common governance semantics can support reproducibility, auditability, and interoperability independently of the native execution abstractions within the evaluated scope.

\end{abstract}

\section{Introduction}
\label{sec:introduction}
Agentic AI systems are rapidly evolving from isolated large language model (LLM) applications into complex software systems composed of interacting autonomous components. Their behaviour increasingly depends on configurable artifacts that evolve throughout both development and execution, making configuration governance substantially more challenging than in traditional software systems \cite{bersoff1978software, bersoff1984elements, whitgift1991methods}. As these systems become larger, more dynamic, and more heterogeneous, ensuring their reproducibility, auditability, and controlled evolution has emerged as a fundamental engineering challenge \cite{Hughes03072025, pandey2025agentic}.

Although current agentic ecosystems provide powerful execution capabilities, the representation of system configurations remains largely framework-specific \cite{bandi2025rise}. Configuration information is distributed across heterogeneous abstractions, lifecycle conventions, and runtime representations that differ from one framework to another \cite{acharya2025agentic, Inioluwa2020}. As a consequence, comparing, auditing, evolving, or reproducing agentic systems independently of their execution environment remains difficult.

This situation exposes a gap between the maturity of Software Configuration Management (SCM) principles and their application to modern agentic systems. Established SCM concepts such as immutable revisions, controlled baselines, change management, provenance, and configuration auditing have proven their effectiveness for conventional software engineering \cite{ieee2012ieee, leon2015software, amershi2019}, yet they have not been systematically adapted to the governance of agentic configurations.

This paper introduces \textbf{Agentic Configuration Management (ACM)}, a framework-independent governance representation for governing the configuration of agentic systems. ACM provides a common configuration representation that complements existing execution frameworks by supporting configuration governance, traceability, reproducibility, and controlled evolution across heterogeneous implementations.

A central design principle of ACM is the explicit separation between governed configurations and their operational realization. Configuration describes the intended, versioned, and governed structure of an agentic system independently of any execution framework, whereas runtime represents one particular execution of that configuration under specific operational conditions. ACM governs the former and provides semantic mechanisms for relating it to the latter without conflating the two. This distinction constitutes the conceptual foundation of ACM and motivates the separation between configuration governance and execution semantics developed throughout the remainder of the paper.

\begin{figure*}[t]
	\centering
	\includegraphics[width=0.7\textwidth]{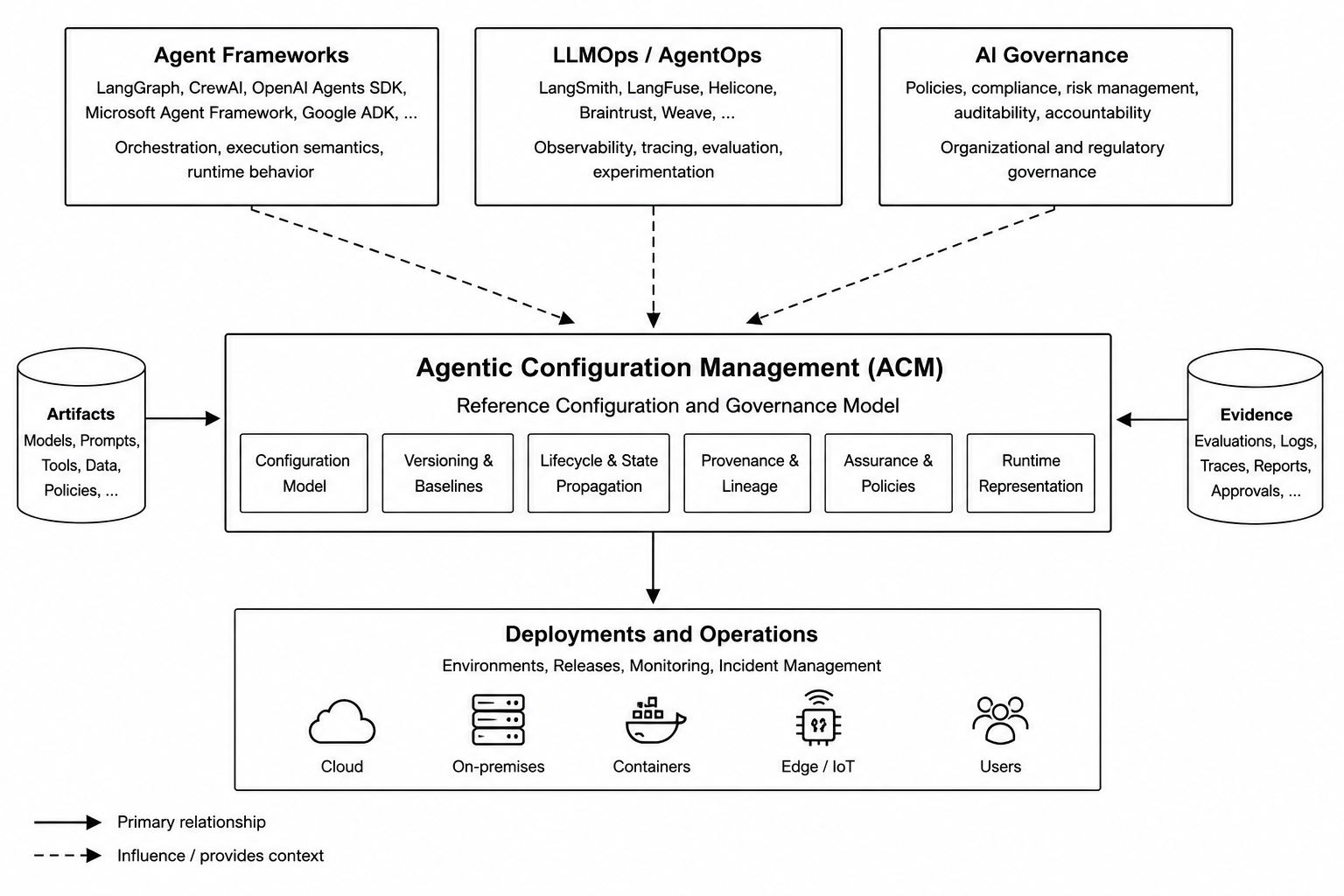}
	\caption{Positioning of ACM. ACM acts as a framework-independent configuration layer that connects agent frameworks (orchestration), LLMOps platforms (observability and evaluation) and deployment environments (execution).}
	\label{fig:ACM_positioning}
\end{figure*}

ACM focuses on governing the former while providing formal semantics to interpret the latter relative to the governed configuration. The contributions of this paper are fourfold:

\begin{itemize}
    \item We propose a \textbf{framework-independent governance representation} that represents heterogeneous agentic configuration artifacts as typed and independently versioned Agentic Configuration Items (ACIs), connected through typed relationships that make their configuration dependencies explicit for governance.
    
    \item We formalize the \textbf{governance semantics} of agentic configurations through lifecycle models, state propagation, assurance semantics, and runtime governance mechanisms.
    
    \item We present a \textbf{reference operationalization} of the model, including a Python reference implementation together with projection mechanisms for heterogeneous agentic frameworks.
    
    \item We evaluate the \textbf{representational coverage and operational feasibility} of the proposed model through normative scenarios, automated validation, quantitative impact analysis, and cross-framework projection experiments.
\end{itemize}
These contributions are intended as an integrated configuration-governance architecture rather than as claims of novelty for each constituent mechanism in isolation. ACM builds on established mechanisms including immutable revisions, baselines, provenance, and fixed-point computation, but combines them with agentic configuration typing, explicit configuration--runtime separation, semantic projection, and framework-independent governance semantics. The resulting property investigated in this work is cross-framework governance equivalence after semantic projection: configurations expressed through heterogeneous execution abstractions can be normalized into governance-equivalent representations on which the same governance semantics operate, within the evaluated scope.

The remainder of this paper is organized as follows. Section~\ref{sec:related_work} reviews the foundations of Software Configuration Management, AI governance, LLMOps, AgentOps, and agentic frameworks, and identifies the gap addressed by the Agentic Configuration Management (ACM). Section~\ref{sec:design_principles} derives the design principles of the proposed reference model. Section~\ref{sec:reference_model} introduces the ACM reference model, while Section~\ref{sec:governance_semantics} formalizes its governance semantics. Section~\ref{sec:operationalization} describes its operationalization through a reference architecture, framework adapters, governance algorithms, and a reference implementation. Section~\ref{sec:evaluation} evaluates the proposed model. Finally, Sections~\ref{sec:discussion}, \ref{sec:threats_validity}, \ref{sec:future_work}, and \ref{sec:conclusion} discuss the implications, limitations, future research directions, and conclusions of this work.

\section{Related Work}
\label{sec:related_work}

\subsection{Software Configuration Management}
\label{sec:scm}

Software Configuration Management (SCM) provides the engineering foundations for identifying, controlling, and auditing the evolution of software systems. Rather than focusing solely on version control, SCM defines a disciplined process for establishing configuration baselines, managing change, maintaining traceability, and ensuring that released systems remain reproducible and auditable throughout their lifecycle \cite{whitgift1991methods, conradi1998version, leon2015software, bersoff1984elements, bersoff1978software}.

These principles have been formalized by long-established standards and bodies of knowledge, including IEEE~828, ISO/IEC/IEEE~12207, and the \emph{Software Engineering Body of Knowledge} (SWEBOK) \cite{swebok40, ieee2012ieee, wohlinsnowballing}. Although these references differ in scope, they consistently describe configuration management as a combination of configuration identification, change control, configuration status accounting, and configuration verification and audit.

Classical SCM has proved effective for conventional software systems composed of relatively stable software artifacts such as source code, libraries, documentation, build configurations, and release packages \cite{chacon2014pro}. Its central concepts---configuration items, immutable baselines, controlled revisions, dependency management, and change auditing---provide the foundations required to ensure consistency and reproducibility across software releases \cite{bersoff1984elements}.

\begin{figure*}[t]
	\centering
	\includegraphics[width=0.7\textwidth]{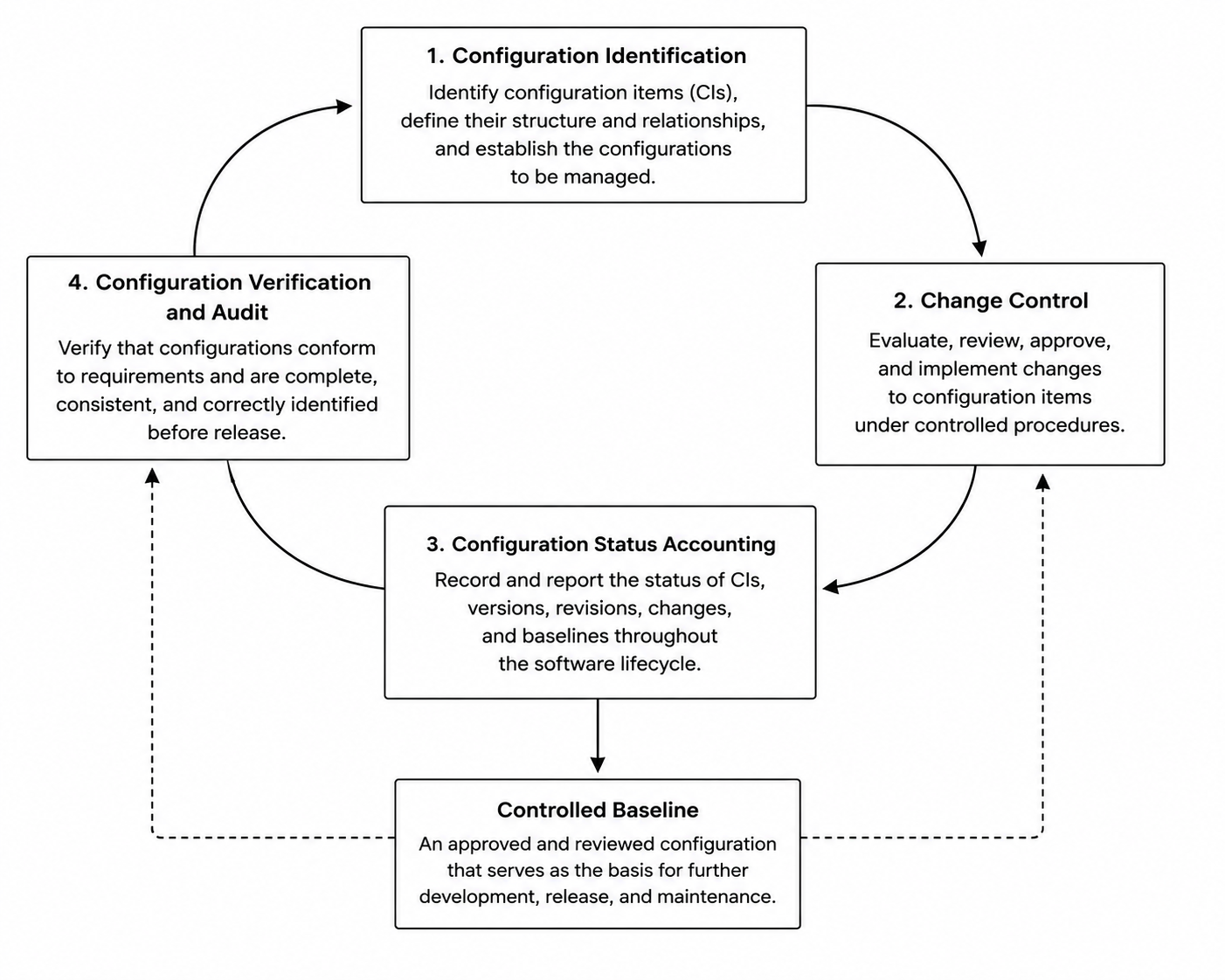}
	\caption{Fundamental Software Configuration Management (SCM) process,
		comprising configuration identification, change control, configuration status accounting, and configuration verification and audit.}
	\label{fig:SCM_framework}
\end{figure*}

However, applying these principles directly to agentic systems is not straightforward. In addition to executable software components, agentic systems rely on heterogeneous configurable artifacts, dynamic execution structures, and runtime-generated elements that are not explicitly addressed by traditional SCM models. Consequently, while SCM provides the conceptual foundations for configuration governance, additional abstractions are required to represent and govern the configuration of modern agentic systems.

\subsection{AI Governance}
\label{sec:ai_governance}

The rapid adoption of AI systems has expanded governance concerns beyond traditional software engineering. Recent standards, regulatory frameworks, and technical initiatives emphasize accountability, transparency, risk management, and auditability throughout the AI lifecycle \cite{pandey2025agentic, Engin_Hand_2026, Gangavarapu2025, shavit2023practices, IMDA2026AgenticMGF}. Although these initiatives differ in scope, they share the objective of ensuring that AI systems remain trustworthy, traceable, and subject to appropriate organizational oversight.

From the perspective of agentic systems, three complementary areas are particularly relevant: governance frameworks defining organizational requirements, provenance mechanisms ensuring artifact traceability, and emerging approaches dedicated to the governance of autonomous agents. Together, these perspectives establish important governance objectives but do not define a common configuration representation for heterogeneous agentic systems.

\subsubsection{Governance Frameworks}
\label{sec:governance_frameworks}

Several governance frameworks have established principles for the responsible development and operation of AI systems. Examples include the NIST AI Risk Management Framework (AI RMF), ISO/IEC~42001, ISO/IEC~23894, the OECD AI Principles, and the European AI Act \cite{ai2023artificial, benraouane2024ai, simonetta2024iso, yeung2020recommendation, edwards2021eu}. While their objectives and intended audiences differ, they consistently emphasize governance practices such as accountability, documentation, risk management, human oversight, traceability, and auditability.

These frameworks define organizational objectives and governance processes rather than implementation models. They describe \emph{what} should be governed and documented but intentionally remain independent of the internal representation adopted by AI platforms or execution frameworks. Consequently, they do not provide a framework-independent configuration model capable of representing the complete structure and evolution of an agentic system.

\subsubsection{Provenance and Software Supply Chain}
\label{sec:provenance_supply_chain}

Provenance has become a fundamental requirement for software and AI governance. Existing standards and specifications, including W3C PROV, Software Bills of Materials (SBOMs), SPDX, and the Supply-chain Levels for Software Artifacts (SLSA) \cite{missier2013w3c, SBOM10336262, stewart2010software, okafor2022sok, cobbe2023understanding}, improve traceability by documenting artifact origins, dependencies, integrity, and production processes. These mechanisms strengthen reproducibility, software supply-chain security, and compliance by making software artifacts and their relationships explicitly traceable.

Although highly valuable, these approaches primarily describe software packages, build processes, datasets, or deployment artifacts. They do not represent the semantic configuration of agentic systems, whose behavior depends on heterogeneous configuration artifacts such as agents, prompts, models, skills, composite subsystems, tools, policies, and dynamically evolving execution structures. Consequently, provenance alone is insufficient to govern the configuration of an entire agentic system.

\subsubsection{Governance of Agentic Systems}
\label{sec:agentic_governance}

The emergence of agentic AI has stimulated research on governance mechanisms specifically targeting autonomous and multi-agent systems. Existing approaches typically address runtime supervision, policy enforcement, authorization, monitoring, safety constraints, or operational accountability. Likewise, current AgentOps and orchestration platforms increasingly provide tracing, evaluation, and operational monitoring capabilities for agent execution but are framework-dependent (see Section~\ref{sec:llmops_agentops}).

This observation complements the limitations identified in classical Software Configuration Management and motivates the need for a unified configuration reference model that separates configuration governance from execution concerns. This gap forms the foundation of the Agentic Configuration Management (ACM) reference model introduced in the following sections.

\subsection{LLMOps and AgentOps}
\label{sec:llmops_agentops}

The emergence of large language models has also given rise to a new generation of operational platforms dedicated to managing AI applications throughout their lifecycle. Inspired by established MLOps practices, LLMOps extends operational support to applications built around foundation models by providing capabilities for prompt management, experimentation, evaluation, deployment, monitoring, and observability. As agentic systems have become increasingly dynamic and autonomous, these practices have naturally evolved toward AgentOps, which addresses the operational management of multi-agent applications and long-running agentic workflows \cite{cesarano2026grand, shavit2023practices, pandey2025agentic, dumas2026}.

Unlike traditional software configuration management, LLMOps and AgentOps focus primarily on the operational lifecycle of AI applications rather than on the governance of their underlying configurations \cite{pahune2025transitioning, amershi2019, dong2024agentops}. Their objective is to facilitate development, deployment, monitoring, experimentation, and continuous improvement while maintaining visibility over system behavior during execution.

Although terminology varies across platforms, most LLMOps and AgentOps solutions converge toward a common set of operational capabilities, including prompt management, execution tracing, evaluation workflows, observability, deployment management, and experiment tracking. These capabilities provide the operational evidence required to understand how an AI application behaves in production and to support iterative improvements of prompts and workflows.

From the perspective of configuration governance, several concepts introduced by these platforms are particularly relevant. 

\begin{itemize}
	\item Configuration management concerns the explicit representation of configurable artifacts and their versions \cite{liu2025towards, Hughes03072025, langfusedocs, heliconedocs}. 
	\item Provenance records the origin and derivation history of these artifacts.
	\item Lineage captures the dependency relationships connecting successive versions, execution traces, and derived artifacts \cite{cesarano2026grand, rajath2025review}. 
	\item Tracing records the sequence of execution events occurring during system operation.
	\item Observability \cite{langsmithdocs} aggregates telemetry that enables operators to inspect the internal behavior of deployed applications \cite{opentelemetry}. 
	\item Deployment state characterizes the operational status of a configuration once deployed within a target environment \cite{braintrustdocs}.
\end{itemize}	

These concepts constitute essential building blocks for governing modern AI applications. However, they generally remain attached to operational artifacts such as prompts, traces, evaluations, or deployments, rather than providing a unified representation of the complete configuration of an agentic system. Relationships among agents, prompts, models, tools, skills, composite subsystems, policies, workflows, and runtime-generated components are typically represented through platform-specific abstractions, preventing the establishment of a framework-independent configuration model.

Representative platforms illustrate this evolution. Prompt management solutions increasingly support prompt versioning, labels, rollback, experimentation, and deployment workflows. Observability platforms provide execution traces, telemetry, evaluation pipelines, and runtime analytics. More recent AgentOps environments extend these capabilities to multi-agent executions through richer runtime monitoring, workflow inspection, and execution replay \cite{sahir2026agentops}. 

Langfuse~\cite{langfusedocs} is an open-source observability platform providing prompt management, execution tracing, evaluation, experimentation, and prompt versioning. It enables developers to associate prompt revisions with runtime traces, datasets, and evaluation results, thereby improving reproducibility and
experimentation across LLM-based applications.
Another observability platform is Helicone ~\cite{heliconedocs}. It supports request tracing, prompt versioning, experimentation, evaluation, and deployment management. Its primary focus is operational monitoring and prompt-centric development workflows, providing valuable runtime telemetry and performance insights. 

Together, these platforms substantially improve the operational governance of AI applications but remain centered on execution management rather than on configuration governance. They do not aim to represent the complete configuration of heterogeneous agentic systems (see Table~\ref{tab:llmops_comparison} for a comparison of governance capabilities across current LLMOps and AgentOps platforms).

\begin{table*}[t]
\centering
\caption{Representative capabilities of current LLMOps and AgentOps platforms. The comparison highlights their primary focus on operational lifecycle management rather than framework-independent configuration governance.}
\label{tab:llmops_comparison}

\renewcommand{\arraystretch}{1.15}
\begin{tabular}{p{4.2cm}ccccc}
\toprule
\textbf{Capability} &
\textbf{LangSmith} &
\textbf{Langfuse} &
\textbf{Helicone} &
\textbf{Braintrust} &
\textbf{PromptLayer} \\
\midrule

Prompt versioning                & \checkmark & \checkmark & \checkmark & \checkmark & \checkmark \\

Prompt deployment / labels       & $\sim$ & \checkmark & \checkmark & $\sim$ &  \checkmark \\

Execution tracing                & \checkmark & \checkmark & \checkmark & \checkmark & \checkmark \\

Observability / telemetry        & \checkmark & \checkmark & \checkmark & \checkmark & $\sim$ \\

Evaluation workflows             & \checkmark & \checkmark & \checkmark & \checkmark & $\sim$ \\

Experiment tracking              & \checkmark & \checkmark & \checkmark & \checkmark & $\sim$ \\

Prompt provenance                & $\sim$ & \checkmark & \checkmark & $\sim$ & \checkmark \\

Execution lineage                & \checkmark & \checkmark & \checkmark & \checkmark & $\sim$ \\

Deployment state                 & $\sim$ & \checkmark & \checkmark & $\sim$ & \checkmark \\

Runtime replay                   & \checkmark & $\sim$ & $\sim$ & $\sim$ & $\sim$ \\

Framework-specific integrations  & \checkmark & \checkmark & \checkmark & \checkmark & \checkmark \\

\midrule

Framework-independent configuration model
                                 & -- & -- & -- & -- & -- \\

Configuration graph              & -- & -- & -- & -- & -- \\

Immutable configuration baseline & -- & -- & -- & -- & -- \\

Configuration lifecycle          & -- & -- & -- & -- & -- \\

Configuration dependency model   & -- & -- & -- & -- & -- \\

Governed configuration revisions & -- & -- & -- & -- & -- \\

Configuration state propagation  & -- & -- & -- & -- & -- \\

Configuration governance semantics
                                 & -- & -- & -- & -- & -- \\

\bottomrule
\end{tabular}

\vspace{1mm}

\footnotesize
$\checkmark$ Supported \hspace{1em}
$\sim$ Partial support \hspace{1em}
-- Not provided as a first-class capability.

\end{table*}

\subsection{Agentic Frameworks}
\label{sec:agentic_frameworks}

The rapid adoption of agentic AI has led to the emergence of numerous frameworks dedicated to designing and executing autonomous or multi-agent applications. Although their architectures differ, these frameworks share a common objective: orchestrating the execution of agents, coordinating interactions among heterogeneous components, and integrating large language models with external tools, memory, and user-defined workflows, as illustrated by representative frameworks \cite{crewaidocs, langgraphdocs, googleadkdocs, microsoftadkdocs, openaiddocs, claudedocs, mistraldocs}).

Current frameworks cover a broad spectrum of orchestration paradigms. Some adopt explicit graph-based execution models, while others organize applications around agents, tasks, crews, or conversational interactions. Despite these architectural differences, they generally provide abstractions for defining execution flows, invoking tools, managing shared or local state, and coordinating agent interactions during runtime (see Table~\ref{tab:framework_comparison} in Appendix~\ref{appendix:related_work}).

Beyond execution, modern frameworks increasingly incorporate operational capabilities such as checkpointing, persistence, human-in-the-loop interactions, execution replay, and integration with observability platforms. These features considerably simplify the development and deployment of agentic applications and complement the operational services provided by LLMOps and AgentOps platforms.

However, the internal representations adopted by these frameworks remain closely coupled to their respective execution models. Configuration artifacts, dependency structures, lifecycle conventions, and runtime abstractions are expressed through framework-specific concepts that are not directly interoperable. Consequently, while these frameworks provide rich execution semantics, they do not define a common representation suitable for governing configurations independently of a particular execution environment.

\subsection{Gap Analysis}
\label{sec:gap_analysis}

The preceding sections have examined complementary perspectives on the governance of AI systems, including Software Configuration Management, AI governance frameworks, LLMOps and AgentOps platforms, and agentic execution frameworks. Although these approaches address different aspects of the AI lifecycle, they collectively contribute to the governance of modern AI applications. However, they cannot simply be combined to obtain a unified configuration-governance solution. Software Configuration Management governs software artifacts but has no framework-independent representation of agentic abstractions. AI governance frameworks define organizational objectives and compliance requirements without specifying a configuration model. LLMOps and AgentOps platforms provide observability, evaluation, and operational tooling, while execution frameworks define runtime semantics through framework-specific abstractions. 

Recent work has begun to address configuration as a first-class concern in agentic systems. Alsegier~\cite{alsegier2026agency} extends Software Product Line Engineering through an implementation-independent Agency Configuration Schema that captures variability in models, prompts, tools, authority, governance constraints, observability, and evolution, with constraint-based validation of agent variants. Madatha~\cite{madatha2026controlplane} proposes a deterministic control plane for LLM coding agents, combining canonical agent definitions, content-addressed configuration, permission enforcement, audit logging, and compilation to multiple execution environments. These approaches establish complementary mechanisms for governing agent configuration and narrow the configuration-management gap addressed in this work. ACM addresses a distinct configuration-management concern by representing heterogeneous agentic systems through typed, independently revisioned configuration items and immutable resolved baselines, relating configuration to assurance, evolution, and runtime observations, and providing formal dependency-aware impact propagation over governed revisions.

Despite these advances, existing approaches address complementary subsets of the configuration-governance problem rather than providing an integrated model for revision-aware configuration management, governance evaluation, dependency-aware impact analysis, and runtime provenance across heterogeneous agentic systems. This remaining gap motivates a framework-independent reference configuration model that integrates these concerns while preserving a stable governance representation across heterogeneous agentic ecosystems.

Table~\ref{tab:gap_analysis} synthesizes the coverage of representative governance capabilities across these complementary families of approaches. Rather than comparing the approaches themselves, the analysis evaluates the extent to which each family addresses the governance capabilities required throughout the lifecycle of agentic systems.

\begin{table*}[t]
\centering
\caption{Coverage of representative governance capabilities across complementary families of approaches. A detailed comparison is provided in Appendix Table~\ref{tab:gap_analysis_detailed}.}
\label{tab:gap_analysis}

\renewcommand{\arraystretch}{1.15}

\begin{tabular}{p{5.2cm}ccccc}
\toprule
\textbf{Governance Capability}
&
\textbf{SCM}
&
\textbf{AI Gov.}
&
\textbf{LLMOps / AgentOps}
&
\textbf{Frameworks}
\\
\midrule

Configuration identification          & \checkmark & $\sim$ & $\sim$ & $\sim$ \\

Version management                    & \checkmark & -- & $\sim$ & $\sim$ \\

Configuration provenance              & $\sim$ & \checkmark & \checkmark & $\sim$ \\

Dependency traceability               & \checkmark & $\sim$ & \checkmark & $\sim$ \\

Execution observability               & -- & $\sim$ & \checkmark & \checkmark \\

Lifecycle governance                  & $\sim$ & \checkmark & $\sim$ & $\sim$ \\

Framework-independent representation  & -- & -- & -- & -- \\

Unified configuration model           & -- & -- & -- & -- \\

\bottomrule
\end{tabular}

\vspace{1mm}

\footnotesize
$\checkmark$ Fully addressed \hspace{1em}
$\sim$ Partially addressed \hspace{1em}
-- Not addressed as a primary capability.

\end{table*}

The limitation of combining these approaches is therefore not the absence of governance, observability, execution, or compliance mechanisms. Rather, these capabilities remain distributed across complementary technologies that operate on different abstractions and representations. Consequently, none of these families alone provides a common semantic representation for identifying, versioning, relating, and governing heterogeneous agentic configurations independently of their execution framework. ACM addresses this specific representation gap while remaining complementary to SCM, organizational AI governance, LLMOps and AgentOps platforms, and agentic execution frameworks. The design principles derived from this gap are presented in Section~\ref{sec:design_principles}.

\section{Design Principles}
\label{sec:design_principles}

Rather than extending an existing orchestration framework or operational platform, the objective of ACM is to establish a reference model that complements these ecosystems while remaining independent of their implementation technologies. ACM addresses therefore a complementary concern by representing governed configurations independently of execution platforms, enabling immutable baselines, dependency-aware configuration management, and explicit relationships between configuration revisions, runtime observations, and assurance evidence.

The design of such a reference model is guided by a set of requirements derived from the limitations identified in the preceding analysis. These principles define the properties that any framework-independent configuration governance model should satisfy. They intentionally describe design requirements rather than implementation choices; the conceptual realization of these principles is introduced in Section~\ref{sec:reference_model}.

Table~\ref{tab:design_principles} summarizes the seven principles that guide the design of ACM.
\begin{table*}[t]
\centering
\caption{Design principles guiding the ACM reference model.}
\label{tab:design_principles}

\renewcommand{\arraystretch}{1.15}

\begin{tabular}{p{1.0cm}p{4.0cm}p{9.2cm}}
\toprule
\textbf{ID} &
\textbf{Principle} &
\textbf{Objective}
\\
\midrule

P1 &
Framework Independence &
Remain independent of execution frameworks, orchestration engines, and AI providers.\\

P2 & Explicit Configuration Representation & Represent every configurable artifact explicitly under configuration governance.\\

P3 &  Immutable Versioning & Ensure deterministic evolution through immutable revisions and controlled baselines. \\

P4 & Configuration–Execution Separation & Clearly distinguish governed configurations from runtime execution.\\

P5 & Explicit Traceability & Preserve provenance and dependency relationships throughout the configuration lifecycle.\\

P6 & Governance by Design & Support lifecycle governance, policy enforcement, assurance, and auditability.\\

P7 & Extensibility & Allow the reference model to evolve without changing its governance foundations.
\\
\bottomrule
\end{tabular}

\end{table*}

\paragraph{P1 -- Framework Independence}

A configuration governance model shall remain independent of any specific orchestration framework, execution engine, or AI provider. Heterogeneous agentic systems should therefore be representable through a common abstraction that preserves governance semantics without exposing framework-specific implementation details. This independence enables interoperability across evolving ecosystems while avoiding dependence on a particular execution technology.

\paragraph{P2 -- Explicit Configuration Representation}

Every artifact whose definition influences the behaviour of an agentic system shall be represented explicitly as a governed configuration object. The model should support heterogeneous artifacts while remaining independent of their implementation and execution semantics.

\paragraph{P3 -- Immutable Versioning}

Configuration evolution shall rely on immutable revisions. Every modification shall produce a new governed revision while preserving previous versions to ensure reproducibility, auditability, and deterministic reconstruction of historical configurations.

\paragraph{P4 -- Configuration--Execution Separation}

Configuration governance shall remain conceptually independent from runtime execution. Runtime observations may enrich governance information but shall not directly modify governed configurations.

\paragraph{P5 -- Explicit Traceability}

The model shall preserve explicit traceability among governed artifacts throughout their lifecycle. Structural dependencies, provenance relationships, and configuration evolution should remain observable and auditable independently of any execution framework.

\paragraph{P6 -- Governance by Design}

Governance shall be an intrinsic property of the configuration model rather than an external operational concern. The model should support lifecycle control, policy evaluation, assurance, auditability, and deterministic governance decisions.

\paragraph{P7 -- Extensibility}

The reference model shall remain extensible to accommodate future categories of configuration artifacts, governance policies, and execution paradigms without modifying its fundamental governance principles.

The following section introduces the ACM reference model, which operationalizes these principles through a normalized representation of governed configuration artifacts and their relationships.

\section{ACM Reference Model}
\label{sec:reference_model}

The ACM is made to offer a common governance layer capable of representing heterogeneous configuration artifacts independently of their execution environment.

The model is designed around three complementary objectives. First, it provides a unified representation of configurable artifacts participating in an agentic system. Second, it explicitly captures the structural and governance relationships among these artifacts. Third, it establishes the conceptual foundations required to support lifecycle management, traceability, provenance, auditability, and configuration evolution.

Section~\ref{sec:conceptual_overview} introduces the conceptual organization of the model. Section~\ref{sec:aci_model} defines the abstract representation of Agentic Configuration Items (ACIs). The following subsections specify the relationships between ACIs, baseline composition, and the structural properties of the reference model.

\subsection{Conceptual Overview}
\label{sec:conceptual_overview}

An agentic system is composed of numerous configurable artifacts originating from heterogeneous technologies, including orchestration frameworks, language models, prompts, tools, skills, composite subsystems, policies, execution environments, and governance mechanisms. Although these artifacts collectively determine the behaviour of the system, they are generally represented through framework-specific abstractions and governed independently.

The ACM reference model introduces a unified conceptual representation in which configurable artifacts are represented as governed configuration items, regardless of their implementation technology or execution framework, while runtime observations remain represented as distinct runtime entities linked to the governed configuration. This abstraction enables heterogeneous artifacts to be managed according to a common governance model while preserving their individual semantics.

Rather than organizing governed artifacts according to implementation
technologies, ACM structures them according to the governance role they fulfil within the system. The reference model distinguishes three complementary categories of governed artifacts and observations:

\begin{itemize}

\item \textbf{Structural items}, which describe the static composition of the system and its configurable artifacts.

\item \textbf{Governance items}, which define policies, assurance evidence, compliance constraints, and lifecycle governance.

\item \textbf{Runtime entities}, which represent execution-related information generated during system operation while remaining linked to the governed configuration.

\end{itemize}

Although these families are conceptually independent, they remain connected through explicit relationships that preserve traceability across the complete lifecycle of the system. This organization separates structural configuration, governance information, and runtime observations while allowing each perspective to evolve independently.

\begin{figure*}[t]
	\centering
	\includegraphics[width=0.6\textwidth]{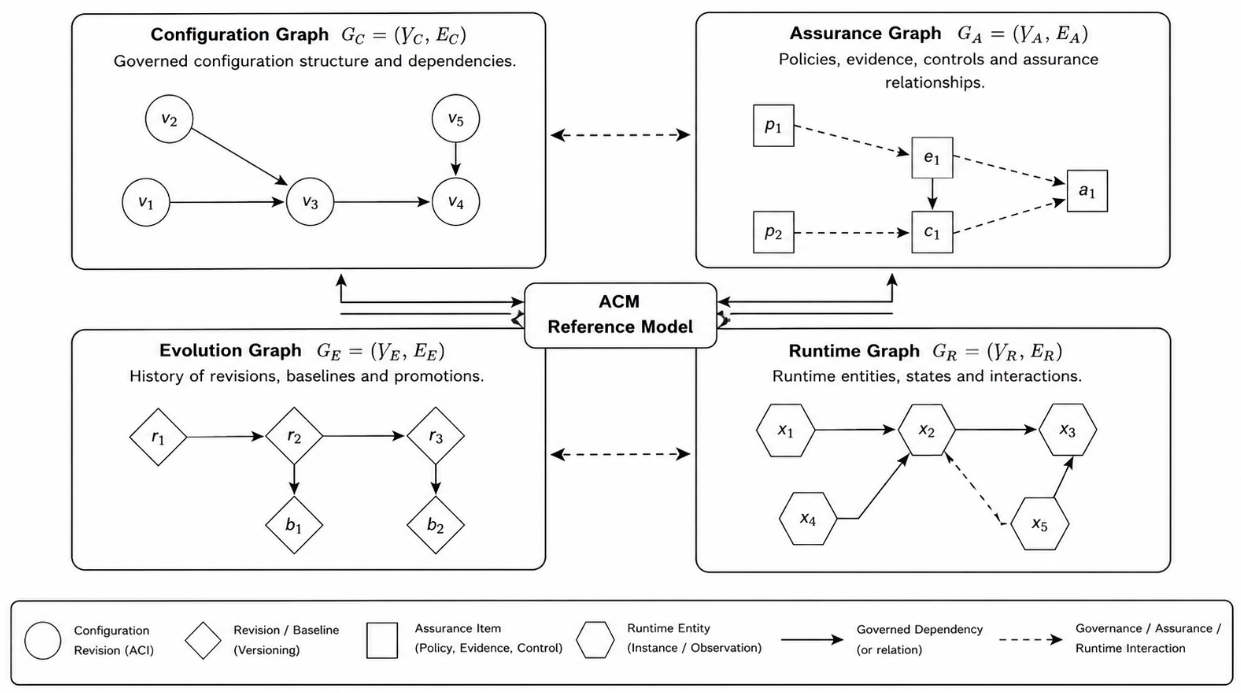}
	\caption{Four-graph organization of the ACM reference model. The Configuration, Evolution, Assurance, and Runtime Graphs represent distinct governance concerns while preserving explicit traceability across the model}
	\label{fig:acm_reference_model}
\end{figure*}

To represent these complementary perspectives, ACM organizes configuration information into four interconnected conceptual graphs, illustrated in Figure~\ref{fig:acm_reference_model}.

Formally, the ACM reference model is represented as the ordered multi-graph structure

\begin{equation}
	\mathcal{G}_{\mathrm{ACM}}
	=
	\left(
	G_C,
	G_E,
	G_A,
	G_R
	\right),
	\label{eq:acm_multigraph_structure}
\end{equation}

where $G_C$, $G_E$, $G_A$, and $G_R$ denote the Configuration, Evolution, Assurance, and Runtime Graphs, respectively. The four graphs represent distinct governance views and remain connected through explicit relationships and revision-level provenance.

\begin{itemize}

\item The \textbf{Configuration Graph} represents the governed configuration of the agentic system and the structural relationships among configuration items.

\item The \textbf{Assurance Graph} captures governance policies, assurance evidence, compliance constraints, and policy enforcement.

\item The \textbf{Runtime Graph} represents execution observations generated during system operation while maintaining explicit links to governed configuration items.

\item The \textbf{Evolution Graph} provides an integrated view of configuration evolution and history from a baseline.

\end{itemize}

The Assurance Graph records the governance information used to evaluate and
justify conclusions about governed configuration revisions. It relates
configuration to applicable policies, assurance evidence, compliance
constraints and results, and governance evaluations. These elements support
the derivation and traceability of governance conclusions without becoming
part of the configuration topology itself.

The Evolution Graph relates successive immutable revisions and baselines through
lineage, replacement, and derivation relationships. It records how governed
configurations evolve over time without modifying previously identified
revisions. The Configuration Graph therefore describes what constitutes a
governed configuration, whereas the Evolution Graph preserves how successive
governed configurations are historically related.

The decomposition into four complementary graphs is a deliberate design decision rather than a modeling convenience. Each graph addresses a distinct governance concern that follows its own semantics and lifecycle. Separating these orthogonal concerns preserves conceptual clarity, avoids mixing static and dynamic governance information, and enables each graph to evolve independently while remaining linked through explicit traceability relationships. 

\subsection{Agentic Configuration Item}
\label{sec:aci_model}

The Agentic Configuration Item (ACI) is the fundamental abstraction of the ACM reference model. An ACI represents any configurable artifact whose definition contributes to the specification, governance, or execution of an agentic system. By introducing a common abstraction independent of implementation technologies, ACM enables heterogeneous artifacts originating from different frameworks to be represented and governed uniformly.

\begin{figure*}[t]
	\centering
	\includegraphics[width=\textwidth]{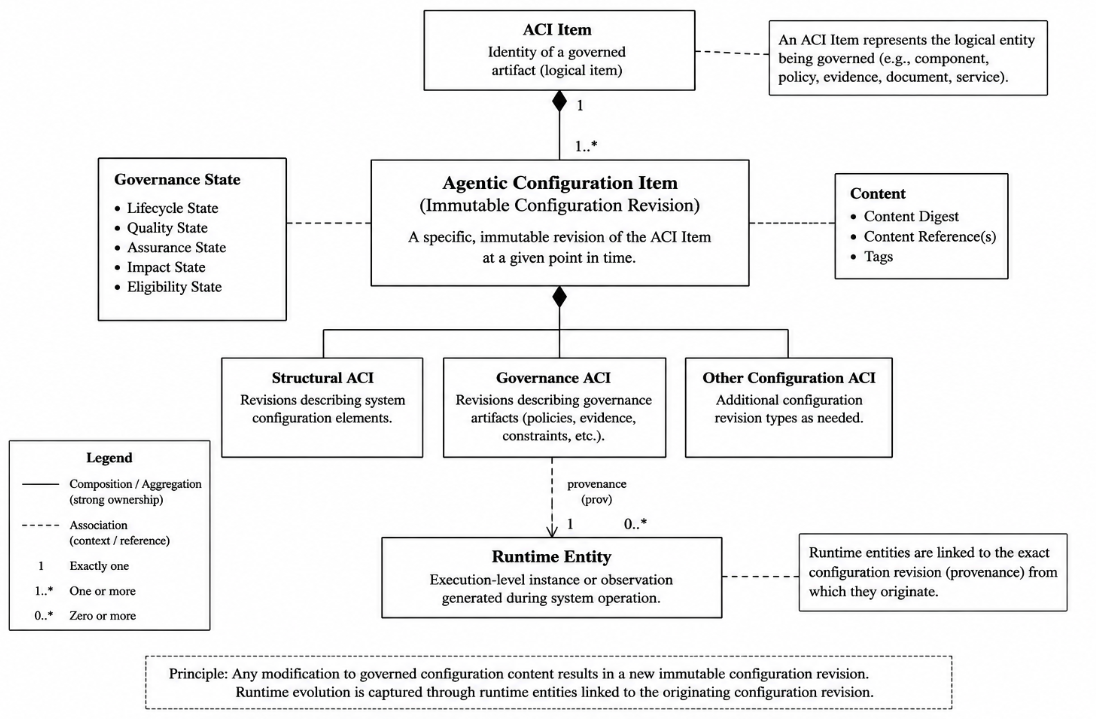}
	\caption{Metamodel of the Agentic Configuration Item (ACI), showing immutable configuration revisions, their specialization, governance metadata, and the association between governed configuration revisions and runtime entities.}
	\label{fig:aci_metamodel}
\end{figure*}

Unlike traditional software configuration items, which primarily describe software artifacts, ACIs encompass the broader range of configurable elements required by modern agentic systems. Examples include agents, prompts, skills, composite subsystems, workflows, tools, language models, policies, assurance evidence, runtime observations, and other governance artifacts. Although these artifacts differ in purpose and lifecycle, they share a common representation within the ACM reference model (see Figure~\ref{fig:aci_metamodel}).

Every configuration ACI is represented through immutable revisions. Once created,
the governed content associated with a revision never changes; any modification
to that content results in the creation of a new revision while preserving
previous versions for reproducibility, auditability, and deterministic
reconstruction of historical configurations. Configuration evolution is
therefore represented explicitly through successive governed revisions rather
than mutable objects.

A configuration ACI combines three complementary dimensions:

\begin{itemize}

\item an immutable configuration content describing the governed artifact;

\item an identity and revision model ensuring uniqueness, version management, and reproducibility;

\item governance metadata describing lifecycle information, provenance, integrity, and other management attributes required for configuration governance.

\end{itemize}

This representation intentionally separates the conceptual identity of a configuration item from the successive revisions describing its evolution. A configuration item therefore exists independently of any particular revision, while each revision represents a complete immutable state of that item at a specific point in its lifecycle.

To accommodate the diversity of governed artifacts encountered in agentic
systems, configuration ACIs specialize the common immutable revision model
according to the role performed by each artifact. Structural ACIs represent
system configuration, whereas Governance ACIs represent governed policies,
evidence, and related governance artifacts. Runtime entities are instantiated
or observed during execution and remain explicitly linked to the exact
configuration revisions from which they originate. Consequently, a change to
governed configuration content produces a new immutable ACI revision, whereas
a runtime-state transition records execution evolution without modifying the
originating revision.

Figure~\ref{fig:aci_metamodel} presents the abstract representation of an Agentic Configuration Item. The following subsections specify its attributes, relationships, and composition rules.

\subsection{ACI Representation}
\label{sec:configuration_item_representation}

The abstract representation introduced in the previous section is realized through a normalized set of attributes shared by all Agentic Configuration Items, regardless of their specialization. This common representation provides a consistent governance model while allowing individual artifact categories to define their own domain-specific content.

The representation of an ACI is organized into four complementary components:

\begin{itemize}

\item \textbf{Identity}, uniquely identifying both the configuration item and the immutable revision to which the representation corresponds.

\item \textbf{Governance state}, describing the lifecycle, quality, assurance,
impact, and eligibility states associated with the revision.

\item \textbf{Configuration content}, containing the governed artifact together with integrity information and extensible metadata.

\item \textbf{Governance metadata}, recording provenance, user-defined properties, and additional information required for lifecycle management and auditing.

\end{itemize}

This separation isolates the immutable content of a configuration revision from the metadata required to govern its evolution. Consequently, governance information may evolve independently from the business semantics of individual artifact categories while preserving the integrity of immutable revisions.

Table~\ref{tab:aci_attributes} summarizes the attributes composing the abstract representation of an ACI.

\begin{table*}[t]
\centering
\caption{Abstract attributes of an Agentic Configuration Item revision.}
\label{tab:aci_attributes}

\renewcommand{\arraystretch}{1.15}

\begin{tabular}{p{3.3cm}p{3.4cm}p{8.3cm}}

\toprule

\textbf{Category} &
\textbf{Representative Attributes} &
\textbf{Purpose}

\\
\midrule

Identity &
item\_id, revision\_id &
Uniquely identify the governed item and its immutable revision.

\\

Governance State &
lifecycle\_state, quality\_state, assurance\_state, impact\_state, eligibility\_state &
Describe the governance status of the revision.

\\

Configuration Content &
content, digest, metadata &
Represent the governed artifact together with integrity and extensible descriptive information.

\\

Governance Metadata &
provenance, tags, custom properties &
Support traceability, auditing, provenance, and organization-specific extensions.

\\

\bottomrule

\end{tabular}

\end{table*}

Although configuration ACIs share this common representation, the interpretation
of their governed content depends on their specialization. Runtime entities
remain distinct execution-level objects linked to the governed configuration
from which they originate.

\subsection{Relationship Model}
\label{sec:configuration_relationships}

Individual configuration items do not exist in isolation. The complete configuration of an agentic system emerges from the explicit relationships connecting governed artifacts across structural, governance, and runtime perspectives. These relationships constitute the semantic backbone of the ACM reference model by preserving dependency information independently of any execution framework.

Unlike conventional orchestration frameworks, where relationships are often embedded within implementation-specific structures, ACM represents relationships as explicit governed entities. This representation enables dependency analysis, impact assessment, provenance tracking, and configuration validation to be performed independently of execution technologies.

ACM distinguishes several categories of relationships according to their governance semantics:

\begin{itemize}

\item \textbf{Configuration relationships} describe the composition and dependency structure of governed configuration items.

\item \textbf{Assurance relationships} associate configuration items with policies, assurance evidence, compliance requirements and ownership.

\item \textbf{Runtime relationships} connect execution observations to the governed configuration from which they originate.

\item \textbf{Evolution relationships} preserve provenance, revision lineage, and derivation history.

\end{itemize}

Each relationship is explicitly typed and directional. Relationship types define their admissible source and target configuration items together with the semantic constraints governing their interpretation. This explicit typing allows validation rules to be evaluated independently from artifact implementations while ensuring consistent dependency semantics across heterogeneous agentic systems.

\begin{figure*}[t]
	\centering
	\includegraphics[width=0.7\textwidth]{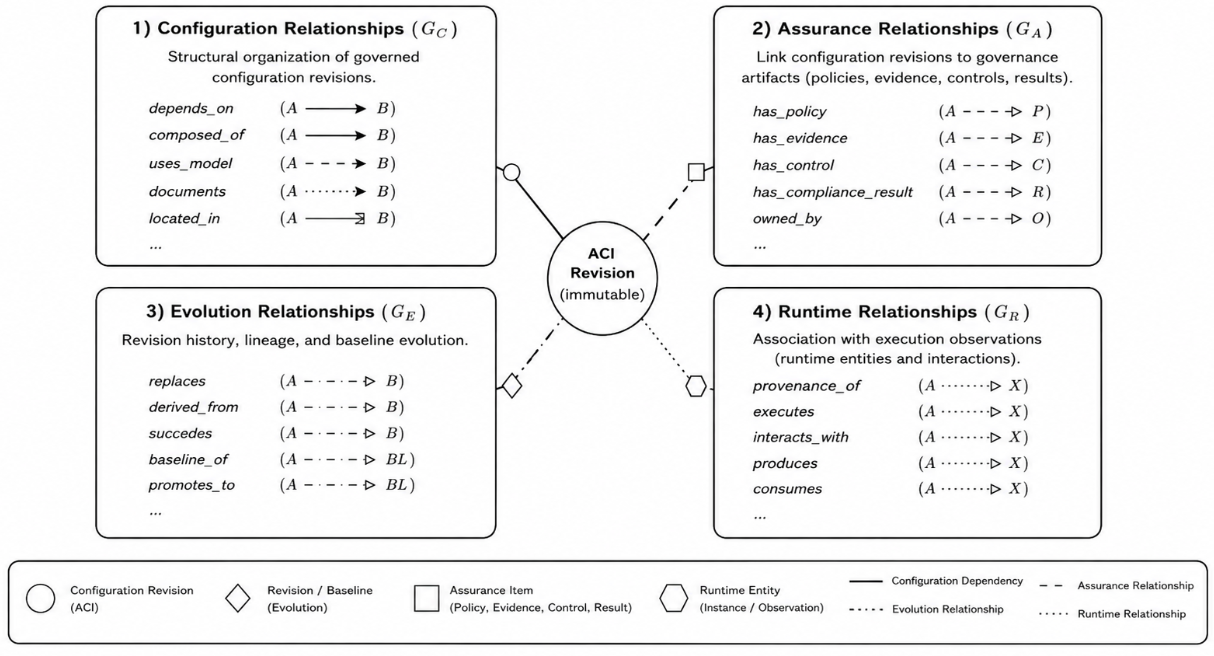}
	\caption{Taxonomy of relationship types in ACM. Relationships are classified
		according to the governance concern represented by the Configuration,
		Evolution, Assurance, and Runtime Graphs. The relation types shown are
		illustrative and not exhaustive.}
	\label{fig:relationship_model}
\end{figure*}

Figure~\ref{fig:relationship_model} illustrates the conceptual organization of relationship categories within the ACM reference model.

\subsection{Release Baselines}
\label{sec:release_baselines}

While Agentic Configuration Items represent individual governed artifacts, operational deployments require a consistent representation of complete system configurations. ACM introduces the notion of a \emph{Release Baseline} to capture a coherent and reproducible snapshot of an agentic system at a specific point in its evolution.

A Release Baseline is an immutable collection of ACI revisions that collectively define the governed configuration of an agentic system. Rather than referencing configuration items independently of their evolution, a baseline always references specific immutable revisions, ensuring that the complete configuration can be reconstructed \emph{deterministically} at any time.

The purpose of a Release Baseline extends beyond simple version aggregation. It establishes the governance boundary within which configuration consistency, traceability, compliance, and assurance can be evaluated. Consequently, governance decisions are performed on complete configurations rather than on isolated artifacts whenever system-wide consistency is required.

A Release Baseline is characterized by the following properties:

\begin{itemize}

\item it references immutable revisions rather than mutable configuration items;

\item it represents a complete and internally consistent governed configuration;

\item it preserves the configuration relationships and governance information
required to reconstruct and assess the released configuration, while runtime
observations remain linked to, but distinct from, the baseline;

\item it constitutes the primary unit for reproducibility, audit, release management, and compliance assessment.

\end{itemize}

Baselines are themselves governed configuration objects. They possess their own identity, lifecycle, provenance, and governance metadata while remaining immutable after publication. The evolution of a system is therefore represented as a sequence of immutable Release Baselines, each corresponding to a governed configuration state.

Because every baseline references immutable ACI revisions, historical configurations remain reproducible even when individual configuration items continue to evolve. This property enables deterministic reconstruction, historical audits, rollback, and longitudinal analysis without modifying previously released configurations.

\begin{figure*}[t]
	\centering
	\includegraphics[width=\textwidth]{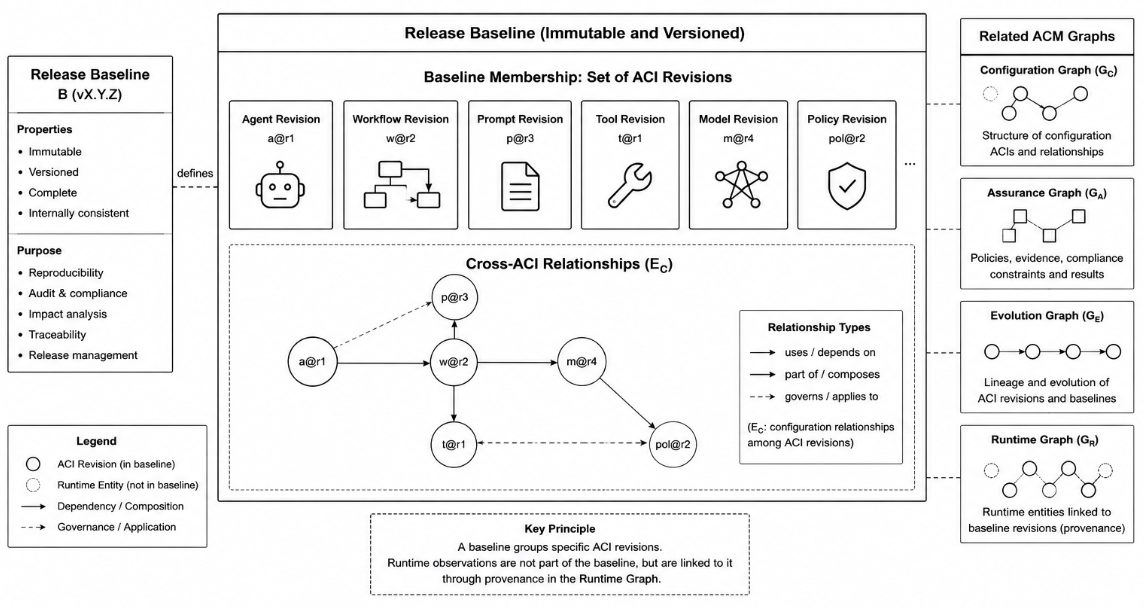}
	\caption{Organization of an ACM Release Baseline. A baseline identifies an
		immutable set of exact ACI revisions composing a governed release, while
		cross-ACI relationships preserve the structural dependencies required for
		traceability and impact analysis.}
	\label{fig:release_baseline}
\end{figure*}

Figure~\ref{fig:release_baseline} illustrates the conceptual organization of a Release Baseline and its relationship with the governed configuration items composing an agentic system.

\section{Governance Semantics}
\label{sec:governance_semantics}

The ACM reference model specifies the structural organization of governed agentic systems through the Configuration, Evolution, Assurance, and Runtime Graphs. This section complements the metamodel by defining the governance semantics associated with these structures while preserving their separation of concerns.

The governance semantics are intentionally defined independently from any execution framework. Configuration-state evaluation and dependency propagation operate on the Configuration Graph, whereas runtime observations remain represented separately through the Runtime Graph.

Let

\begin{equation}
	G_C=(V_C,E_C)
	\label{eq:configuration_graph}
\end{equation}

denote the ACM Configuration Graph, where $V_C$ is the finite set of immutable ACI revisions represented in the governed configuration and
$E_C$ is the finite set of typed configuration relationships between these revisions.

Each configuration revision $v\in V_C$ is associated with a configuration governance descriptor

\begin{equation}
	\Gamma_C(v)=
	\left(
	L(v),
	Q(v),
	A(v),
	I(v),
	El(v)
	\right),
	\label{eq:configuration_governance_descriptor}
\end{equation}

which aggregates the lifecycle, quality, assurance, impact, and eligibility states associated with the governed configuration revision.

Runtime information is represented separately through the runtime view

\begin{equation}
	\rho(v)
	=
	\left\{
	(x,Rt(x))
	\mid
	x\in V_R,\;
	prov(x)=v
	\right\},
	\label{eq:runtime_view}
\end{equation}

where $V_R$ denotes the runtime entities represented in the Runtime Graph and $prov:V_R\rightarrow V_C$ associates each runtime entity with the exact configuration revision from which it originates.

The complete governance view associated with a configuration revision is therefore

\begin{equation}
	\Gamma(v)
	=
	\left(
	\Gamma_C(v),
	\rho(v)
	\right).
	\label{eq:complete_governance_descriptor}
\end{equation}

Neither $\Gamma_C$, $\rho$, nor $\Gamma$ introduces computational semantics. Configuration governance states are evaluated according to the semantics defined below, whereas runtime states are reconstructed from execution observations.

This separation between configuration governance, runtime observations, evaluation functions, and propagation semantics ensures that each semantic component is defined once and at the appropriate modeling level.

This separation also defines the boundary between framework-specific
projection and framework-independent governance processing. For a framework $F$, semantic projection maps an extracted native model $M_F$ to the normalized Configuration Graph $G_C$ through $\mathrm{Proj}_F$. Once $G_C$ has been obtained, the governance semantics defined in this section operate exclusively on this normalized representation. The resulting processing chain is therefore
$M_F \xrightarrow{\mathrm{Proj}_F} G_C
\xrightarrow{\text{governance semantics}}
(\iota^*,\Gamma_C,\text{governance outcomes})$.
Framework-specific interpretation is confined to the projection boundary;
equivalent governance outcomes therefore do not require native configurations or their projections to be informationally identical, but require preservation of the governance-relevant information consumed by the ACM semantics.

\subsection{Governance Lifecycle}
\label{sec:lifecycle_semantics}

The lifecycle dimension represents the intrinsic governance maturity of an immutable ACI revision independently of any dependency or runtime consideration. Lifecycle states are explicitly assigned through governance actions, whereas the other configuration governance dimensions are evaluated or derived according to their respective semantics.

Each revision progresses monotonically through the lifecycle. Once a revision reaches a given state, it cannot transition back to a previous lifecycle state.

This monotonic property preserves lifecycle consistency for each immutable revision while maintaining governance traceability across successive revisions.

Lifecycle transitions are therefore constrained by a partial order

\begin{equation}
	Draft
	<
	Validated
	<
	Approved
	<
	Released
	<
	Deprecated
	<
	Archived,
	\label{eq:lifecycle_order}
\end{equation}

where each transition represents an explicit governance decision.

The lifecycle state is denoted

\begin{equation}
	L(v)\in\mathcal{L},
	\label{eq:lifecycle_state}
\end{equation}

where $\mathcal{L}$ denotes the lifecycle state space formally defined in Appendix~\ref{appendix:state_spaces}.

Lifecycle states never result from dependency propagation. They constitute primary governance information subsequently used, together with the other configuration governance states, in local eligibility evaluation.

\subsection{Governance State Spaces}
\label{sec:governance_state_spaces}

The governance semantics distinguish five configuration governance state spaces associated with immutable ACI revisions and one runtime state space associated with runtime entities. These state spaces represent complementary semantic dimensions but operate at two distinct levels of the ACM reference model.

Within the configuration governance descriptor, lifecycle is explicitly assigned through governance decisions, while quality, assurance, impact, and eligibility are evaluated according to their respective semantics. Runtime state is reconstructed separately from execution observations in the Runtime Graph.

Figure~\ref{fig:governance_state_machine} summarizes the configuration
governance state dimensions used throughout the remainder of this section.
Runtime states are defined separately on runtime entities.

The configuration governance dimensions are:

\begin{itemize}
	\item \textbf{Lifecycle ($\mathcal{L}$):} governance maturity of a revision;
	\item \textbf{Quality ($\mathcal{Q}$):} validation status of the revision itself;
	\item \textbf{Assurance ($\mathcal{A}$):} completeness of governance evidence;
	\item \textbf{Impact ($\mathcal{I}$):} propagation status resulting from dependency analysis;
	\item \textbf{Eligibility ($\mathcal{E}$):} authorization for operational use.
\end{itemize}

The runtime state space $\mathcal{R}$ instead describes the observed execution
state of runtime entities in $V_R$. The five configuration governance state
spaces and the distinct runtime state space are formally defined in
Appendix~\ref{appendix:state_spaces}; their respective evaluation,
propagation, and reconstruction semantics are specified separately.

\subsection{Propagation Policies}
\label{sec:propagation_policies}

Governance propagation is intentionally separated from state evaluation. While governance state spaces define \emph{what} is evaluated, propagation policies determine how configuration relationships are interpreted for governance propagation. They are represented as governance policies in the Assurance Graph $G_A$ and govern relationships of the Configuration Graph $G_C$ without modifying its topology.

Each configuration relationship is governed by a propagation policy determined by its relationship type. The policy assignment is represented in the Assurance Graph $G_A$ and remains independent of the current governance states of the connected revisions.

A propagation policy is defined as a mapping

\begin{equation}
	\Pi:E_C\rightarrow\mathcal P,
	\label{eq:policy_mapping}
\end{equation}

where $\Pi$ is the propagation-policy view resolved from the governance policies represented in $G_A$. For each configuration relationship $e\in E_C$, $\Pi(e)$ denotes the propagation policy governing that relationship.

\begin{equation}
	\mathcal{P}=
	\left\{
	\textit{Blocking},
	\textit{Warning},
	\textit{Informational},
	\textit{None}
	\right\}.
	\label{eq:policy_space}
\end{equation}

A propagation policy is declarative: it does not itself modify an impact state. Instead, it selects the impact-transfer semantics applied when propagation traverses a governed configuration relationship.

\begin{itemize}
	\item \textbf{Blocking} indicates that an impact affecting a revision shall be
	propagated across the governed relationship to dependent revisions. The resulting impact may subsequently restrict operational eligibility until reassessment is completed.
	
	\item \textbf{Warning} propagates an impact indication without directly
	determining operational eligibility. The resulting impact records that the
	dependent revision requires governance attention.
	
	\item \textbf{Informational} preserves the governed dependency for traceability
	without contributing a propagated impact.
	
	\item \textbf{None} disables propagation across the relationship while preserving the structural dependency.
\end{itemize}

In the current ACM semantics, the policy assigned to a relationship is determined by its relationship type. Consequently, relationships of the same type induce the same propagation policy independently of their endpoint revisions.

\begin{equation}
	type(e_1)=type(e_2)
	\Longrightarrow
	\Pi(e_1)=\Pi(e_2).
	\label{eq:policy_type_consistency}
\end{equation}

When several propagation paths converge on the same revision, multiple propagation policies may simultaneously apply. Their effective policy is resolved
deterministically according to the precedence rules defined in Appendix~\ref{appendix:propagation_policies}. Policy composition remains distinct
from the aggregation of propagated impact contributions defined by the propagation semantics.

\subsection{Governance Evaluation Functions}
\label{sec:evaluation_functions}

The governance state spaces introduced previously define the possible states associated with an ACI revision. Governance evaluation functions determine how these states are assigned from governance evidence while remaining independent of dependency propagation.

Three evaluation functions are defined. They respectively evaluate quality,
governance assurance, and operational eligibility. Quality and assurance are evaluated locally and independently of impact propagation, whereas eligibility is evaluated locally after the impact valuation has
stabilized.

ACM specifies the interfaces and semantic roles of the governance evaluation functions rather than universal domain-specific decision rules. For a fixed governance context, these functions are deterministic local mappings over the state spaces defined above.

\subsubsection{Quality Evaluation}

The quality evaluation function assigns the quality state associated with an ACI revision according to the applicable quality criteria in the governance context.

\begin{equation}
	f_{\mathrm{quality}} : V_C \rightarrow \mathcal{Q},
	\label{eq:f_quality}
\end{equation}

where $V_C$ denotes the set of immutable ACI revisions represented in the Configuration Graph and $\mathcal{Q}$ is the quality state space introduced in Section~\ref{sec:governance_state_spaces}.

The concrete quality criteria are supplied by the deployed governance context; ACM requires only that the resulting evaluation be deterministic for a fixed configuration and fixed evaluation context.

The resulting quality state is denoted

\begin{equation}
	Q(v)=f_{\mathrm{quality}}(v).
	\label{eq:quality_assignment}
\end{equation}

Quality evaluation depends exclusively on the considered revision and is therefore independent of dependency propagation.

\subsubsection{Assurance Evaluation}

The assurance evaluation function determines the completeness of the governance
evidence required for a revision under the applicable governance context.

\begin{equation}
	f_{\mathrm{assurance}} : V_C \rightarrow \mathcal{A},
	\label{eq:f_assurance}
\end{equation}

where $\mathcal{A}$ denotes the assurance state space.

The concrete assurance criteria are supplied by the deployed governance context; ACM specifies the evaluation interface and resulting assurance state but does not prescribe domain-specific evidence requirements.

The resulting assurance state is

\begin{equation}
	A(v)=f_{\mathrm{assurance}}(v).
	\label{eq:assurance_assignment}
\end{equation}

Assurance reflects the completeness of governance evidence independently of the intrinsic quality of the revision.

\subsubsection{Eligibility Evaluation}

Operational eligibility depends on the combined interpretation of the governance dimensions. Unlike quality and assurance, eligibility is therefore derived from multiple governance states.

The eligibility evaluation function is defined as

\begin{equation}
	\label{eq:eligibility_function}
	f_{\mathrm{elig}}
	:
	\mathcal{L}
	\times
	\mathcal{Q}
	\times
	\mathcal{A}
	\times
	\mathcal{I}
	\rightarrow
	\mathcal{E}
\end{equation}

The eligibility state associated with revision $v$ is therefore

\begin{equation}
	El(v)=
	f_{\mathrm{elig}}
	\left(
	L(v),
	Q(v),
	A(v),
	I(v)
	\right).
	\label{eq:eligibility_assignment}
\end{equation}

\begin{figure*}[t]
	\centering
	\includegraphics[width=0.7\textwidth]{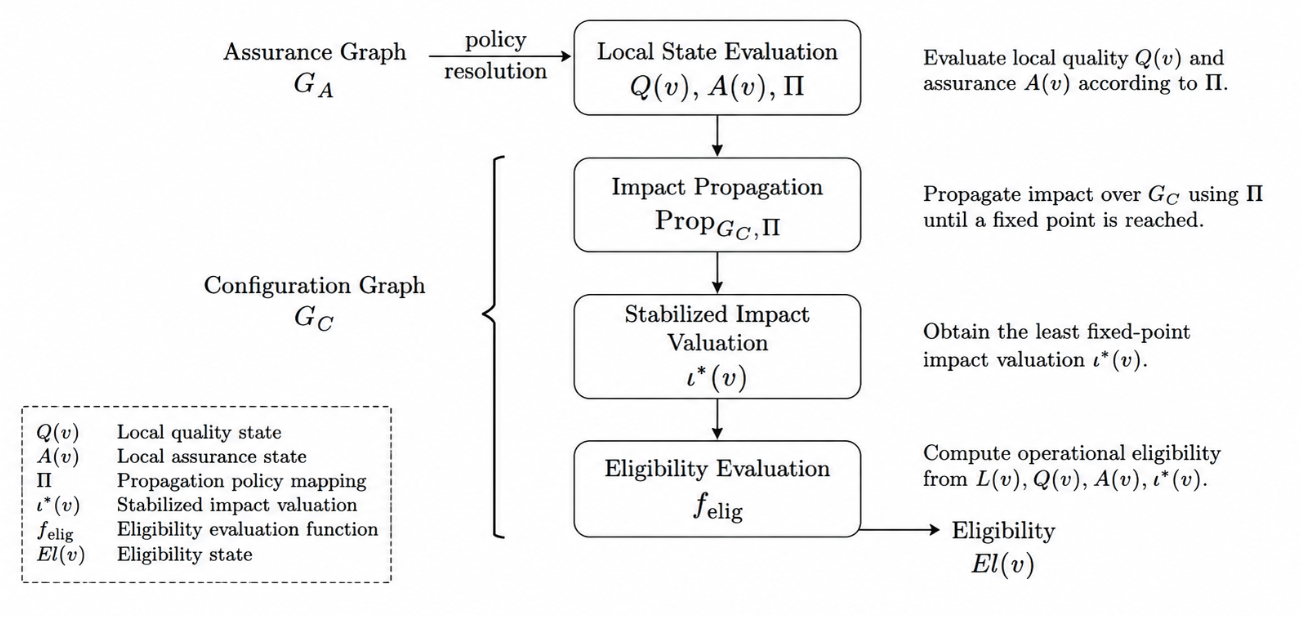}
	\caption{Governance evaluation flow. Local quality and assurance states are evaluated independently of impact propagation. The propagation operator computes the stabilized impact valuation over the Configuration Graph using the resolved propagation-policy mapping, after which eligibility is evaluated from lifecycle, quality, assurance, and stabilized impact states.}
	\label{fig:governance_evaluation_flow}
\end{figure*}

Eligibility summarizes the governance status required for operational use (see Figure~\ref{fig:governance_evaluation_flow} for more details). The evaluation remains local to the considered revision and is performed after impact propagation has stabilized. Propagation policies therefore affect eligibility only indirectly through the stabilized impact state and are not arguments of $f_{\mathrm{elig}}$.

For a fixed governance context, $f_{\mathrm{elig}}$ is total, deterministic, and local. ACM does not prescribe a universal mapping from lifecycle, quality, assurance, and impact states to eligibility; the concrete mapping is supplied by the deployed governance context.

\subsection{Propagation Semantics}
\label{subsec:propagation_semantics}

The governance dimensions introduced in the previous subsections define the local state associated with each ACI revision. Configuration dependencies may, however, propagate the impact of a local change throughout the Configuration Graph. The propagation semantics therefore define how impact valuations are iteratively updated until a stable valuation is reached.

Propagation follows the directed relationships of the configuration graph and is controlled by the propagation policies introduced in
Section~\ref{sec:propagation_policies}. These policies determine how impact is transferred across each governed relationship, including whether an impact is propagated or suppressed according to the applicable policy.

The propagation process is defined by a global operator, denoted
$\widehat{\mathrm{Prop}}_{G_C,\Pi}$, which applies the resolved propagation policies over the fixed Configuration Graph to compute the propagated impact valuation. Only impact states evolve during propagation; $G_C$ and $\Pi$ remain unchanged throughout the computation. Once the impact valuation has stabilized, eligibility is evaluated locally as defined in the previous subsection.

Formally, let

\begin{equation}
	\label{eq:prop_operator}
	\widehat{\mathrm{Prop}}_{G_C,\Pi}
	:
	\mathcal{D}_{G_C}
	\rightarrow
	\mathcal{D}_{G_C}.
\end{equation}

where $\mathcal{D}_{G_C}=\mathcal{I}^{V_C}$ denotes the impact-valuation domain associated with the fixed Configuration Graph $G_C$, while $G_C$ and the propagation-policy mapping $\Pi$ remain constant throughout one propagation computation.

Starting from the initial impact valuation $\iota^{(0)}$, successive
applications of the propagation operator generate the sequence

\begin{equation}
	\label{eq:prop_iteration}
	\iota^{(k+1)}
	=
	\widehat{\mathrm{Prop}}_{G_C,\Pi}
	\!\left(
	\iota^{(k)}
	\right).
\end{equation}

where $k$ denotes the propagation iteration applied to the impact valuation.

Propagation terminates when the impact valuation becomes stable, that is, when an additional application of the propagation operator produces no further change.

\begin{equation}
	\label{eq:fixed_point}
	\iota^{*}
	=
	\widehat{\mathrm{Prop}}_{G_C,\Pi}
	\!\left(
	\iota^{*}
	\right).
\end{equation}

The stabilized valuation $\iota^{*}$ is the least fixed point of the
propagation semantics above the initial valuation $\iota^{(0)}$ and assigns the final stabilized impact state to each configuration revision.

Because impact propagation is performed before eligibility evaluation, the eligibility of each revision remains a purely local computation. Once the impact valuation has stabilized, eligibility is evaluated from the lifecycle, quality, assurance, and stabilized impact states as defined in Section~\ref{sec:evaluation_functions}.

The propagation semantics are independent of any implementation strategy. A compliant implementation may therefore employ different execution algorithms, provided that they compute the same least fixed point above the initial impact valuation. The reference implementation adopts a worklist-based evaluation strategy because it avoids re-evaluating unaffected revisions while preserving the normative semantics. The complete worklist algorithm, together with the proofs of monotonicity, termination

\subsubsection{Unified Governance State Diagram}

\begin{figure*}[t]
	\centering
\includegraphics[width=0.8\textwidth]{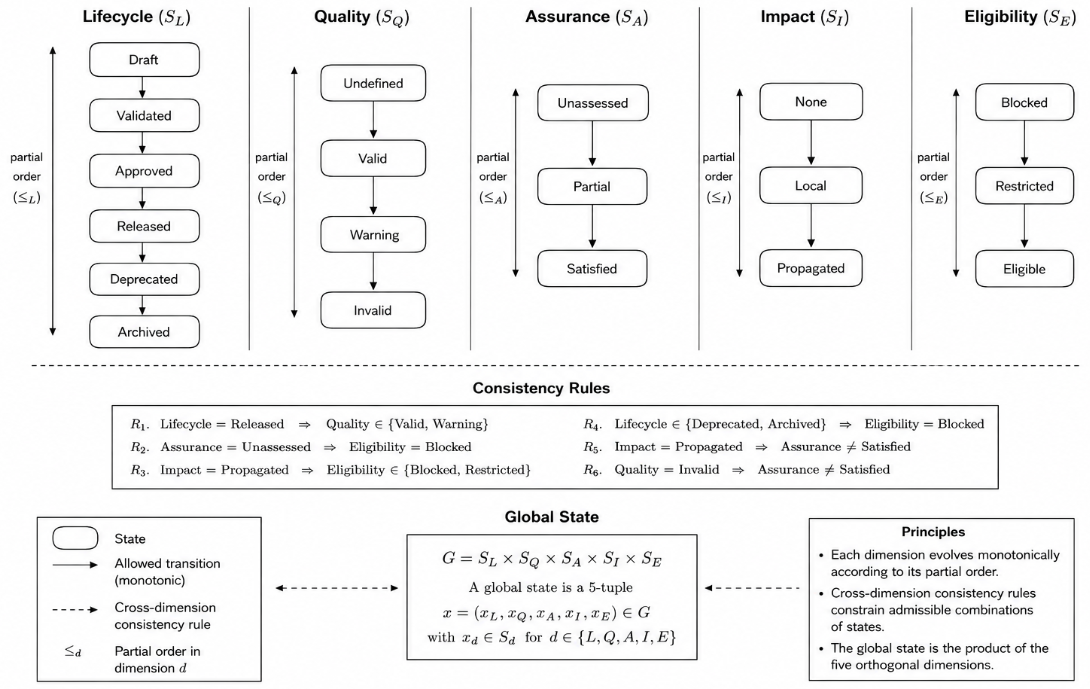}
	
	\caption{Four configuration governance state spaces---lifecycle $L$, quality $Q$, assurance $A$, and impact $I$---characterize the governance status of the revision. The eligibility state $El$ is derived locally from these four states by a deterministic evaluation function $f_{elig}$. Runtime states ($R_t \in \{$Created, Ready, Running, Waiting, Completed, Failed, Cancelled, Terminated$\}$) are defined runtime entities ($V_R$) and are orthogonal to the configuration governance semantics.}
	\label{fig:governance_state_machine}
\end{figure*}

Figure~\ref{fig:governance_state_machine} illustrates the unified governance state machine for an ACI revision $v\in V_C$.

\subsubsection{Transition Properties}
\label{subsec:transition_properties}
The governance state machines satisfy the following properties.

\paragraph{Property G1 (Determinism).}

For fixed inputs and governance context, each governance semantic function or operator produces a uniquely determined result.

\paragraph{Property G2 (Orthogonality).}

The governance dimensions remain semantically distinct and are represented in separate state spaces. Lifecycle, quality, assurance, impact, and eligibility belong to the configuration governance descriptor, while runtime state is defined separately on runtime entities.

\paragraph{Property G3 (Composability).}

The global governance state is obtained through the composition of independent semantic dimensions rather than by constructing a single global automaton.

\paragraph{Property G4 (Extensibility).}

Additional governance dimensions may be integrated through explicitly defined state spaces and evaluation semantics, while preserving the separation of concerns among existing governance dimensions.

These properties provide a structured basis for extending ACM while
preserving the semantic separation of its governance dimensions.

The next subsection formalizes how local governance changes propagate through dependency relationships, enabling global reasoning over complete agentic configuration graphs.

\begin{figure*}[t]
	\centering
	\includegraphics[width=0.6\textwidth]{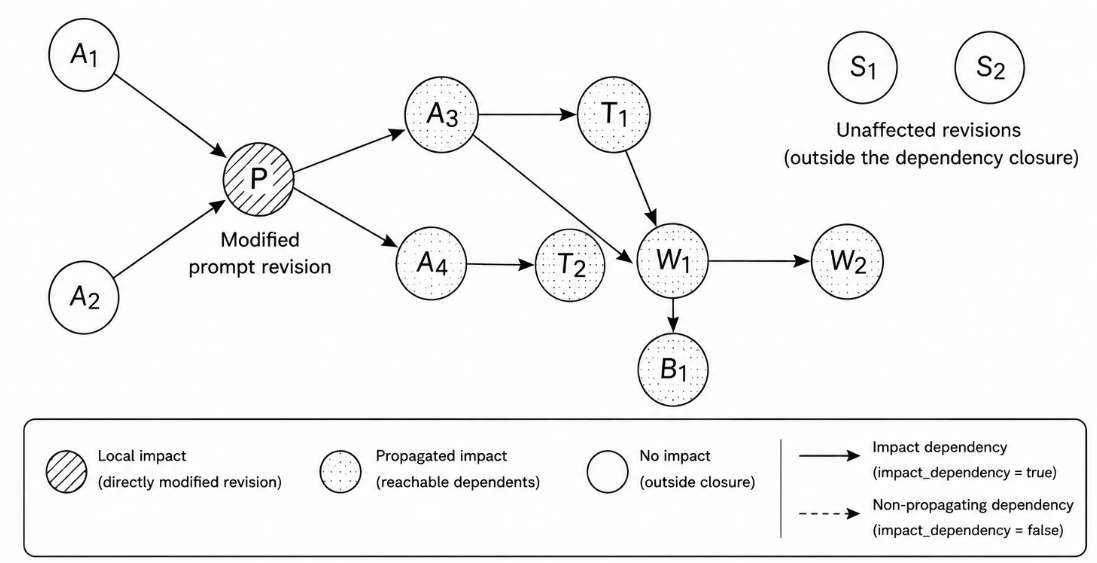}
	\caption{
		Impact propagation following the modification of prompt revision $P$.
		The directly modified revision receives the \textsc{Local} impact state,
		reachable dependent revisions receive the \textsc{Propagated} state, and
		revisions outside the dependency closure remain in the \textsc{None} state.
	}
	\label{fig:impact-propagation}
\end{figure*}

Figure~\ref{fig:impact-propagation} illustrates the propagation of a local configuration change through governance dependencies. The directly modified revision receives the \textit{Local} impact state, while reachable dependent revisions receive the \textit{Propagated} state. Revisions outside the resulting impact closure remain in the \textit{None} state.

\subsubsection{Illustrative Example}

Assume that a workflow

\[
W
\]

contains an agent

\[
A_1
\]

whose prompt revision is modified. The prompt immediately becomes

\[
Local.
\]

The dependent configuration revisions, including the associated agent and
workflow, receive the $Propagated$ impact state because they depend on the modified prompt revision.

Importantly, propagation alone does not imply execution failure.

It merely indicates that governance evaluation should be recomputed before the corresponding artifacts are considered trustworthy.


This fixed-point formulation establishes the formal semantics of governance propagation independently of any particular implementation. Whether ACM is implemented through recursive graph traversal, work-list evaluation, or incremental update algorithms, every compliant implementation is required to compute the same least fixed point above the initial impact valuation, thereby guaranteeing deterministic,
reproducible, and implementation-independent governance evaluation.


\subsection{Runtime Semantics}
\label{subsec:runtime_semantics}

The governance semantics defined so far describe governance states associated
with immutable revisions in the Configuration Graph. Runtime execution
introduces an additional dimension that cannot be represented solely by static
configuration artifacts. Agent executions create transient observations whose
purpose is not to modify the governed configuration but to record how a released
baseline is instantiated during execution.

Accordingly, ACM distinguishes the \emph{Configuration Graph} $G_C$, which
contains immutable governed revisions, from the \emph{Runtime Graph} $G_R$,
which contains the runtime entities $V_R$ reconstructed for a specific
execution. Runtime observations never alter immutable configuration revisions.
Runtime state is assigned to runtime entities through
$Rt:V_R\rightarrow\mathcal R$, while provenance relates these entities to their
originating configuration revisions through $prov:V_R\rightarrow V_C$.

The Runtime Graph is constructed from an ordered sequence of normalized runtime
events. Each event captures an observable execution transition affecting a
runtime entity together with the configuration provenance required to relate
that entity to its originating governed revision. Because runtime observations
are represented independently from framework-specific execution traces, the same
runtime semantics can be applied to heterogeneous execution frameworks after
normalization into the ACM runtime representation.

Rather than storing successive runtime snapshots, ACM reconstructs runtime states
by replaying the normalized event sequence with respect to a released baseline.
The baseline provides the immutable configuration reference for the initial
Runtime Graph, after which runtime events are deterministically applied in their
defined total order to reconstruct the Runtime Graph corresponding to any
observation point.

This separation provides two complementary benefits. First, released
configurations remain immutable throughout execution, preserving configuration
integrity and provenance. Second, runtime reconstruction becomes independent of
the execution framework, provided that equivalent normalized events are
available. Consequently, runtime analyses, governance verification, and
conformance checking operate on a canonical Runtime Graph rather than on
framework-specific execution representations.

The replay semantics are intentionally defined independently of the propagation
operator introduced in Section~\ref{subsec:propagation_semantics}. Propagation
computes the stabilized impact valuation over immutable configuration revisions,
whereas replay reconstructs the execution state generated from a released
baseline. These two mechanisms therefore address complementary concerns: governance evaluation determines the governance states and operational eligibility of the governed configuration, while runtime replay reconstructs how the released configuration was actually executed.

The formal definition of normalized runtime events, the replay function, the
incremental reconstruction procedure, and the associated correctness properties
are presented in Appendix~\ref{appendix:runtime_semantics}.

\subsection{Formal Properties}
\label{sec:formal_properties}

The governance semantics defined in the previous subsections satisfy a number of formal properties that ensure the consistency, reproducibility, and predictability of ACM evaluations. These properties follow directly from the state spaces, evaluation functions, propagation semantics, and runtime replay model introduced throughout Section~\ref{sec:governance_semantics}.

First, the impact propagation semantics are monotone over the finite impact lattice defined in Appendix~\ref{appendix:finite_lattice}. Consequently, iterative propagation converges toward the least fixed point above the initial impact valuation, while the reference worklist algorithm computes the same stabilized result under any fair evaluation strategy. The corresponding proofs are provided in Appendix~\ref{appendix:formal_properties}.

Second, runtime replay is deterministic. Given an immutable released baseline and an identical ordered sequence of normalized runtime events, replay always reconstructs the same Runtime Graph. This property guarantees that runtime observations remain reproducible independently of the underlying execution framework once configuration projection and runtime-event normalization have produced the corresponding ACM representations. The formal replay model and proof of determinism are presented in Appendix~\ref{appendix:runtime_semantics}.

Together, these properties establish reproducibility of governance evaluation and runtime reconstruction for a fixed governance context. For identical governed configurations, identical propagation policies, identical evaluation contexts, and equivalent normalized runtime observations, ACM produces the same
stabilized impact valuation, the same locally evaluated governance states, and the same reconstructed Runtime Graph. These properties provide the foundation for reproducible governance verification, impact analysis, runtime conformance checking, and assurance evaluation across heterogeneous agentic frameworks.

The complete proofs, together with the complexity analysis and additional formal results, are presented in Appendix~\ref{appendix:complexity_analysis}.

\section{Operationalization}
\label{sec:operationalization}

The purpose of this section is not to define another execution framework but to demonstrate that the proposed reference model is operationalizable without introducing framework-specific concepts into its governance semantics. The operationalization deliberately separates governance semantics from implementation details. The ACM metamodel, lifecycle semantics, and validation rules are independent of programming language, persistence technology, communication protocol, or execution framework. The Python implementation therefore serves only as one concrete realization of the reference model, demonstrating implementability rather than prescribing a specific software architecture.

\subsection{Reference Architecture}
\label{subsec:reference_architecture}

The ACM reference model is intentionally independent of any particular agentic framework. Its operationalization therefore adopts a layered architecture that cleanly separates framework-specific extraction from framework-independent governance. This separation preserves the conceptual independence of the reference model while allowing heterogeneous agentic systems to be represented through a common governed configuration.

In our reference architecture, framework-specific adapters extract native
configuration constructs and project them into the canonical Configuration
Graph $G_C$. Subsequent configuration validation and governance evaluation
operate on this normalized graph, while runtime reconstruction separately
integrates normalized execution observations into $G_R$.

The governance kernel constitutes the core of the operationalization. It implements the normative semantics introduced in Section~\ref{sec:governance_semantics}, including structural validation, lifecycle management, governance state evaluation, impact propagation, assurance evaluation, and eligibility assessment. Since these operations manipulate only canonical ACM representations, their behavior remains independent of the execution framework.

The runtime interface complements the governance kernel by exposing governed configurations together with normalized execution observations. Runtime events enrich the operational view of the system without modifying immutable configuration revisions, thereby preserving the separation between configuration governance and execution behavior.

The reference implementation should not be interpreted as a production platform or a preferred execution environment. Its purpose is to demonstrate that the ACM reference model can be operationalized through a framework-independent governance kernel while allowing the evaluated execution frameworks to be governed through the same ACM semantics after projection.

The operationalization is organized into five successive stages:

\begin{enumerate}
	\item \textbf{Framework Extraction}, which retrieves governance-relevant native configuration artifacts exposed by the execution framework;
	
	\item \textbf{Semantic Projection}, which maps these native artifacts onto ACM concepts through framework-specific projection rules;
	
	\item \textbf{Canonical Normalization}, which constructs immutable ACI revisions, resolves references, assigns stable identities, computes canonical digests, and builds the governed configuration graph;
	
	\item \textbf{Governance Processing}, which applies the framework-independent governance semantics defined in Section~\ref{sec:governance_semantics};
	
	\item \textbf{Governed Representation}, which exposes the validated and governed
	Configuration Graph for auditing, impact analysis, lifecycle management, and
	subsequent runtime integration.
\end{enumerate}

This organization deliberately isolates framework-dependent concerns from governance semantics. Native frameworks remain responsible for execution, whereas ACM provides a common governance layer operating on a unified configuration representation.

\subsection{Projection Pipeline}
\label{subsec:semantic_projection_pipeline}

The reference architecture introduced in Section~\ref{subsec:reference_architecture} is realized through a semantic projection pipeline, including canonical normalization, that transforms heterogeneous framework configurations into ACM representation. Rather than reproducing the internal execution model of each framework, the pipeline identifies only the constructs carrying governance significance and projects them onto the concepts defined by the ACM reference model.

The projection process consists of four successive stages:

\begin{enumerate}
	\item native structure extraction;
	\item semantic classification;
	\item canonical normalization;
	\item projection validation and coverage assessment.
\end{enumerate}

This pipeline establishes a common projection contract shared by all framework
adapters. Although extraction mechanisms differ across execution frameworks,
each adapter ultimately produces a normalized ACM Configuration Graph $G_C$,
enabling the governance kernel to operate independently of framework-specific
implementation details.

The complete projection formalism is provided in Appendix~\ref{appendix:projection_formalism}.

\subsubsection{Framework-specific Instantiations}

The prototype implements this projection contract through dedicated adapters for LangGraph, CrewAI, and the OpenAI Agents SDK. These adapters differ only in the extraction phase, reflecting the distinct introspection mechanisms provided by each framework, while sharing the same normalization rules and governance semantics.

For each adapter, the evaluation reports:

\begin{itemize}
	\item the number of governance-relevant native entities;
	\item the number of completely, partially, and unsuccessfully projected entities;
	\item the number of governance-relevant native relationships;
	\item the corresponding entity, relationship, aggregate, and strict coverage scores.
\end{itemize}

The measured values are reported in Section~\ref{sec:evaluation}. They are computed from the explicit inventory of governance-relevant native constructs defined by the experimental protocol rather than inferred from the mere existence of a functioning adapter. Consequently, projection coverage evaluates the expressiveness of the ACM reference model independently of the implementation itself.

\subsection{Framework Adapters}
\label{sec:framework_adapters}

The semantic projection pipeline is realized through framework-specific adapters responsible for extracting governance-relevant configuration artifacts from native execution environments. Although the extraction mechanisms differ across frameworks, every adapter produces a normalized ACM Configuration Graph before governance processing begins. Consequently, configuration validation, governance evaluation, and impact propagation operate independently of the originating framework. Runtime replay follows a separate normalization path for execution observations.

The current reference implementation includes adapters for LangGraph, CrewAI, and the OpenAI Agents SDK. These three frameworks intentionally represent complementary introspection regimes rather than alternative implementations of the same architecture, providing a broader validation of the projection approach.

\paragraph{LangGraph.}

LangGraph exposes an explicit execution graph whose nodes, edges, routing conditions, prompts, tools, and models can be extracted directly through framework introspection. Consequently, the projection is predominantly structural and requires only limited semantic reconstruction. Governance-relevant relationships are preserved almost directly before canonical normalization.

\paragraph{CrewAI.}

CrewAI organizes execution around crews, tasks, flows, and associated metadata. While the primary governance entities remain directly identifiable, part of the execution topology must be reconstructed from declarative configuration metadata and execution conventions. The adapter therefore combines explicit extraction with semantic reconstruction while preserving the governance semantics required by ACM.

\paragraph{OpenAI Agents SDK.}

The OpenAI Agents SDK represents agent interactions through explicit handoff relationships. Unlike workflow-oriented frameworks, delegation topology is reconstructed by introspecting agent definitions and their declared handoffs rather than by traversing an explicit workflow graph. This demonstrates that ACM does not depend on a single representation of execution topology and can normalize heterogeneous delegation mechanisms into the same governance model.

Table~\ref{tab:framework_projection_summary} summarizes the principal characteristics of the three adapters.

\begin{table}[t]
	\centering
	\caption{Comparison of the projection mechanisms used for the three evaluated frameworks and their complementary introspection regimes.}
	\label{tab:framework_projection_summary}
	\small
	\begin{tabular}{p{3.0cm}p{2.6cm}p{2.8cm}p{2.8cm}}
		\toprule
		\textbf{Characteristic} &
		\textbf{LangGraph} &
		\textbf{CrewAI} &
		\textbf{OpenAI Agents SDK} \\
		\midrule
		Primary topology &
		Explicit execution graph &
		Flow and task structure &
		Agent handoffs \\
		Governance entities &
		Direct extraction &
		Direct extraction &
		Direct extraction \\
		Relationship recovery &
		Structural &
		Hybrid reconstruction &
		Delegation introspection \\
		Semantic normalization &
		Direct &
		Hybrid &
		Direct \\
		\bottomrule
	\end{tabular}
\end{table}

Together, these adapters cover three complementary introspection regimes while preserving a common governance representation. The objective of the adapters is therefore not to reproduce framework-specific execution semantics but to expose a normalized configuration model suitable for framework-independent governance.
These three adapters demonstrate that the operationalization is not tied to a particular representation of agentic systems. Instead, heterogeneous native abstractions are normalized into a common governance representation that is subsequently processed by the same framework-independent governance algorithms.

\subsection{Governance Algorithms}
\label{sec:governance_algorithms}

The reference implementation realizes the ACM governance semantics through four framework-independent algorithms operating exclusively on normalized ACM representations. Their purpose is not to redefine the formal semantics introduced in Section~\ref{sec:governance_semantics}, but to provide a deterministic operational realization suitable for heterogeneous execution frameworks. 

\paragraph{Algorithm 1 --- Semantic Projection}

The projection algorithm transforms native framework artifacts into immutable
ACM revisions and relationships while preserving governance-relevant semantics.

\begin{algorithm}[t]
	\caption{Semantic Projection}
	\begin{enumerate}
		\item \textbf{Input:} native framework configuration.
		\item Extract governance-relevant native entities.
		\item Identify relationships and associated metadata.
		\item Map native constructs to ACM concepts.
		\item Normalize identities and references.
		\item Compute canonical digests.
		\item Construct immutable ACM revisions and relationships.
		\item \textbf{Output:} normalized ACM Configuration Graph.
	\end{enumerate}
\end{algorithm}

\paragraph{Algorithm 2 --- Configuration Validation}

Structural validation verifies the integrity of the normalized configuration
before governance evaluation.

\begin{algorithm}[t]
	\caption{Configuration Validation}
	\begin{enumerate}
		\item \textbf{Input:} normalized ACM Configuration Graph.
		\item Verify structural consistency.
		\item Resolve cross-references.
		\item Validate lifecycle constraints.
		\item Check applicable governance invariants.
		\item Record validation results.
		\item \textbf{Output:} validated Configuration Graph.
	\end{enumerate}
\end{algorithm}

\paragraph{Algorithm 3 --- Governance Evaluation}

Governance evaluation combines local state evaluation with deterministic impact
propagation until stabilization. It operates on a validated Configuration Graph
together with the propagation-policy mapping $\Pi$ resolved from the Assurance
Graph. The algorithm is a reference operational realization of the semantics
defined in Section~\ref{sec:governance_semantics} and does not redefine their
formal specification.

\begin{algorithm}[t]
	\caption{Governance Evaluation}
	\begin{enumerate}
		\item \textbf{Input:} validated Configuration Graph, resolved propagation-policy mapping $\Pi$, and initial impact valuation $\iota^{(0)}$.
		\item Evaluate the local quality and assurance states and retain the assigned lifecycle states.
		\item Initialize the impact valuation from $\iota^{(0)}$.
		\item Initialize the propagation worklist.
		\item While the worklist is not empty:
		\begin{enumerate}
			\item select one pending ACI;
			\item propagate impact through applicable relationships according to $\Pi$;
			\item update impacted dependents when their state increases;
			\item enqueue dependents whose state has changed.
		\end{enumerate}
		\item Evaluate eligibility from the stabilized local and impact states.
		\item \textbf{Output:} stabilized impact valuation and evaluated configuration governance states associated with the validated Configuration Graph.
	\end{enumerate}
\end{algorithm}

\paragraph{Algorithm 4 --- Runtime Integration}

Runtime observations are normalized into governance events that incrementally
reconstruct the Runtime Graph without modifying immutable configuration
revisions.

\begin{algorithm}[t]
	\caption{Runtime Integration}
	\begin{enumerate}
		\item \textbf{Input:} normalized runtime event stream.
		\item For each event:
		\begin{enumerate}
			\item validate event consistency and configuration provenance;
			\item identify or instantiate the referenced runtime entity in $V_R$;
			\item preserve its provenance through $prov:V_R\rightarrow V_C$;
			\item append the event to the ordered event log;
			\item update the corresponding runtime state $Rt(x)$ and runtime relationships.
		\end{enumerate}
		\item Replay the ordered event log when reconstruction is requested.
		\item \textbf{Output:} updated or reconstructed Runtime Graph.
	\end{enumerate}
\end{algorithm}

Table~\ref{tab:algorithm_complexity} summarizes the asymptotic complexity of the principal operational algorithms. For Governance Evaluation, the reported bound corresponds to the reference
impact-propagation worklist, whose complexity is established formally in
Appendix~\ref{appendix:complexity_analysis}; local governance-state evaluations are evaluated independently of this propagation bound.

\begin{table}[t]
	\centering
	\caption{Asymptotic complexity of the reference algorithms.}
	\label{tab:algorithm_complexity}
	\small
	\begin{tabular}{lc}
		\toprule
		\textbf{Algorithm} & \textbf{Complexity} \\
		\midrule
		Semantic Projection & $O(|N_F|+|R_F|)$ \\
		Configuration Validation & $O(|V_C|+|E_C|)$ \\
		Governance Evaluation & $O(h_{\mathcal I}|E_C|)=O(|E_C|)$ \\
		Runtime Replay & $O(|\Sigma|)$ \\
		\bottomrule
	\end{tabular}
\end{table}

The reference implementation emphasizes deterministic behavior and semantic
reproducibility rather than execution performance. The detailed implementation
architecture is presented in the following subsection, while the formal
properties of the propagation semantics are established in
Section~\ref{sec:governance_semantics} and its appendix.

\subsection{Reference Implementation}
\label{sec:reference_implementation}

The reference implementation provides an executable operationalization of the ACM reference model. It is intended to demonstrate the feasibility and reproducibility of the proposed governance semantics rather than to serve as a production-ready orchestration platform.

The prototype is implemented in Python~3.11--3.13 using Pydantic~v2 to represent
the normative ACM concepts through strict data contracts and schema validation.
Immutable configuration revisions, graphs, and governance states remain concepts
of the reference model; Pydantic provides their concrete representation in this
implementation. The modular organization separates framework-specific extraction
from framework-independent governance, allowing new execution frameworks to be
integrated through an additional projection adapter while leaving the governance
kernel unchanged.

The implementation is organized into five principal modules:

\begin{itemize}
	\item \textbf{Core models}, providing immutable ACI revisions, configuration graphs, release baselines, governance descriptors, runtime objects, and canonical serialization;
	\item \textbf{Governance engine}, implementing validation, governance state
	evaluation, impact propagation, and lifecycle management;
	\item \textbf{Runtime layer}, implementing normalized runtime-event integration,
	provenance preservation, and deterministic Runtime Graph reconstruction;
	\item \textbf{Framework adapters}, responsible for semantic projection from native framework representations into canonical ACM objects;
	\item \textbf{Scenario harness}, providing reproducible experimental scenarios, semantic-preservation analyses, quantitative impact evaluation, and report generation;
	\item \textbf{Verification suite}, comprising unit tests, property-based tests, and regression tests validating both the governance semantics and framework projections.
\end{itemize}

For a fixed governance context, canonical serialization, immutable revisions, and
deterministic governance evaluation ensure that identical normalized
configurations produce reproducible governance results. Normalized runtime
events, replayable execution histories, and deterministic impact propagation
further support reproducible analyses independently of the originating execution
framework.

The modular organization also facilitates extensibility. Supporting an
additional framework primarily requires implementing a semantic projection
adapter satisfying the common projection contract introduced in
Section~\ref{subsec:semantic_projection_pipeline}; framework-specific extraction
or reconstruction may vary according to the available introspection mechanisms,
while the governance kernel and normative semantics remain unchanged.

\paragraph{Algorithm Overview}
Table~\ref{tab:governance_algorithms} summarizes the governance of the implemented algorithms by the ACM reference prototype.

\begin{table}[t]
	\centering
	\caption{Core governance algorithms implemented by the ACM reference prototype.}
	\label{tab:governance_algorithms}
	\small
	\renewcommand{\arraystretch}{1.15}
	\begin{tabularx}{\columnwidth}{>{\raggedright\arraybackslash}p{2.8cm}XX>{\raggedright\arraybackslash}p{2.6cm}}
		\toprule
		\textbf{Algorithm} &
		\textbf{Input} &
		\textbf{Output} &
		\textbf{Main Purpose} \\
		\midrule
		
		Semantic Projection &
		Native framework configuration &
		Canonical ACM configuration graph &
		Semantic normalization \\
		
		Configuration Validation &
		ACM configuration graph &
		Validated ACM configuration graph &
		Structural and governance integrity verification \\
		
		Governance Evaluation &
		Validated ACM configuration graph and resolved policies &
		Stabilized impact valuation and evaluated governance states &
		Governance state evaluation \\
		
		Runtime Integration &
		Normalized runtime events &
		ACM runtime graph &
		Replay and execution reconstruction \\
		
		\bottomrule
	\end{tabularx}
\end{table}

\paragraph{Experimental Instrumentation.}

The reference implementation additionally includes an experimental instrumentation layer used exclusively for the evaluation reported in Section~\ref{sec:evaluation}. This layer collects projection inventories, semantic-preservation outcomes, normalized impact-propagation results, reproducibility metadata, and comparison data used by the experimental protocol. It remains separated from the governance kernel and does not participate in the computation of normative ACM states. This separation ensures that the evaluation infrastructure observes and measures the operationalization without altering the semantics being evaluated.

\section{Evaluation}
\label{sec:evaluation}

This section evaluates the ACM reference model with respect to the research questions introduced in Section~\ref{sec:introduction}. The objective is to assess whether the proposed reference model preserves governance semantics across heterogeneous agentic frameworks, supports deterministic governance evaluation, and provides a framework-independent foundation for agentic configuration management. The evaluation combines two complementary experimental campaigns. The first assesses reference-model coverage, conformance of the operationalization to the specified governance semantics, and cross-framework applicability through twenty-seven governance scenarios
executed on LangGraph, CrewAI, and the OpenAI Agents SDK. The second provides a dedicated quantitative evaluation of impact propagation across nine additional cases. Together, both campaigns provide evidence for the four research questions addressed in this work.

\subsection{Research Questions}
\label{sec:research_questions}

The evaluation addresses the following research questions.

\begin{itemize}
	
	\item \textbf{RQ1 -- Reference Model Coverage.}
	Can the ACM reference model represent the governance-relevant configuration concepts required by heterogeneous agentic systems, independently of differences in native framework introspectability?
	
	\item \textbf{RQ2 -- Governance Semantics.}
	Does the operationalization faithfully realize the governance semantics defined in Section~\ref{sec:governance_semantics}, including lifecycle management, governance evaluation, impact propagation, release governance, and runtime reconstruction?
	
	\item \textbf{RQ3 -- Framework Independence.}
	Can heterogeneous native configurations be projected into governance-equivalent
	ACM representations without requiring framework-specific governance semantics?

	\item \textbf{RQ4 -- Quantitative Impact Analysis.}
	Can the governance kernel produce reproducible quantitative impact analyses across heterogeneous frameworks while reducing the configuration inspection scope under the experimental protocol?
	
\end{itemize}

The first three research questions evaluate the reference model itself, whereas the fourth assesses the operational behavior of the governance kernel through a dedicated quantitative impact analysis.
The following sections describe the experimental methodology adopted to answer these research questions and present the empirical evidence collected using the reference implementation.

\subsection{Experimental Design}
\label{sec:experimental_protocol}

\begin{figure*}[t]
	\centering
	\includegraphics[width=0.5\textwidth]{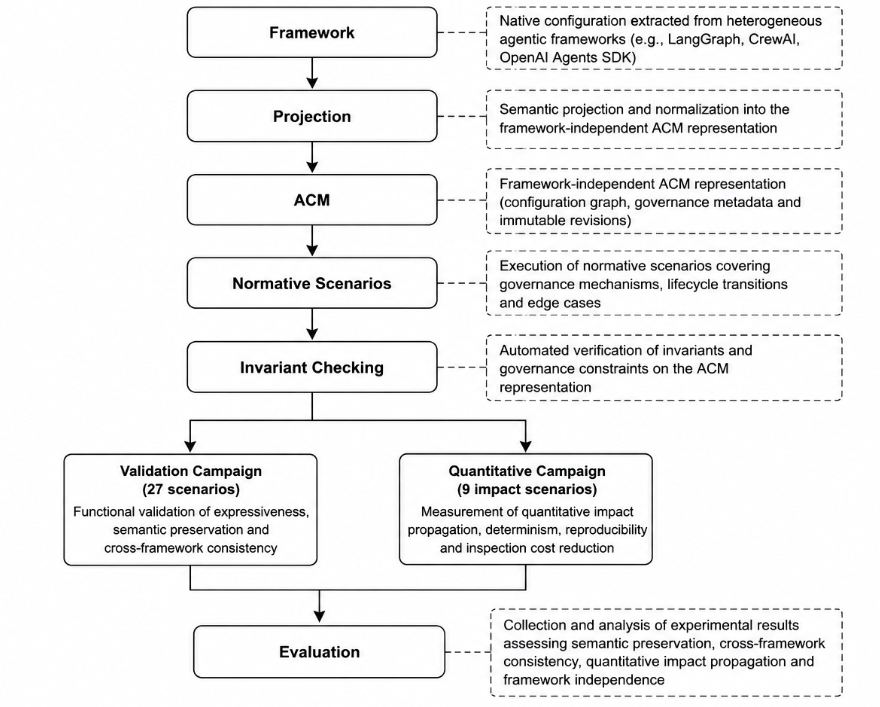}
	\caption{Overview of the experimental methodology. Native framework configurations are projected into the ACM reference model through semantic projection, validated using normative governance scenarios and automated invariant checking, and finally evaluated.}
	\label{fig:experimental_protocol}
\end{figure*}

The evaluation follows two complementary experimental campaigns designed to assess distinct properties of the ACM reference model. Figure~\ref{fig:experimental_protocol} provides an overview of the experimental
methodology used to evaluate the ACM reference model.

\subsubsection{Campaign A -- Reference Model Coverage and Semantic Conformance}

The first campaign evaluates the representational coverage of the reference model, the conformance of the operationalization to the specified governance semantics, and its applicability across the three evaluated agentic frameworks.

It consists of twenty-seven controlled governance scenarios covering the principal concepts introduced throughout the reference model, including immutable revisions, lifecycle evolution, governance states, dependency propagation, release baselines, runtime governance, dynamic agents, and semantic projection. The complete scenario suite is executed independently on LangGraph, CrewAI, and the OpenAI Agents SDK through their respective projection adapters. Each native configuration is projected into the canonical ACM representation before governance processing, while the same framework-independent governance kernel is used for all three frameworks. This design makes it possible to distinguish framework-specific extraction from the semantics evaluated by the ACM core. Each scenario specifies an initial governed configuration, the sequence of configuration or runtime events applied to the system, and the expected governance outcomes. 

The resulting governed representations are compared against pre-specified expected semantic properties derived from the reference model. The campaign therefore provides scenario-based evidence of model coverage and implementation conformance rather than an independent empirical validation of the universal correctness of the ACM semantics. Collectively, the twenty-seven scenarios provide evidence for RQ1 and RQ2, while their execution across the three frameworks directly supports RQ3 by testing whether heterogeneous native representations converge toward equivalent governed ACM representations (Table~\ref{tab:qualitative_validation} summarizes the main results associated to the Research Questions). 

\begin{table}[t]
	\caption{Coverage of the qualitative validation campaign. The 27 governance scenarios collectively exercise the main semantic capabilities of the ACM reference model across the supported execution frameworks.}
	\label{tab:qualitative_validation}
	\centering
	\footnotesize
	\begin{tabular}{p{3.2cm}p{4.5cm}p{1.8cm}c}
		\toprule
		\textbf{Validation aspect} &
		\textbf{Representative capabilities} &
		\textbf{Research question(s)} &
		\textbf{Evidence} \\
		\midrule
		
		Configuration model
		&
		ACI revisions, configuration relationships, release baselines, governance descriptors
		&
		RQ1
		&
		Supported
		\\
		
		Governance semantics
		&
		Lifecycle transitions, quality assessment, assurance evaluation, impact propagation, eligibility computation
		&
		RQ2
		&
		Supported
		\\
		
		Framework-independent projection
		&
		Semantic projection from LangGraph, CrewAI, and OpenAI Agents SDK into a common ACM representation
		&
		RQ3
		&
		Supported
		\\
		
		Runtime governance
		&
		Runtime event representation, execution replay, provenance reconstruction, deterministic state reconstruction
		&
		RQ2, RQ3
		&
		Supported
		\\
		
		Cross-framework consistency
		&
		Equivalent governance outcomes obtained from semantically equivalent native implementations
		&
		RQ3
		&
		Supported
		\\
		
		\bottomrule
	\end{tabular}
\end{table}

\subsubsection{Campaign B -- Quantitative Impact Evaluation} 
\begin{figure*}[t]
	\centering
    \includegraphics[width=\textwidth]{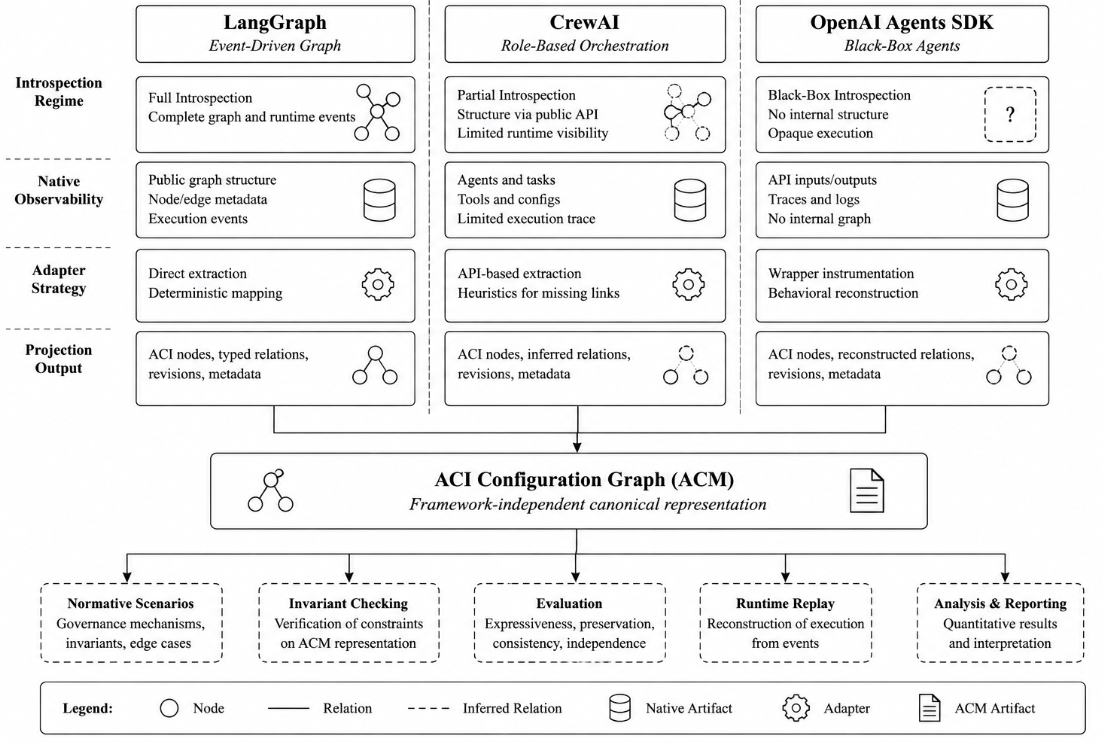}
	\caption{Experimental methodology adopted to evaluate the ACM reference model. Campaign A (Reference Model Coverage and Semantic Conformance) assesses model coverage and implementation conformance through 27 governance scenarios and semantic projection, whereas Campaign B (Quantitative Campaign) evaluates quantitative impact propagation across nine cross-framework cases. Together, both campaigns provide complementary evidence for RQ1--RQ4.}
	
	\label{fig:evaluation_methodology}
\end{figure*}

The second campaign focuses specifically on the quantitative behavior of impact propagation. Three representative configuration changes are evaluated for each of the same three frameworks, corresponding to local, intermediate, and global propagation scopes, resulting in nine additional experimental cases. Each case is executed five consecutive times to assess deterministic behavior and reproducibility. The same framework-independent fixed-point propagation engine is used throughout the evaluation. For every experimental case, the evaluation records the governed impact set, impact size, impact depth, impact ratio, fixed-point convergence behavior, reproducibility across repeated executions, and residual inspection-scope reduction. The experiment additionally compares transitive ACM propagation with a naive one-hop dependency inspection baseline. This second campaign provides dedicated quantitative evidence for RQ4 while complementing the broader semantic-conformance evidence provided by Campaign~A (see Figure~\ref{fig:evaluation_methodology}). 

Table~\ref{tab:rq_validation} summarizes how the two complementary experimental campaigns collectively address the four research questions investigated in this work.
\begin{table}[t]
	\caption{Relationship between the research questions, the experimental evidence, and the corresponding validation results.}
	\label{tab:rq_validation}
	\centering
	\footnotesize
	\begin{tabular}{p{1.0cm}p{4.0cm}p{4.2cm}p{3.0cm}}
		\toprule
		\textbf{RQ} &
		\textbf{Research objective} &
		\textbf{Experimental evidence} &
		\textbf{Main result} \\
		\midrule
		
		RQ1 & Does the reference model provide the abstractions required to represent governance-relevant configuration concepts? & Scenario-based coverage across twenty-seven governance scenarios, complemented by projection inventories for the three evaluated frameworks. & All governance concepts required by the evaluated scenarios were representable; framework introspection limitations were characterized separately. 
		\\ 
		RQ2 & Does the operationalization conform to the specified ACM governance semantics? & Comparison of observed scenario outcomes with the pre-specified lifecycle, quality, assurance, impact, eligibility, runtime, and baseline semantics. & The operationalization conformed to the specified semantics for all evaluated scenario outcomes. 
		\\ 
		RQ3 & Can heterogeneous frameworks yield governance-equivalent ACM representations? & Semantic projection of LangGraph, CrewAI, and OpenAI Agents SDK under three complementary introspection regimes. & Governance-equivalent representations and outcomes were obtained for the evaluated configurations without requiring framework-specific governance logic. 
		\\ 
		RQ4 & Can ACM reproducibly characterize configuration impact and reduce the inspection scope under the experimental protocol? & Nine quantitative impact cases across three frameworks, each repeated five times, with impact extent, propagation depth, fixed-point convergence, and inspection-scope measurements. & Reproducible impact analyses were obtained across all evaluated cases; transitive propagation recovered impacts missed by one-hop inspection and reduced the residual inspection scope. \\
		
		\bottomrule
	\end{tabular}
\end{table}

\subsection{Validation of the Reference Model}
\label{sec:reference_model_validation}

Campaign~A evaluates the ACM reference model through a suite of twenty-seven controlled governance scenarios executed independently on LangGraph, CrewAI, and the OpenAI Agents SDK. The objective is not to benchmark execution performance or establish universal model validity, but to assess representational coverage and verify that the operationalization conforms to the governance semantics defined in Section~\ref{sec:governance_semantics} for the evaluated scenarios.

The scenario suite was designed to exercise the principal concepts introduced throughout the reference model. Collectively, the scenarios cover immutable ACI revisions, lifecycle evolution, governance states, structural relationships, semantic projection, release baselines, runtime governance, dependency propagation, and dynamic agent evolution. Rather than validating isolated software components, each scenario evaluates the interaction between several governance mechanisms under representative configuration changes.

For every scenario, equivalent native configurations were implemented for the three supported frameworks. After semantic projection, the resulting ACM representations were processed by the same framework-independent governance kernel. The observed governed representations were then compared against the expected governance outcomes defined by the experimental specification. This protocol evaluates both the correctness of semantic projection and the preservation of governance semantics independently of the originating execution framework.

Table~\ref{tab:scenario_categories} summarizes the categories covered by the experimental campaign.

\begin{table}[t]
	\centering
	\caption{Coverage of the reference model validation scenarios.}
	\label{tab:scenario_categories}
	\small
	\begin{tabular}{p{3.2cm}p{3.8cm}}
		\toprule
		\textbf{Category} & \textbf{Representative governance aspects} \\
		\midrule
		Revision management &
		Immutable revisions, promotion, lifecycle evolution \\
		Dependency management &
		Structural relationships and impact dependencies \\
		Governance evaluation &
		Eligibility assessment, governance states, policy evaluation \\
		Release governance &
		Baseline composition, revision consistency, release evolution \\
		Runtime governance &
		Execution events and runtime reconstruction \\
		Dynamic evolution &
		Dynamic agents and evolving configurations \\
		Semantic projection &
		Cross-framework normalization and representation consistency \\
		\bottomrule
	\end{tabular}
\end{table}

Across the complete campaign, the twenty-seven scenarios produced governed representations consistent with the expected semantic properties on all three execution frameworks. For the evaluated scenarios, no framework-specific governance behavior was introduced after semantic projection: the same governance kernel operated on the normalized ACM representations independently of the originating execution framework.

These observations provide complementary evidence for the first three research questions. The scenario suite provides evidence of representational coverage for the evaluated governance concepts (RQ1), conformance between the operationalization and the specified governance semantics (RQ2), and the ability to obtain governance-equivalent representations from the three heterogeneous execution frameworks through semantic projection (RQ3).

\subsection{Cross-framework Semantic Projection}
\label{sec:cross_framework_projection}

The objective of this evaluation is to determine whether heterogeneous execution frameworks can be projected into governance-equivalent ACM representations while preserving the information required by the evaluated governance semantics. Governance equivalence denotes equivalence with respect to these governance requirements and does not imply complete informational identity between native or projected representations.

This experiment addresses RQ3 by evaluating the semantic projection implemented for LangGraph, CrewAI, and the OpenAI Agents SDK.

Although the three frameworks support comparable agentic applications, they expose substantially different introspection mechanisms. LangGraph provides an explicit workflow graph whose nodes and structural relationships can be reconstructed directly through framework introspection. CrewAI separates orchestration into Crew and Flow abstractions, requiring part of the execution topology to be reconstructed from adapter metadata because the complete graph is not statically introspectable. In contrast, the OpenAI Agents SDK exposes delegation explicitly through agent handoffs, allowing the delegation graph to be reconstructed directly while agent-level references remain normalized through adapter metadata. These three frameworks therefore represent complementary introspection regimes while sharing the same ACM governance model after semantic projection.

To evaluate information preservation, representative native workflows were extracted independently from each framework and compared with manually established golden representations over the normative ACM perimeter. Each governance-relevant property was classified as \emph{preserved}, \emph{declared\_by\_adapter}, \emph{approximated}, or \emph{unsupported}. The \emph{preserved} status denotes information recovered from native introspection without semantic loss; \emph{declared\_by\_adapter} identifies information supplied explicitly by the adapter when native introspection is insufficient; \emph{approximated} denotes an intentional canonical abstraction; and \emph{unsupported} denotes information for which no representation is provided within the evaluated projection perimeter.

These classifications and the associated coverage measures characterize projection fidelity with respect to the predefined governance-relevant perimeter of the experiment; they are not interpreted as universal measures of ACM expressiveness.

Table~\ref{tab:projection_results} summarizes the principal information-preservation metrics obtained across the three frameworks.

\begin{table*}[t]
	\centering
	\caption{Summary of information preservation after semantic projection.}
	\label{tab:projection_results}
	\small
	\begin{tabular}{lccc}
		\toprule
		\textbf{Metric}
		&
		\textbf{LangGraph}
		&
		\textbf{CrewAI}
		&
		\textbf{OpenAI Agents SDK}
		\\
		\midrule
		Node coverage & 100\% & 100\% & 100\% \\
		Relationship coverage & 100\% & 100\% & 100\% \\
		Branch coverage & 100\% & 100\% & 100\% \\
		Entry point preserved & Yes & Yes & Yes \\
		Terminal nodes preserved & Yes & Yes & Yes \\
		Agent--prompt references & 100\% & 100\% & 100\% \\
		Agent--tool references & 100\% & 100\% & 100\% \\
		Normative properties preserved & 10 & 9--10 & 10 \\
		Approximated properties & 1 & 1 & 1 \\
		Unsupported properties & 0 & 0--1 & 0 \\
		Declared by adapter & 4 & 2 (Flow only) & 1 (Agent only) \\
		         &   & 3 (Crew only) & 5 (Agent + Graph) \\
		         &   & 5 (Flow + Crew) & \\
		\bottomrule
	\end{tabular}
\end{table*}

While the overall preservation results are comparable, the extraction process differs significantly between frameworks. LangGraph exposes its execution topology directly through the native workflow graph. CrewAI illustrates the limits of static introspection: workflow nodes are extracted from the Flow definition, whereas part of the orchestration topology must be supplied through adapter metadata because it is resolved dynamically at execution time. Conversely, the OpenAI Agents SDK exposes the delegation topology explicitly through handoff relationships, allowing the corresponding graph to be reconstructed entirely by introspection without unresolved structural elements.

For the evaluated workflows, these observations show that ACM projection can accommodate distinct execution abstractions and introspection mechanisms. Semantic projection normalizes the governance-relevant native information into a common configuration model while explicitly documenting its extraction status. Differences in framework introspectability therefore remained confined to the projection stage for the evaluated governance scenarios and did not require framework-specific changes to the governance kernel.

Collectively, these results support RQ3 by showing that the three evaluated native representations yield governance-equivalent ACM configurations despite relying on different extraction mechanisms.

\subsection{Quantitative Impact Evaluation}
\label{sec:quantitative_impact}

Campaign~B complements the qualitative validation performed in the previous sections by evaluating the quantitative behavior of the ACM impact propagation semantics. Whereas Campaign~A verifies that the governance semantics are correctly preserved, Campaign~B measures the properties of the resulting propagation across representative configuration changes.

Nine experimental cases were constructed by combining the three supported execution frameworks with three representative change classes corresponding to local, intermediate, and global propagation scopes. Every experimental case was executed five consecutive times using the same framework-independent governance kernel.

The quantitative evaluation measures five complementary properties:

\begin{itemize} 
	\item impact size, measuring the number of governed ACIs reached by propagation; 
	\item impact depth, measuring the maximum propagation distance from the changed ACI; 
	\item impact ratio, normalizing the impacted set size by the configuration-graph size; 
	\item fixed-point convergence and reproducibility across repeated executions; 
	\item reduction of the residual configuration inspection scope. \end{itemize}

The reported inspection reduction values characterize the residual configuration inspection scope defined by the experimental protocol rather than measured human effort. They compare the number of configuration elements considered by the exhaustive native inspection procedure with the residual elements requiring verification after ACM has produced the governed impact set. They therefore do not quantify inspection time, operational cost, or human productivity.

Table~\ref{tab:impact_summary} summarizes the principal quantitative results.

\begin{table*}[t]
	\centering
	\caption{Summary of the quantitative impact evaluation. Impact depth and fixed-point convergence are reported from the experimental measurements; inspection reduction denotes residual configuration-inspection scope rather than measured human effort.}
	\label{tab:impact_summary}
	\small
	\begin{tabular}{llccccc}
		\toprule
		Framework &
		Change &
		Impact Size &
		Impact Ratio &
		\makecell{Impact\\Depth} &
		\makecell{Fixed-point\\Iterations} &
		\makecell{Inspection scope\\reduction}
		\\
		\midrule
		LangGraph &
		Local &
		2 &
		0.182 &
		2 &
		3 &
		0.846
		\\
		&
		Intermediate &
		2 &
		0.182 &
		2 &
		3 &
		0.692
		\\
		&
		Global &
		5 &
		0.455 &
		2 &
		3 &
		0.615
		\\
		\midrule
		CrewAI &
		Local &
		2 &
		0.182 &
		2 &
		3 &
		0.846
		\\
		&
		Intermediate &
		2 &
		0.182 &
		2 &
		3 &
		0.692
		\\
		&
		Global &
		5 &
		0.455 &
		2 &
		3 &
		0.615
		\\
		\midrule
		OpenAI Agents &
		Local &
		2 &
		0.182 &
		2 &
		3 &
		0.846
		\\
		&
		Intermediate &
		2 &
		0.182 &
		2 &
		3 &
		0.692
		\\
		&
		Global &
		5 &
		0.455 &
		2 &
		3 &
		0.615
		\\
		\bottomrule
	\end{tabular}
\end{table*}

The impact metrics remained identical across the three execution frameworks for every change class. Local and intermediate changes affected two governed ACIs, whereas global model updates propagated to five governed ACIs. The corresponding impact ratios remained consistent across frameworks, reflecting the governance-equivalent dependency structures used by the impact propagation experiment after semantic projection. The number of iterations remains constant across change classes, empirically confirming the scope-independent complexity bound established in Appendix~\ref{appendix:formal_properties}.

The evaluation also shows a reduction in the residual configuration inspection scope ranging from 61.5\% for global changes to 84.6\% for local changes under the experimental protocol. These values quantify the reduction in configuration elements requiring human verification after ACM has identified the governed impact set; they do not represent execution-time performance measurements. 

The quantitative campaign reported above is conducted on an eleven-item, thirteen-relation configuration. A complementary baseline, conducted separately on a dedicated thirteen-item, eighteen-relation configuration constructed to exhibit a two-level transitive dependency, evaluates whether direct dependency inspection alone is sufficient to recover the propagated scope. On that configuration, a naive one-hop traversal recovered six of the nine affected items, whereas ACM fixed-point propagation recovered all nine, including three items reached transitively through an intermediate revision. This comparison isolates the contribution of transitive governed propagation from dependency visibility alone.

Collectively, these observations provide quantitative evidence supporting RQ4. Across the evaluated cases, the reference implementation produced reproducible and cross-framework-consistent impact analyses, converged deterministically under repeated execution, recovered transitive impacts omitted by one-hop inspection, and reduced the residual configuration inspection scope under the experimental protocol. These empirical observations complement the formal propagation properties established in Section~\ref{sec:governance_semantics}.

Across the twenty-seven scenarios, all three adapters successfully produced governed ACM representations compatible with the common governance kernel. Differences remained limited to framework-specific extraction mechanisms and did not affect the normalized governance representation used for subsequent evaluation.

These observations support RQ3 by showing that, for the evaluated
configurations, framework-specific extraction can terminate at the semantic projection boundary, after which the same ACM governance semantics operate on the normalized Configuration Graph. The resulting governed representations therefore provide a common basis for lifecycle evaluation, impact propagation, release governance, and runtime reconstruction across the three evaluated frameworks.

\subsection{Reproducibility}
\label{subsec:eval-reproducibility}

Beyond semantic correctness, the evaluation assesses the reproducibility of the proposed governance semantics. All experimental campaigns were executed using the same framework-independent governance kernel, while semantic projection was performed independently for LangGraph, CrewAI, and the OpenAI Agents SDK.

Campaign~A executed the complete suite of twenty-seven governance scenarios on each framework, producing consistent governed ACM representations and identical validation outcomes independently of the originating execution environment. Campaign~B complemented this validation through nine quantitative impact scenarios, each executed five consecutive times. Across all repetitions, the governance kernel produced identical impacted sets, identical governance states, and identical quantitative metrics.

The fixed-point propagation algorithm consistently converged after the same number of iterations for identical configuration graphs and produced identical impact sets across repeated executions. These observations provide empirical evidence that the reference implementation realizes the specified propagation semantics reproducibly once the evaluated configurations have been normalized into ACM representations. The formal properties of impact propagation are established separately in Section~\ref{sec:governance_semantics} and its appendix.

The results therefore show that, for the evaluated governance computations, reproducibility is maintained independently of the stochastic outputs of the underlying language models and of framework-specific execution mechanisms.

This distinction is essential because ACM governs configurations and their relationships rather than the probabilistic outputs generated during agent execution.

\subsection{Summary with Respect to the Research Questions}
\label{sec:rq_summary}

The experimental results collectively support the four research questions introduced in Section~\ref{sec:research_questions}.

RQ1 is addressed by the qualitative campaign, which provides scenario-based evidence that the ACM reference model contains the abstractions required to represent the governance concepts exercised by the evaluated configurations. Differences in native introspectability are treated separately as properties of the projection process rather than as direct limitations of model expressiveness.

RQ2 is supported by the governance scenarios as conformance evidence: for the evaluated cases, the operationalization produced outcomes consistent with the lifecycle, quality, assurance, impact, eligibility, runtime, and release semantics specified by ACM. This evidence establishes implementation conformance for the tested scenarios rather than universal empirical validity of the normative semantics. 

RQ3 is supported by the semantic projection of LangGraph, CrewAI, and OpenAI Agents SDK. Despite differences in native abstractions and introspection capabilities, the evaluated configurations produced governance-equivalent ACM representations and governance outcomes without requiring framework-specific governance semantics. This equivalence does not imply complete informational identity between projections. 

RQ4 is addressed by the quantitative impact evaluation. Across the nine representative change scenarios, the reference implementation produced reproducible impact sets and consistent quantitative results across the three evaluated frameworks. The experiment further showed deterministic fixed-point convergence, recovery of transitive impacts missed by one-hop inspection, and a reduced residual inspection scope under the experimental protocol.

Overall, the experimental results support the central hypothesis of this work within the evaluated scope: heterogeneous agentic configurations can be projected into a common immutable configuration model and governed through common governance semantics independently of their native execution abstractions. The limitations of the present evaluation and the implications for broader adoption are discussed in the following section.

\section{Discussion}
\label{sec:discussion}
\subsection{Positioning with Respect to Existing Agentic Frameworks}
\label{sec:positioning}

The rapid emergence of agentic frameworks has significantly simplified the development of LLM-based applications by providing increasingly sophisticated abstractions for workflow orchestration, agent coordination, tool integration, and execution management. Frameworks such as LangGraph, CrewAI, and the OpenAI Agents SDK illustrate complementary execution and introspection paradigms, ranging from explicit workflow graphs to metadata-assisted orchestration and delegation-oriented agent interactions. These differences provide representative examples of the heterogeneous configuration abstractions that ACM aims to govern through a common semantic representation.

The objective of ACM is fundamentally different. Rather than providing execution capabilities, ACM introduces a framework-independent governance layer dedicated to the management of agentic configurations throughout their lifecycle. Consequently, ACM should not be interpreted as an alternative orchestration framework, but as a complementary governance model capable of operating across heterogeneous execution environments.

This distinction is reflected in the responsibilities addressed by each approach. Agentic frameworks primarily define how agents execute, communicate, and coordinate during runtime. ACM, by contrast, governs what is being executed by providing immutable configuration revisions, provenance tracking, lifecycle management, dependency analysis, baseline management, assurance evaluation, and runtime traceability. Execution and governance therefore constitute complementary concerns operating at different abstraction levels.

The experimental evaluation further indicates that this separation remains applicable across heterogeneous introspection regimes. While LangGraph exposes an explicit execution graph, CrewAI requires partial semantic reconstruction from declarative metadata, and the OpenAI Agents SDK derives governance topology from explicit agent handoff relationships. Despite these differences, semantic projection produces governance-equivalent ACM representations that can be processed by the same framework-independent governance kernel.

The semantic projection mechanism provides the interface between these two layers (see Appendix~\ref{appendix:projection_formalism}). By translating framework-specific configurations into normalized ACM artifacts, it confines framework-specific interpretation to the projection boundary while preserving the governance-relevant information required by the common governance semantics. This separation allows the same governance mechanisms to operate on configurations originating from heterogeneous orchestration paradigms without introducing framework-specific logic into the governance kernel.

More generally, the proposed architecture aligns with the long-established separation between execution infrastructures and configuration management observed in traditional software engineering. Source code management systems, build systems, deployment platforms, and runtime environments have historically evolved as complementary but distinct components of the software lifecycle. ACM extends this principle to agentic systems by introducing an explicit governance layer dedicated to the management of agentic configurations rather than their execution.

The experimental evaluation presented in Section~\ref{sec:evaluation} provides empirical evidence supporting this positioning. Across twenty-seven governance scenarios and a complementary quantitative impact evaluation, heterogeneous configurations originating from LangGraph, CrewAI, and the OpenAI Agents SDK were successfully normalized into governance-equivalent ACM representations. These frameworks represent three complementary introspection regimes, providing evidence that, for the evaluated configurations, the same governance semantics can operate on the normalized configuration model independently of the originating execution abstraction or introspection mechanism. Although additional validation on other frameworks remains necessary, these results provide encouraging evidence that semantic projection constitutes a viable foundation for framework-independent governance of heterogeneous agentic systems.

\subsection{Contributions to Agentic Configuration Governance}
\label{sec:governance_contributions}

The contribution of ACM lies neither in introducing a new orchestration mechanism nor in extending the execution capabilities of existing agentic frameworks. Its primary contribution is to establish configuration governance as an explicit and framework-independent concern for agentic systems. From this perspective, ACM contributes at four complementary levels: representation, semantics, interoperability, and operational validation.

\subsubsection{A Framework-Independent Governance Representation}

The first contribution is a reference model that represents an agentic system as a governed configuration rather than as a collection of framework-specific runtime objects. The model introduces explicit configuration items, immutable revisions, typed relationships, provenance information, and release baselines. These abstractions provide a common vocabulary for describing the constituent elements of an agentic system and the dependencies that determine its effective configuration.

This representation extends established configuration-management principles to artifacts that are specific to agentic systems. In particular, it treats agents, prompts, tools, policies, models, workflows, and dynamically instantiated components as governable objects whose identity and evolution can be tracked independently. The resulting configuration is therefore not reduced to source code, deployment metadata, or execution traces. It is represented as an explicit and auditable object in its own right.

A central consequence of this model is the ability to distinguish logical identity from immutable revision identity. Successive revisions of the same configuration item preserve a stable conceptual identity while remaining individually addressable through exact revision identifiers and content digests. This distinction supports reproducibility, comparison, controlled replacement, and historical reconstruction without conflating an evolving artifact with one of its concrete states.

\subsubsection{Formal Governance Semantics}

The second contribution is the formalization of governance semantics over the reference model. ACM does not merely describe which artifacts exist; it defines how their governance states evolve and how local changes affect dependent configurations.

The lifecycle model specifies admissible transitions for configuration revisions, baselines, and runtime instances. Quality, assurance, impact, and eligibility are maintained as distinct dimensions rather than being collapsed into a single status. This separation prevents ambiguous interpretations such as treating a lack of evidence as proof of poor quality or propagating the lifecycle state of a dependency directly to its parent.

The propagation semantics define how an initial impact valuation is propagated through governed configuration relationships according to the resolved propagation policies. Propagation acts exclusively on the impact dimension and does not modify lifecycle, quality, or assurance states; operational eligibility is subsequently evaluated from the stabilized impact state together with the local governance states.

The model thereby preserves the provenance of governance decisions while supporting deterministic and explainable state computation.

The assurance model complements these semantics by linking governance conclusions to explicit evidence. Assurance is evaluated locally for each exact revision according to the evidence required by the applicable governance context. Governance states can therefore be interpreted together with the evidence and rules from which they were derived.

\subsubsection{Framework-Independent Semantic Projection}

The third contribution is the separation between framework-specific extraction and framework-independent governance processing. Native configurations are translated into ACM through semantic projection, while the governance kernel operates exclusively on normalized ACM artifacts.

This architecture allows heterogeneous orchestration paradigms to be governed through a common representation. A graph-oriented framework may expose nodes and edges directly, whereas a role- and task-oriented framework may require part of its topology to be reconstructed from assignments and dependencies. ACM does not require these native representations to be structurally identical. It requires the projection to preserve the information needed for configuration governance and to make approximated or unavailable information explicit.

Framework independence should therefore be understood as independence of the governance model and processing semantics from a specific execution framework. The adapters remain framework-specific because they interpret native concepts. Once configuration projection is complete, lifecycle validation, baseline construction, dependency analysis, assurance evaluation, and impact propagation operate on the normalized ACM configuration independently of the source framework. Runtime integration follows a separate normalization path for execution observations while preserving provenance to the governed configuration.

This separation provides a basis for interoperability without imposing a common execution model. Frameworks retain their native orchestration semantics, while ACM supplies a shared governance representation above them.
The experimental evaluation further demonstrates that this interoperability extends beyond heterogeneous orchestration models to heterogeneous introspection regimes. The evaluated frameworks expose governance-relevant information through explicit execution graphs, metadata-assisted reconstruction, and delegation-oriented agent handoffs, respectively. By normalizing these complementary representations into a common ACM configuration graph, semantic projection isolates framework-specific extraction from governance processing, allowing identical governance semantics to be applied independently of the native introspection mechanism.

\subsubsection{Reference Operationalization and Evaluation}

The fourth contribution is the operationalization of the reference model
through an executable implementation together with its experimental evaluation across three representative agentic frameworks.
Beyond demonstrating implementability, the reference implementation provides a reproducible operational realization of the proposed governance semantics, including semantic projection, deterministic governance evaluation, runtime reconstruction, and quantitative impact analysis.

The LangGraph, CrewAI, and OpenAI Agents SDK adapters instantiate the semantic-projection boundary for three complementary execution and introspection paradigms. Their diversity does not establish universal applicability, but provides representative evidence that the ACM reference model accommodates heterogeneous configuration abstractions while preserving common governance semantics. 

The evaluation also contributes a structured evaluation methodology combining normative governance scenarios, automated tests, and semantic-preservation analysis. This methodology distinguishes successful representation from exact structural equivalence and records which governance-relevant properties are preserved, reconstructed, approximated, or unavailable. It therefore provides a more appropriate evaluation criterion for a reference model than conventional runtime-performance benchmarking. The experimental protocol further complements this qualitative validation with a dedicated quantitative assessment of deterministic impact propagation. Together, the two experimental campaigns demonstrate not only that heterogeneous configurations can be governed through a common reference model, but also that the resulting governance analyses remain reproducible and quantitatively consistent across complementary execution frameworks.

Taken together, these four contributions establish ACM as a configuration-governance abstraction positioned between agentic execution environments and higher-level operational or organizational governance processes. The reference model defines what is governed, the formal semantics define how governance states are computed, semantic projection connects heterogeneous frameworks to the model, and the reference implementation demonstrates that these mechanisms can be operationalized consistently.

\subsubsection{Integration Rationale and Complementary Roles}
\label{subsec:ablation_mechanism}

The contribution of ACM does not rely on the novelty of each constituent mechanism in isolation. Immutable revisions, baselines, provenance, dependency models, and fixed-point computation have established foundations in software engineering. ACM combines these mechanisms with agentic configuration typing, configuration--runtime separation, semantic projection, and explicit governance semantics because they address complementary requirements of framework-independent configuration governance. The following analysis clarifies the distinct role of each mechanism and the dependencies among them. It is an architectural rationale rather than an experimental ablation: the study does not claim that the proposed decomposition is the unique architecture capable of satisfying these requirements.

\paragraph{Agentic typing of configuration items}. Removing the agentic specialization of the ACI taxonomy reduces the model to conventional software configuration management operating over opaque artifacts. 
The immediate consequence is the loss of relationship typing: without
distinguishing agents, prompts, tools, models, and policies, the typed
configuration relationships used to associate dependencies with their
applicable propagation policies (Section~\ref{sec:propagation_policies}) can no longer be assigned, and impact propagation loses the semantic distinctions encoded by those policies. The typing is therefore not merely descriptive convenience: within ACM, it provides the semantic basis required to associate heterogeneous configuration relationships with governance policies and impact semantics.

\paragraph{Configuration–runtime separation}. Collapsing the distinction between immutable configuration revisions and runtime entities destroys the property that makes governance evaluation reproducible. If runtime observations were permitted to enrich the revisions they instantiate, every execution would mutate the governed configuration, and identical configurations executed under different stochastic model outputs would no longer yield identical governance states. The reproducibility results of Section~\ref{subsec:eval-reproducibility} depend entirely on this separation: governance evaluation is reproducible because it is defined over the immutable configuration graph and is insulated from the probabilistic behavior of the underlying language models. Remove the separation and this insulation, together with deterministic replay, no longer holds.

\paragraph{Four-graph decomposition}. The decomposition into Configuration, Evolution, Assurance, and Runtime graphs separates structural dependencies, revision lineage, governance evidence, and runtime observations into distinct semantic views. In particular, impact propagation is defined exclusively over the Configuration Graph, allowing the fixed-point semantics of Appendix~\ref{appendix:formal_properties} to operate over a well-defined dependency structure without mixing runtime, assurance, or evolution relationships. The four-graph decomposition is therefore a design choice supporting separation of concerns and a precisely scoped propagation domain; this work does not claim that it is the only architecture capable of providing these properties.

\paragraph{Deterministic relation governance and propagation}. If relationships were preserved for traceability but not governed by explicit propagation policies, ACM would retain dependency visibility while losing the ability to compute a stable, policy-differentiated impact set. The consequence is observable in the comparative impact study: a dependency-visible but ungoverned traversal that stops at direct dependents identifies only six of the nine genuinely affected items on a 13-item, 18-relation configuration, missing three items affected transitively through an intermediate revision. Deterministic fixed-point propagation over governed relationships is precisely what recovers the three transitively affected items that a naïve one-level traversal omits. The mechanism therefore contributes an effect that dependency visibility alone does not produce.

\paragraph{Semantic projection}. Without semantic projection, governance would have to be redefined per framework, and cross-framework governance-equivalence — the emergent property of the model — could not exist. The necessity of projection is not merely that it enables extraction, but that it makes the boundary of extractability explicit rather than silent. The preservation study makes this concrete: across the three frameworks, the same normative perimeter is reached through three distinct introspection regimes, and projection records where information is extracted, where it is supplied by adapter metadata, and where it is approximated or unsupported — for instance, the opaque conditional-branch semantics preserved only as topological abstraction, and the CrewAI Flow state schema reported as unsupported rather than silently dropped. A model lacking projection could still be instantiated on a single framework; it could not normalize heterogeneous native topologies into one governed representation while documenting its own losses. This documented-loss behavior is what distinguishes projection from ad hoc per-framework extraction.

\paragraph{Immutable revisions, baselines, and provenance}. These three mechanisms are the ones ACM inherits most directly from established configuration management, and individually none is novel. Their necessity within ACM is nonetheless structural: immutable revisions are the objects over which the propagation lattice is defined and without which deterministic reconstruction is impossible; baselines are the immutable reference against which runtime replay is anchored (Appendix~\ref{appendix:runtime_semantics}; and provenance is what preserves the configuration-to-runtime link that the config–runtime separation would otherwise sever. The point is not that these mechanisms are new, but that removing any of them breaks a property that the specifically agentic mechanisms above depend upon.

\paragraph{Summary}. Taken together, these mechanisms address distinct but interdependent requirements of agentic configuration governance. Agentic typing provides a common vocabulary for heterogeneous configuration artifacts; configuration--runtime separation preserves immutable governed configurations; the four-graph decomposition separates configuration, evolution, assurance, and runtime concerns; deterministic propagation evaluates governed dependencies;
and semantic projection maps heterogeneous native representations onto the configuration model on which these semantics operate. Their integration enables the property evaluated in RQ3: cross-framework governance equivalence after semantic projection for governance-equivalent configurations within the evaluated scope. This property is not attributed to any constituent mechanism in isolation. The present analysis establishes the architectural rationale for
their combination but does not constitute an experimental ablation study.

\subsection{Practical Implications}
\label{sec:practical_implications}

Beyond its conceptual contribution, ACM has practical implications for the engineering and governance of agentic systems. By introducing an explicit and framework-independent configuration representation, the proposed model enables governance activities that are difficult to perform when configuration information remains fragmented across framework-specific abstractions.

The experimental evaluation indicates that these governance capabilities remain applicable across heterogeneous execution environments despite substantial differences in their native introspection mechanisms. Consequently, governance policies can be formulated over a stable configuration representation rather than being coupled to framework-specific execution abstractions. This separation is particularly important as agentic ecosystems continue to diversify and evolve.

A first implication concerns reproducibility. In current agentic ecosystems, reproducing a previous execution often requires reconstructing configurations from multiple heterogeneous artifacts, including prompts, workflow definitions, deployment parameters, runtime traces, and external resources. ACM addresses this fragmentation by associating executions with immutable configuration revisions and controlled release baselines. A governed execution can therefore be traced back to the exact configuration from which it originated, supporting reproducible experiments, regression analyses, and long-term maintenance. The experimental campaigns further indicate that reproducibility extends beyond immutable configuration snapshots. Equivalent governed representations, deterministic impact propagation, and identical governance outcomes were obtained across repeated executions and heterogeneous execution frameworks. These observations suggest that governance reproducibility depends primarily on the normalized ACM representation rather than on the native execution environment.

A second implication concerns auditability and traceability. Since governance decisions are explicitly linked to identified configuration revisions and supporting evidence, the provenance of a validation result or lifecycle transition becomes inspectable rather than implicit. This capability facilitates post-deployment analysis, compliance reporting, and forensic investigation by making configuration evolution transparent throughout the lifecycle of the system.

The explicit representation of dependencies also supports systematic change management. Instead of reasoning independently about prompts, agents, tools, or models, ACM represents these artifacts as interconnected configuration items whose relationships can be analyzed before modifications are promoted to a released baseline. Dependency-aware impact analysis provides early visibility into the potential consequences of configuration changes, reducing the likelihood of unintended regressions. The quantitative evaluation additionally shows a reduction in the residual configuration inspection scope under the experimental protocol. Rather than characterizing human effort, the reported metric compares the exhaustive native inspection scope with the residual configuration elements requiring verification after ACM has produced the governed impact set. It therefore quantifies inspection scope rather than inspection time, operational cost, or human productivity.

The proposed model also provides a common governance abstraction that may facilitate interoperability across heterogeneous execution environments. Organizations frequently combine multiple orchestration frameworks to address different operational requirements. By projecting these heterogeneous configurations into a shared governance representation, ACM enables governance policies to be expressed independently of the underlying execution technology while allowing each framework to preserve its native execution semantics. 
The evaluation also indicates that interoperability should not be interpreted solely as compatibility between orchestration frameworks. The successful projection of explicit workflow graphs, metadata-assisted orchestration, and delegation-based agent systems suggests that governance interoperability can also be achieved across heterogeneous introspection regimes, provided that governance-relevant semantics are preserved during semantic projection.

ACM may therefore contribute to the emerging AgentOps ecosystem by complementing existing observability and orchestration platforms with explicit configuration governance. Current operational platforms primarily focus on execution monitoring, tracing, evaluation, and operational metrics. ACM addresses a complementary concern by governing the configuration artifacts that determine how those executions are produced. Consequently, the model should be viewed as a governance layer that can coexist with, rather than replace, existing AgentOps and LLMOps infrastructures. 
From this perspective, ACM complements operational observability rather than competing with it. Execution-oriented platforms primarily describe what occurred during runtime, whereas ACM governs the configuration artifacts, dependencies, and assurance evidence that explain why a governed execution was possible. The combination of both perspectives provides a more complete governance foundation spanning configuration, execution, and evolution.

\subsection{Limitations of the Current Reference Model}
\label{subsec:limitations}

The objective of ACM is to provide a framework-independent governance representation for the governance of agentic system configurations rather than an execution platform or an autonomous reasoning architecture. Accordingly, the current version focuses on representing governed configuration artifacts, their lifecycle, dependencies, provenance, assurance, and runtime conformance, while intentionally leaving the internal execution logic of agentic systems outside the scope of the reference model.

Table~\ref{tab:limitations_scope} summarizes the current coverage of ACM and the principal capabilities intentionally deferred to future work.

\begin{table}[t]
	\caption{Current scope of the ACM reference model. The two columns summarize capabilities currently covered by ACM and capabilities intentionally outside the scope of the present model; entries are not paired row by row.}
	\label{tab:limitations_scope}
	\centering
	\begin{tabular}{ll}
		
		\toprule
		
		Currently covered &
		Not yet covered \\
		
		\midrule
		
		Semantic projection &
		Distributed execution \\
		
		Configuration governance &
		Multi-agent planning strategies \\
		
		Immutable revision management &
		Learning and adaptation mechanisms \\
		
		Lifecycle semantics &
		Autonomous reasoning policies \\
		
		Dependency propagation &
		Self-modifying execution strategies \\
		
		Runtime provenance &
		Collective agent negotiation \\
		
		Assurance evaluation &
		Long-term memory management \\
		
		Deterministic replay &
		Continual learning \\
		
		Framework-independent governance &
		Native communication protocols (e.g., MCP, A2A) \\
		
		Cross-framework governance &
		Large-scale industrial validation \\
		
		\bottomrule
		
	\end{tabular}
	
	\end{table}

These limitations result from deliberate modeling decisions rather than technical constraints. ACM intentionally separates governance concerns from execution concerns in order to establish a stable and framework-independent governance layer. Integrating planning algorithms, reasoning strategies, learning mechanisms, or execution optimizations into the reference model would both increase its complexity and compromise the conceptual independence required for long-term interoperability.

Consequently, ACM treats reasoning engines, orchestration frameworks, planning modules, learning components, communication protocols, and other execution technologies as external systems whose observable configuration and runtime behavior can be represented and governed without prescribing their internal operation. ACM therefore specifies \emph{what} is governed---configuration revisions, baselines, dependencies, runtime observations, and assurance evidence---rather than \emph{how} autonomous agents reason, plan, negotiate, or learn.

This separation is intended to preserve the stability of the governance model as agentic technologies evolve. New execution frameworks, orchestration paradigms, communication protocols, or reasoning techniques may be incorporated through semantic projection without modifying the governance semantics, provided that their governance-relevant abstractions can be represented through the existing ACM concepts and relationships.

Although the present evaluation demonstrates consistent governance semantics across three representative frameworks, these results should not be interpreted as establishing universal framework coverage. Additional validation across emerging execution environments, communication protocols, and industrial-scale deployments remains necessary to assess the broader applicability of the proposed governance model.

The experimental validation remains limited to the representative orchestration frameworks and configuration scales considered by the reference implementation. Although the reported results provide encouraging evidence regarding the generality of the proposed governance semantics across complementary introspection regimes, they remain confined to controlled experimental scenarios. Broader empirical validation involving additional execution frameworks, industrial-scale configurations, human-centered governance workflows, and emerging agent communication protocols will be required before stronger claims regarding general applicability can be made.

\subsection{Section Summary}

This discussion has positioned ACM as a framework-independent governance layer dedicated to the management of agentic system configurations throughout their lifecycle. Rather than introducing another orchestration framework or execution model, ACM separates governance from execution by providing a common semantic representation upon which lifecycle management, assurance evaluation, dependency analysis, provenance tracking, runtime traceability, and controlled evolution can be performed independently of the originating execution framework.

The experimental evaluation strengthens this positioning by demonstrating that the proposed governance semantics remain applicable across heterogeneous execution frameworks and complementary introspection regimes. Native configurations originating from explicit workflow graphs, metadata-assisted orchestration, and delegation-oriented agent systems are normalized through semantic projection into governance-equivalent ACM representations. 

Together with the quantitative evaluation of deterministic impact propagation, these results indicate that governance can be expressed over a stable and framework-independent configuration model despite substantial differences in native execution abstractions.

The proposed reference model should therefore be viewed as a governance abstraction complementing existing AgentOps and agentic execution ecosystems rather than replacing them. By combining a unified configuration representation with formal governance semantics and reproducible operationalization, ACM establishes a common foundation for configuration governance while remaining intentionally independent of framework-specific orchestration mechanisms. The following section discusses the principal threats to the validity of the present study and the factors that should be considered when interpreting these results.

\section{Threats to Validity}
\label{sec:threats_validity}

The preceding sections presented the ACM reference model, its formal governance semantics, its operationalization through a reference implementation, and an experimental evaluation conducted across three heterogeneous agentic frameworks spanning complementary introspection regimes. Together, these elements provide formal and empirical evidence that the proposed model is operationally realizable, that the reference implementation conforms to the specified governance semantics for the evaluated scenarios, and that the model provides the representational coverage required by those scenarios.

As with any reference model, however, these results must be interpreted within the boundaries of the assumptions, implementation choices, and experimental conditions that characterize the present study. The purpose of this section is therefore not to question the conceptual validity of ACM itself, but to identify the principal factors that may influence the interpretation, reproducibility, and generalization of the reported results.

Following established empirical research practice, we distinguish threats related to the scope of the reference model, the validation of its formal semantics, the internal validity of the experimental methodology, the external validity of the reported conclusions, and the reproducibility of the evaluation artifacts. These limitations define the current scope of evidence supporting ACM and naturally motivate the research directions discussed in the following section.

\subsection{Validity of the Reference Model}
\label{sec:validity_reference_model}

The primary objective of this study is to evaluate whether ACM provides a sufficiently expressive and framework-independent governance representation for representing and governing heterogeneous agentic system configurations. Consequently, the reported results should be interpreted as evidence supporting the proposed reference model within the scope of the evaluated frameworks rather than as a demonstration of universal applicability.

The evaluation intentionally considers three representative agentic frameworks exhibiting complementary introspection regimes. LangGraph exposes an explicit execution graph, CrewAI combines declarative orchestration with partial semantic reconstruction, whereas the OpenAI Agents SDK derives governance topology from explicit agent handoff relationships. Together, these frameworks exercise explicit introspection, partial introspection, and delegation-oriented topology reconstruction. Successfully projecting these heterogeneous representations into a common ACM configuration model provides evidence that the proposed governance abstractions are not tied to a single execution paradigm or introspection mechanism. Nevertheless, the evaluation remains limited to these three representative frameworks and should not yet be interpreted as exhaustive validation of the broader agentic ecosystem.

Other categories of agentic systems, including architectures relying on dynamic handoffs, highly decentralized coordination mechanisms, large-scale enterprise orchestration platforms, or future framework abstractions, have not yet been experimentally evaluated. Although the semantic projection architecture was explicitly designed to accommodate heterogeneous native representations, additional adapters and experimental studies are required before broader claims regarding framework independence can be established empirically.

Accordingly, the conclusions drawn from this study should be understood as demonstrating governance portability across three complementary introspection regimes rather than complete universality across every existing or future agentic framework. Additional validation across emerging execution environments, communication protocols, and industrial deployments remains necessary before stronger claims regarding general applicability can be established.

\subsection{Validity of the Formal Semantics}
\label{sec:validity_formal_semantics}

The governance semantics introduced in this work define the formal behavior of the ACM reference model through lifecycle state spaces, local quality and assurance evaluation, deterministic impact propagation, eligibility evaluation, and runtime reconstruction semantics.

Unlike purely conceptual reference models, these semantics are operationalized by a reference implementation that executes the corresponding state computations, validation procedures, and propagation algorithms. Consequently, the reported experimental results validate not only the structural expressiveness of the reference model but also the practical executability of its governance semantics.

The present study does not provide machine-assisted verification of the complete ACM governance model. The mathematical development establishes specific properties of the impact-propagation semantics, including monotonicity, convergence, termination, and least-fixed-point computation under the stated assumptions. The experimental evaluation separately provides conformance evidence showing that the reference implementation realizes the specified semantics for the evaluated scenarios. These results should not be interpreted as a proof of correctness of every component or future extension of the complete governance model.

Although the present work establishes formal properties including monotonicity, convergence, termination, and deterministic fixed-point computation under the assumptions of the ACM governance model, these results are derived for the reference semantics defined in this work. Extending these guarantees to future extensions of the metamodel, additional propagation policies, or independently developed implementations remains outside the scope of the present study.

Accordingly, the formal semantics should be regarded as an executable semantic specification supported by experimental evidence rather than as a mathematically verified formal system. Extending ACM with machine-checked proofs or theorem-based verification constitutes an important direction for future research and would further strengthen the theoretical foundations of the proposed governance model.

\subsection{Generalizability}
\label{sec:generalizability}

The evaluation reported in this study provides initial evidence that the ACM reference model can represent and govern heterogeneous agentic systems originating from multiple orchestration paradigms. Nevertheless, the scope of the experimental validation remains intentionally limited and should not be interpreted as demonstrating universal applicability across the rapidly evolving ecosystem of agentic frameworks.

The experimental study covers three representative frameworks spanning complementary introspection regimes. LangGraph represents explicit graph-based introspection, CrewAI combines declarative orchestration with partial semantic reconstruction, and the OpenAI Agents SDK reconstructs governance topology from explicit delegation relationships. Together, these frameworks provide broader evidence that ACM governance semantics remain applicable despite substantial differences in native observability and execution abstractions.

\begin{table}[t]
\caption{Scope of the current empirical validation and representative future validation targets.}
\label{tab:validation_scope}
\centering
\begin{tabular}{p{4.0cm}cc}
\toprule
\textbf{Agentic System Category} & \textbf{Evaluated} & \textbf{Future Work} \\
\midrule
Graph-based orchestration (LangGraph) & \checkmark & \\
Role/task-based orchestration (CrewAI) & \checkmark & \\
Delegation-based handoff architectures (OpenAI Agents SDK) & \checkmark & \\
Hierarchical multi-agent delegation & & \checkmark \\
Swarm-based coordination & & \checkmark \\
Pipeline-oriented orchestration & & \checkmark \\
Large-scale enterprise deployments & & \checkmark \\
Cross-organizational governance & & \checkmark \\
\bottomrule
\end{tabular}
\end{table}

Nevertheless, the evaluation does not yet include several emerging categories of agentic systems, including decentralized swarms, federated multi-agent ecosystems, large-scale industrial deployments, communication-centric architectures, or future orchestration paradigms that may expose fundamentally different governance-relevant abstractions. Consequently, the present evaluation should be interpreted as representative rather than exhaustive.

Similarly, the current experiments were conducted on normative governance scenarios and a reference implementation intended to evaluate the representational coverage and operational feasibility of the proposed model within the experimental scope.
They do not constitute an evaluation of industrial deployments involving thousands of configuration items, heterogeneous execution infrastructures, long-running production systems, or continuously evolving organizational governance processes.

Consequently, the conclusions of this study should be interpreted as evidence that ACM provides a viable framework-independent governance model for the classes of agentic systems evaluated, rather than as proof of universal applicability to every existing or future orchestration framework. Extending the empirical validation to additional ecosystems and deployment contexts constitutes an important direction for future work.

\subsection{Reproducibility and Replicability}
\label{subsec:reproducibility}

Reproducibility was a primary design objective throughout the development and evaluation of ACM. To facilitate independent verification, the complete experimental protocol is defined through declarative fixtures, normative evaluation scenarios, deterministic governance rules, and an immutable event model. These artifacts enable independent implementations to evaluate the same governance properties under comparable experimental conditions.

An important characteristic of the evaluation methodology is that the
experimental scenarios are \emph{normative} rather than demonstrative. Each scenario specifies a governance property that the reference implementation is expected to verify, regardless of whether the evaluated execution represents a valid or an invalid configuration. Consequently, successful evaluation does not necessarily imply that the tested configuration is accepted by ACM; in many cases, the expected outcome is precisely the detection of a governance
violation.
The normative scenarios and their expected outcomes were defined as part of the same study that developed the ACM reference model and reference implementation. They therefore provide conformance evidence against the specified governance semantics rather than an independent empirical validation of those semantics. This shared authorship introduces a risk of common assumptions between the model, the scenarios, and the operationalization, which is mitigated by declarative fixtures, explicit expected outcomes, separation of projection and governance computation, and reproducible experimental artifacts, but is not eliminated by the present study.

Several scenarios intentionally introduce invalid lifecycle transitions, missing dependencies, unauthorized runtime behavior, or configuration drift. The correct behavior of the governance kernel is therefore to detect, classify, and report these inconsistencies according to the formal semantics defined by the reference model. From this perspective, rejecting an invalid configuration constitutes a successful execution of the corresponding normative scenario.

This distinction proved particularly valuable during the iterative development of the reference implementation. Two scenarios (ACM-S02 and ACM-S20) initially revealed inconsistencies between the normative specification and the prototype implementation. Rather than invalidating the experimental methodology, these results demonstrated its effectiveness as a validation mechanism: the observed discrepancies led to refinements of the governance semantics before completion of the evaluation campaign.

Replicability is supported by two complementary forms of evidence. The formal semantics establish deterministic propagation properties under the assumptions stated in Section~\ref{sec:governance_semantics}, while repeated experimental executions verify that the reference implementation realizes this behavior for the evaluated configurations. Given identical configuration revisions, governance policies, assurance evidence, and runtime events, repeated executions produce equivalent governance states and runtime reconstructions. 
This governance-level reproducibility does not require reproduction of the probabilistic outputs generated by the underlying LLMs, because ACM governs configuration and normalized execution semantics rather than model inference itself.
This deterministic behavior is additionally supported by the quantitative impact evaluation. Nine representative impact scenarios, combining three propagation classes across the three evaluated frameworks, were each executed repeatedly under identical conditions. Across all repetitions, the governance kernel produced identical governed impact sets, identical fixed-point convergence behavior, and identical quantitative metrics. These observations provide empirical evidence that governance reproducibility depends on the normalized ACM representation and deterministic governance semantics rather than on framework-specific execution behavior.

The publication of the reference implementation, the declarative test
fixtures, the normative scenario suite, and the associated evaluation protocol allows future implementations of ACM to reproduce the experimental campaign, compare governance behaviors, and evaluate semantic equivalence independently of the original prototype. The reproducibility objective of this work is therefore not to reproduce a specific software implementation, but to enable independent verification of the governance semantics defined by the ACM reference model.
The evaluation protocol also deliberately separates semantic projection, governance computation, and experimental instrumentation. Measurement components observe the governed representations produced by the governance kernel without participating in governance state computation. This architectural separation reduces the risk that evaluation artifacts influence the semantics being evaluated and contributes to the reproducibility of the reported results.

\subsection{Mitigation Strategies}
\label{sec:mitigation_strategies}

Although the preceding subsections identify several limitations affecting the present study, a number of methodological choices were deliberately adopted to reduce their impact on the reported results. Rather than attempting to maximize the breadth of the evaluation, the experimental methodology emphasizes internal consistency, deterministic governance semantics, and explicit separation between conceptual modeling, operationalization, and experimental validation.

First, the reference model was intentionally designed independently of any particular execution framework. The semantic projection architecture isolates framework-specific extraction mechanisms from the governance kernel, allowing the latter to remain unchanged across heterogeneous orchestration paradigms. This separation reduces implementation bias and facilitates the extension of the evaluation to additional frameworks without modifying the underlying governance semantics.

Second, the evaluation relies on normative governance scenarios rather than application-specific benchmarks. Each scenario targets a well-defined property of the reference model, such as lifecycle management, dependency propagation, assurance computation, baseline validation, or runtime governance. This scenario-driven methodology improves the traceability between the formal specification, the reference implementation, and the experimental evidence.

Third, reproducibility is reinforced through deterministic governance computation, immutable configuration revisions, canonical serialization, replayable runtime evidence, and automated validation procedures. These mechanisms ensure that governance outcomes depend exclusively on governed configurations and observed runtime events rather than on the stochastic behavior of the underlying language models.

In conclusion, this article deliberately distinguishes between the conceptual reference model, its formal governance semantics, the reference implementation, and the experimental evaluation. This separation avoids conflating theoretical contributions with implementation artifacts and allows future implementations to validate the same governance semantics independently of the current prototype.

Table~\ref{tab:mitigation_strategies} summarizes the principal threats identified in this section together with the corresponding mitigation strategies adopted throughout the study.

\begin{table}[t]
\caption{Summary of Threats and Mitigation Strategies}
\label{tab:mitigation_strategies}
\centering
\begin{tabular}{p{4.2cm}p{3.9cm}}
\toprule
\textbf{Threat} & \textbf{Mitigation Strategy} \\
\midrule
Limited coverage of currently evaluated introspection regimes &
Evaluation across heterogeneous orchestration paradigms; framework-independent governance kernel. \\

Incomplete formal verification &
Executable formal semantics supported by deterministic implementation and normative validation scenarios. \\

Limited empirical generalization &
Conservative interpretation of results; explicit delimitation of the evaluation scope. \\

Implementation-dependent bias &
Separation between semantic projection, governance kernel, and framework adapters. \\

Experimental reproducibility &
Immutable revisions, replayable runtime events, canonical serialization, and automated validation suite. \\

Emerging agentic architectures &
Projection architecture designed to accommodate additional adapters without modifying the reference model. \\

Experimental instrumentation bias &
Strict separation between semantic projection, governance computation, and experimental instrumentation; evaluation artifacts remain external to the governance kernel. \\

Experimental instrumentation coupling &
Strict architectural separation between semantic projection, governance computation, and evaluation instrumentation prevents experimental measurements from influencing governed state computation. \\

\bottomrule
\end{tabular}
\end{table}

\subsection{Section Summary}
\label{sec:threats_summary}

The objective of this section has been to delimit the scope within which the conclusions of the present study should be interpreted. The principal limitations concern the breadth and authorship of the experimental validation, the absence of machine-assisted verification of the complete governance model, and the extent to which the reported results can be generalized across the rapidly evolving landscape of agentic systems.

The experimental methodology intentionally favors methodological rigor over exhaustive coverage by combining a framework-independent governance representation, executable governance semantics, deterministic validation procedures, and reproducible experimental artifacts. Within this scope, the reported results provide credible evidence supporting the feasibility and expressiveness of the proposed governance model while avoiding claims that extend beyond the evaluated configurations.

These limitations should therefore be interpreted as defining the current scope of empirical evidence rather than indicating inconsistencies in the proposed governance model. Within this scope, the combination of formal semantics, deterministic operationalization, qualitative validation, and quantitative impact evaluation provides consistent evidence supporting the feasibility of framework-independent governance for heterogeneous agentic systems. The remaining limitations primarily concern the breadth of empirical validation and the progressive extension of the reference model to the rapidly evolving agentic ecosystem.

\section{Future Work}
\label{sec:future_work}

The ACM reference model introduced in this paper constitutes an initial step toward a framework-independent approach to the governance of agentic systems. The proposed metamodel, governance semantics, reference implementation, and experimental evaluation collectively demonstrate the feasibility of this approach within the scope considered in the present study. At the same time, the limitations identified in the previous section highlight several opportunities for extending both the theoretical foundations and the practical applicability of ACM.

Future research therefore naturally follows three complementary directions. The first concerns the evolution of the reference model itself to accommodate emerging abstractions while preserving its governance principles. The second focuses on strengthening the empirical and theoretical validation of the proposed semantics through broader experimental studies and formal verification. The long-term perspective is to position ACM as a common governance layer capable of interoperating with the rapidly expanding ecosystem of agentic development, deployment, and observability platforms.

\subsection{Extension of the Reference Model}
\label{sec:future_reference_model}

The reference model presented in this paper intentionally focuses on the minimal set of governance concepts required to represent heterogeneous agentic systems independently of their execution framework. This design choice favors conceptual stability and limits the number of core abstractions to those introducing distinct governance semantics. Nevertheless, the rapid evolution of agentic ecosystems will inevitably lead to the emergence of new configuration artifacts, interaction patterns, and execution abstractions that may require additional modeling capabilities.

A primary direction for future research is therefore the controlled evolution of the ACM metamodel while preserving its foundational design principles. Rather than continuously introducing framework-specific concepts into the core model, future extensions should remain guided by semantic necessity: a new abstraction should only become part of the reference model if it introduces governance properties that cannot be represented through the existing concepts and relationships. This principle preserves the framework independence of ACM while avoiding unnecessary growth of the metamodel.

The experimental evaluation reported in this work provides an initial empirical basis for this extension strategy. By successfully projecting three representative frameworks spanning complementary introspection regimes into a common governance representation, the current results suggest that future extensions should primarily address genuinely new governance abstractions rather than framework-specific implementation differences.

An important example concerns reusable composite capabilities that are increasingly appearing under different names across modern agentic platforms, such as skills, reusable workflows, capability modules, or packaged execution components. Although these constructs differ in their implementation details, they frequently encapsulate existing governed artifacts—including prompts, tools, models, policies, workflows, and external resources—rather than introducing fundamentally new governance semantics. Consequently, many of these emerging abstractions can already be represented through semantic projection as governed composite configuration items without requiring modifications to the ACM core.

An important direction concerns the governance of increasingly composite AI systems, including Retrieval-Augmented Generation (RAG) pipelines and other retrieval-centric architectures. Rather than introducing a dedicated governance abstraction, ACM can represent RAG pipelines through the composition of existing configuration items, including embedding models, retrieval policies, vector indexes, knowledge repositories, prompt definitions, and retrieval components. This representation enables immutable versioning of the complete retrieval configuration, explicit provenance of derived vector indexes, and governance of retrieval policies independently of the underlying storage technology. Consequently, ACM can govern the evolution of retrieval-augmented systems without embedding retrieval-specific concepts into its core metamodel.

Another important direction concerns self-learning or continuously adaptive agents. Such systems challenge traditional configuration management because parts of their operational behavior may evolve autonomously through accumulated experience, parameter adaptation, or long-term memory updates. Extending ACM to distinguish immutable governed configurations from controlled runtime knowledge evolution would enable adaptive systems to remain auditable while preserving reproducibility of approved configurations. Future work should therefore investigate governance semantics capable of representing learning artifacts, adaptation boundaries, and evidence associated with autonomous configuration evolution.

Finally, the increasing adoption of interoperability protocols and workflow orchestration platforms, including Model Context Protocol (MCP), Agent-to-Agent (A2A) communication, and low-code orchestration environments such as n8n, provides another opportunity for extending the reference model. Rather than incorporating protocol-specific abstractions, ACM can represent these ecosystems through semantic projection of communication endpoints, external services, orchestration workflows, and interaction contracts into governed configuration items. This approach preserves the framework-independent nature of the reference model while enabling configuration governance across heterogeneous execution environments and emerging interoperability standards.

Future extensions may also address governance capabilities that extend beyond the scope of the current reference model. Examples include distributed governance across multiple organizations, federated configuration repositories, cross-system dependency management, richer policy composition mechanisms, and governance of long-lived autonomous ecosystems whose configuration evolves continuously over time. These extensions would broaden the applicability of ACM while preserving the separation between configuration governance and execution that constitutes one of its central architectural principles.

Ultimately, the long-term evolution of ACM should prioritize conceptual stability over feature accumulation. As with mature reference models in software engineering, the objective is not to mirror every innovation introduced by individual frameworks, but rather to identify the stable governance abstractions that remain applicable despite the continuous evolution of execution technologies. This philosophy enables the reference model to evolve incrementally while maintaining interoperability, explainability, and long-term sustainability.

\subsection{Broader Framework Validation}
\label{sec:future_framework_validation}

The empirical evaluation presented in this paper intentionally focuses on three representative frameworks spanning complementary introspection regimes in order to assess the feasibility of framework-independent configuration governance. Although the semantic projection of LangGraph, CrewAI, and the OpenAI Agents SDK provides encouraging evidence supporting the proposed reference model within this scope, broader validation across the rapidly evolving landscape of agentic systems remains an essential research direction.

Future work should therefore prioritize the evaluation of additional classes of agentic architectures rather than merely increasing the number of supported execution frameworks. The objective is to determine whether fundamentally different governance-relevant abstractions require extensions of the semantic projection contract or can be represented using the existing ACM concepts. In particular, architectures relying on dynamic delegation, decentralized collaboration, communication-centric coordination, adaptive execution, or large-scale distributed governance, would provide valuable opportunities to evaluate the expressiveness of the ACM metamodel under substantially different execution models.

Beyond execution frameworks themselves, validation could also incorporate complementary components of the emerging AgentOps and LLMOps ecosystem. Provenance platforms, observability infrastructures, evaluation frameworks, and deployment management systems increasingly expose governance-relevant metadata that could naturally participate in semantic projection. Studying the integration of ACM with such platforms would provide additional evidence regarding the ability of the reference model to serve as a common governance layer across heterogeneous operational environments.

Another important direction concerns empirical validation at industrial scale. The current experiments intentionally emphasize representational coverage, semantic conformance, and reproducibility rather than operational scalability. Future studies should therefore evaluate ACM on large agentic systems involving hundreds or thousands of governed configuration items, long-lived execution histories, continuously evolving baselines, and collaborative governance processes distributed across multiple development teams. Such evaluations would provide quantitative evidence regarding scalability, governance overhead, and operational usability in realistic production environments.

Ultimately, extending the experimental validation is not intended merely to increase the number of supported frameworks. Rather, it aims to progressively strengthen the empirical evidence supporting the central hypothesis of this work: that governance semantics can remain stable despite the continuous evolution of agentic execution technologies.

The current evaluation already indicates that governance portability extends across three complementary introspection regimes. Future validation should therefore progressively investigate whether this portability also extends across new architectural families, organizational scales, and interoperability mechanisms.

\subsection{Formal Verification of Governance Semantics}
\label{sec:future_formal_verification}

The governance semantics introduced in this paper are defined through a formal operational specification and validated through deterministic execution of the reference implementation. While this approach establishes formal properties and demonstrates their operational realizability, it does not yet provide machine-assisted verification of the complete governance model.

One immediate objective concerns machine-assisted verification of the fixed-point propagation semantics already established analytically in this work. The existing proofs of monotonicity, convergence, termination, and least-fixed-point computation could be encoded in a theorem prover or formal specification environment to obtain independently checkable guarantees and verify that future implementations preserve the stated assumptions and semantics.

A second research direction involves machine-assisted analysis of the governance semantics beyond impact propagation. Lifecycle transitions, local quality and assurance evaluation, eligibility evaluation, and runtime reconstruction could be expressed within a formal verification framework to establish additional invariants concerning transition consistency, governance-state compatibility, provenance preservation, and runtime reconstruction.

Beyond individual algorithms, future work may investigate the formal specification of the ACM reference model as a complete governance system. Model-checking techniques, temporal logic, theorem proving, or specification languages dedicated to distributed systems could be employed to verify global properties of the governance semantics and to detect inconsistencies automatically during the evolution of the reference model. Such approaches would complement the current implementation-based validation with machine-verifiable correctness guarantees.

Formal verification would facilitate the development of independent ACM implementations. By providing mathematically specified governance semantics rather than relying solely on behavioral equivalence with the reference implementation, future implementations could demonstrate conformance through proof of semantic equivalence instead of implementation similarity. This evolution would represent an important step toward establishing ACM as a stable and implementation-independent governance specification.

\subsection{Toward an Agentic Configuration Ecosystem}
\label{sec:future_ecosystem}

The rapid maturation of agentic software engineering has led to the emergence of a rich ecosystem of complementary technologies dedicated to development, deployment, observability, evaluation, provenance, and operational governance. Rather than replacing these existing technologies, ACM is intended to provide a framework-independent governance layer capable of connecting heterogeneous execution frameworks, observability platforms, evaluation systems, and software engineering infrastructures through a common semantic representation of governed configuration artifacts. The experimental results presented in this work provide initial evidence that such a governance layer can remain stable despite heterogeneous execution models and complementary introspection regimes.

\begin{figure}[t]
	\centering
	\includegraphics[width=0.8\textwidth]{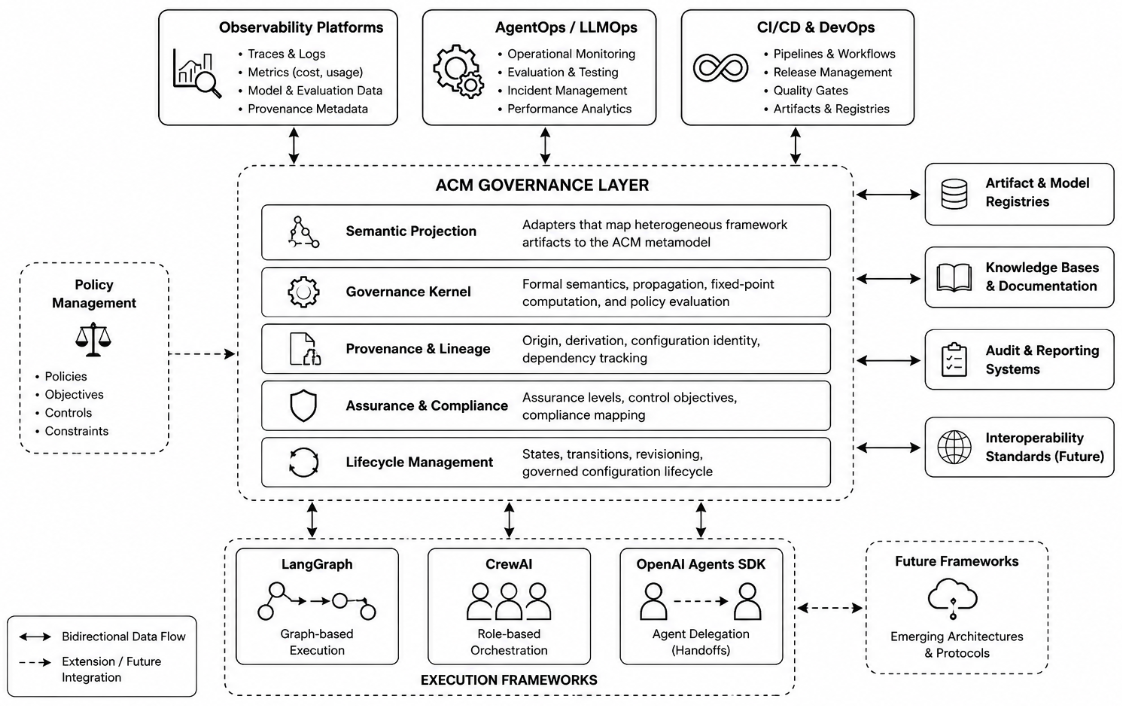}
	\caption{Conceptual positioning of ACM within an agentic configuration ecosystem. Heterogeneous execution frameworks are connected to a common governance layer through semantic projection. This layer provides configuration governance services that interact with AgentOps platforms, software engineering infrastructures, deployment pipelines, observability systems, and future interoperability standards while remaining independent of execution technologies.}
	\label{fig:agentic_configuration_ecosystem}
\end{figure}

A promising direction for future research therefore consists in studying the integration of ACM with broader AgentOps and LLMOps ecosystems. Modern observability platforms increasingly collect execution traces, evaluation results, model usage metrics, cost information, and provenance metadata that constitute valuable inputs for governance processes. Conversely, governance decisions produced by ACM could enrich these platforms with lifecycle status, assurance levels, configuration lineage, dependency analysis, and compliance information, providing a higher semantic level for operational monitoring. This separation also enables governance policies to evolve independently from execution technologies. As orchestration frameworks continue to diversify, governance services built upon ACM could remain stable while semantic projection adapters evolve to accommodate new execution abstractions.

Rather than exchanging framework-specific metadata directly, these heterogeneous systems could interoperate through governed ACM representations that preserve configuration identity, provenance, lifecycle semantics, and assurance information independently of the originating execution technology.

Another important perspective concerns the integration of ACM with software engineering infrastructures. Continuous integration and deployment pipelines, software configuration management systems, artifact repositories, and supply-chain security mechanisms already manage the lifecycle of conventional software assets. Extending these infrastructures to incorporate governed agentic configuration items would enable unified governance across both traditional software components and AI-native artifacts while preserving traceability throughout the complete development lifecycle. Such integration would extend configuration governance beyond software artifacts alone by enabling AI-native configuration items to participate in existing software engineering governance processes while preserving immutable revision identity and reproducible configuration baselines.

Beyond tooling integration, future work may explore the definition of interoperable framework-independent governance services exposing standardized interfaces for semantic projection, configuration validation, governance evaluation, and provenance exchange. Such services would facilitate the development of heterogeneous toolchains in which execution frameworks, observability platforms, deployment systems, and governance engines cooperate while remaining independently evolvable.

Ultimately, the long-term objective is not to establish ACM as another execution platform, but to provide a stable governance substrate upon which heterogeneous execution frameworks, AgentOps platforms, software engineering infrastructures, and future interoperability standards can progressively converge. As the agentic ecosystem continues to evolve, such a framework-independent governance layer may facilitate reproducibility, interoperability, auditability, and controlled evolution without constraining innovation in execution technologies.

\subsection{Standardization Perspectives}
\label{sec:future_standardization}

The absence of a common vocabulary and shared governance abstractions currently represents one of the principal obstacles to interoperability across agentic systems. While execution frameworks continue to evolve rapidly, equivalent concepts are frequently described using framework-specific terminology, making configuration exchange, governance portability, and comparative evaluation unnecessarily difficult.

The ACM reference model represents an initial step toward addressing this fragmentation by proposing a framework-independent governance vocabulary together with formally defined governance semantics and an operational reference implementation. The experimental evaluation provides evidence that these concepts remain applicable across the three evaluated frameworks and complementary introspection regimes without requiring framework-specific governance semantics after projection.
Although the present work does not aim to establish a standard, it provides a structured foundation upon which future community discussions regarding interoperable governance models may build.

Future research may therefore investigate the definition of standardized exchange formats for governed configuration items, semantic projection interfaces, governance evidence, provenance information, and lifecycle metadata. Such specifications would facilitate interoperability between heterogeneous execution frameworks while allowing governance information to remain portable independently of implementation technologies. Likewise, standardized conformance criteria could enable independent implementations to demonstrate governance semantic compatibility with the ACM reference model while preserving implementation freedom. Such specifications would also facilitate independent implementations capable of exchanging governed configuration artifacts while preserving provenance, lifecycle semantics, assurance evidence, and reproducibility across heterogeneous governance environments.

The maturation of ACM may also benefit from collaboration with broader software engineering and artificial intelligence communities working on software configuration management, software supply-chain security, provenance, and AI governance. Aligning ACM with existing standards whenever appropriate would facilitate its integration into established engineering practices while reducing duplication of concepts that are already well understood in adjacent disciplines.

Ultimately, the long-term objective is not to standardize a particular implementation or execution framework, but to encourage convergence toward a common governance reference model for heterogeneous agentic systems. By separating governance semantics from execution technologies while validating this separation across complementary orchestration paradigms and introspection regimes, ACM provides a stable conceptual foundation upon which interoperable governance specifications may progressively emerge. Such convergence could strengthen reproducibility, auditability, configuration portability, and long-term sustainability across an increasingly diverse agentic ecosystem.

\subsection*{Section Summary}

Taken together, these research directions outline a progressive evolution of ACM rather than a redefinition of its core principles. Future work primarily concerns extending empirical validation to additional architectural classes, strengthening formal guarantees, integrating ACM within broader software engineering and AgentOps ecosystems, and fostering interoperable governance specifications. Throughout these developments, the central objective remains unchanged: preserving a stable, framework-independent governance model while enabling the continuous evolution of agentic execution technologies.

\section{Conclusion}
\label{sec:conclusion}

The rapid evolution of agentic software engineering has fundamentally transformed the role of configuration artifacts. Prompts, tools, execution graphs, policies, models, runtime state, and provenance have become first-class engineering assets whose governance can no longer rely solely on traditional software configuration management practices. At the same time, the increasing diversity of orchestration frameworks has introduced heterogeneous configuration abstractions, making governance difficult to reproduce, audit, and compare across platforms.

This paper introduced Agentic Configuration Management (ACM), a framework-independent governance model for heterogeneous agentic systems. Rather than proposing another execution framework, ACM separates governance semantics from execution technologies through a framework-independent reference model combining governed configuration items, immutable revisions, semantic projection, explicit governance semantics, assurance evaluation, and runtime traceability.

Beyond the conceptual contribution, this work demonstrated that the proposed governance semantics can be operationalized through an executable reference implementation and evaluated across three representative agentic frameworks spanning complementary introspection regimes. The evaluation provides scenario-based evidence of representational coverage, conformance of the operationalization to the specified governance semantics, cross-framework governance equivalence after semantic projection, and reproducible quantitative impact behavior. These results support the feasibility of framework-independent configuration governance within the evaluated scope without implying complete informational identity between projections or universal applicability across agentic frameworks.

More broadly, this work argues that governance should become an explicit architectural layer of agentic software engineering rather than an implementation-specific concern embedded within individual execution frameworks. By separating governance from execution through a common reference model, formal governance semantics, and framework-independent operationalization, ACM establishes a foundation upon which interoperability, reproducibility, auditability, controlled evolution, and future governance standards can progressively be built.

Ultimately, ACM does not prescribe how agentic systems should reason or execute; it specifies how their configurations can be governed consistently throughout their lifecycle. 
As execution technologies continue to evolve, the long-term objective of ACM is to provide a stable governance foundation whose semantics can remain decoupled from changing orchestration paradigms while supporting reproducibility, provenance, interoperability, assurance, and controlled configuration evolution. Such a separation between execution and governance may ultimately become a fundamental architectural principle for engineering trustworthy agentic systems.

\appendices

\section{Related Work}
\label{appendix:related_work}
\subsubsection{Framework Comparison}
Table~\ref{tab:framework_comparison} summarizes representative characteristics of several widely used agentic frameworks.
\begin{table*}[t]
	\centering
	\caption{Representative characteristics of widely used agentic frameworks. The comparison illustrates the diversity of execution paradigms while highlighting the absence of a common configuration representation.}
	\label{tab:framework_comparison}
	
	\renewcommand{\arraystretch}{1.15}
	
	\begin{tabular}{p{3.2cm}p{3.1cm}p{3.4cm}p{3.4cm}c}
		\toprule
		\textbf{Framework}
		&
		\textbf{Primary Paradigm}
		&
		\textbf{Execution Model}
		&
		\textbf{Native Configuration Abstraction}
		\\
		\midrule
		
		LangGraph
		&
		State graph
		&
		Explicit graph execution
		&
		Nodes, edges and shared state
		\\
		
		CrewAI
		&
		Role-based multi-agent
		&
		Agents, tasks and crews
		&
		Agents, tasks, crews and flows
		\\
		
		OpenAI Agents SDK
		&
		Delegation-based agents
		&
		Handoffs and tool orchestration
		&
		Agents, tools and handoffs
		\\
		
		Microsoft Agent Framework
		&
		Workflow orchestration
		&
		Agent workflows
		&
		Workflow definitions
		\\
		
		Google ADK
		&
		Hierarchical agents
		&
		Multi-agent orchestration
		&
		Agent hierarchies
		\\
		
		\bottomrule
	\end{tabular}
	
\end{table*}
\subsection{Gap Analysis}
The Table~\ref{tab:gap_analysis_detailed} supports a comparative gap analysis between the different governance aproaches.
\begin{table*}[t]
	\centering
	\caption{Detailed rationale supporting the comparative gap analysis
		presented in Table~\ref{tab:gap_analysis}.}
	\label{tab:gap_analysis_detailed}
	
	\footnotesize
	\setlength{\tabcolsep}{3pt}
	\renewcommand{\arraystretch}{1.18}
	
	\begin{tabularx}{\textwidth}{
			>{\RaggedRight\arraybackslash}p{3.25cm}
			C{1.05cm} 
			C{1.15cm} 
			C{1.45cm} 
			C{1.30cm} 
			Y
		}
		\toprule
		
		\textbf{Governance capability}
		&
		\makecell{\textbf{SCM}}
		&
		\makecell{\textbf{AI}\\\textbf{Gov.}}
		&
		\makecell{\textbf{LLMOps /}\\\textbf{AgentOps}}
		&
		\makecell{\textbf{Agent}\\\textbf{frameworks}}
		&
		\textbf{Remaining gap}
		\\
		
		\midrule
		
		Configuration identification
		&
		$\checkmark$
		&
		$\sim$
		&
		$\sim$
		&
		$\sim$
		&
		SCM explicitly identifies configuration items.
		Other approaches identify only the artefacts relevant to their
		respective governance, operational, or execution objectives.
		\\
		
		Version management
		&
		$\checkmark$
		&
		--
		&
		$\sim$
		&
		$\sim$
		&
		SCM provides mature version control.
		LLMOps platforms and agent frameworks version selected artefacts, such as prompts or workflows, but not complete governed configurations.

		\\
		
		Configuration provenance
		&
		$\sim$
		&
		$\checkmark$
		&
		$\checkmark$
		&
		$\sim$
		&
		SCM preserves software history, while AI governance and LLMOps
		emphasize provenance, accountability, and traceability.
		Frameworks expose partial execution provenance.
		\\
		
		Dependency traceability
		&
		$\checkmark$
		&
		$\sim$
		&
		$\checkmark$
		&
		$\sim$
		&
		SCM manages software dependencies, and LLMOps relates selected
		artefacts such as prompts, traces, and evaluations.
		Frameworks expose dependencies through native abstractions.
		\\
		
		Execution observability
		&
		--
		&
		$\sim$
		&
		$\checkmark$
		&
		$\checkmark$
		&
		Execution monitoring is not a primary SCM capability.
		AI governance recommends monitoring, whereas LLMOps platforms and agent frameworks provide traces and runtime telemetry.
		\\
		
		Lifecycle governance
		&
		$\sim$
		&
		$\checkmark$
		&
		$\sim$
		&
		$\sim$
		&
		SCM governs controlled software evolution, and AI governance defines processes across the AI lifecycle. Operational platforms and frameworks mainly address deployment or
		execution stages.
		\\
		
		Framework-independent representation
		&
		--
		&
		--
		&
		--
		&
		--
		&
		Existing approaches remain centred on software artefacts, organizational processes, operational platforms, or native execution models; none provides a common representation of governed agentic configurations across execution frameworks.
		\\
		
		Unified configuration model
		&
		--
		&
		--
		&
		--
		&
		--
		&
		The capabilities remain distributed across distinct abstractions; no single family provides a common configuration model spanning heterogeneous agentic artefacts, revision identity, typed dependencies, governance semantics, and runtime provenance.
		\\
		
		\bottomrule
	\end{tabularx}
	
	\vspace{0.4em}
	
	{\scriptsize
		$\checkmark$ Fully addressed
		\hspace{1em}
		$\sim$ Partially addressed
		\hspace{1em}
		-- Not provided as a primary capability.
	}
	
\end{table*}

\section{Formal Foundations}
\label{appendix:formal_foundations}

This appendix provides the mathematical definitions supporting the governance semantics introduced in Section~\ref{sec:governance_semantics}. It complements the normative presentation without introducing additional governance concepts.

\subsection{Configuration Graph}
The ACM reference model is organized as the multi-graph structure defined in Equation~(\ref{eq:acm_multigraph_structure}). The governance semantics in this appendix operate on the finite directed Configuration Graph

\begin{equation}
	G_C=(V_C,E_C),
	\label{eq:graph_definition}
\end{equation}

where:

\begin{itemize}
	\item $V_C$ denotes the finite set of immutable ACI revisions represented in the governed configuration;
	\item $E_C\subseteq V_C\times V_C\times T_R$ denotes the finite set of typed configuration relationships;
	\item $T_R$ is the finite set of relationship types defined by the reference metamodel.
\end{itemize}
The graph topology remains immutable for a given revision set. Configuration evolution is represented by introducing new immutable revisions rather than modifying existing vertices.

\subsection{Governance Descriptor}

Each configuration revision $v\in V_C$ is associated with the
configuration governance descriptor

\begin{equation}
	\Gamma_C(v)=
	\left(
	L(v),
	Q(v),
	A(v),
	I(v),
	El(v)
	\right),
	\label{eq:descriptor_definition}
\end{equation}

whose components belong to the five configuration governance state spaces.
Runtime state is defined separately on runtime entities in $V_R$.

The configuration governance descriptor represents the governance status associated with an ACI revision and introduces no computational semantics. The complete governance view additionally includes the runtime view $\rho(v)$ defined in Equation~(\ref{eq:runtime_view}).

All semantic computations are defined through dedicated evaluation functions introduced later in this appendix.

\subsection{Notation}

Throughout the remainder of this appendix, configuration governance state assignments are defined over configuration revisions. Runtime state follows a distinct assignment $Rt:V_R\rightarrow\mathcal{R}$ because runtime entities and configuration revisions belong to different graph domains.

For propagation reasoning, an impact valuation is a state assignment

\begin{equation}
	\iota : V_C \rightarrow \mathcal{I},
	\label{eq:impact_valuation}
\end{equation}

where $\iota(v)$ denotes the impact state associated with revision $v$. The
impact component $I(v)$ of the configuration governance descriptor denotes this
same state; during propagation, its evolving valuation is represented by
$\iota(v)$.

For a fixed Configuration Graph, the set of all impact valuations is

\begin{equation}
	\mathcal{D}_{G_C}
	=
	\mathcal{I}^{V_C}.
	\label{eq:impact_valuation_domain}
\end{equation}

The graph structure and the propagation-policy mapping remain fixed during one propagation computation. Only the impact valuation evolves.

\section{Governance State Spaces}
\label{appendix:state_spaces}

This appendix formally defines the state spaces introduced in
Section~\ref{sec:governance_state_spaces}. Lifecycle, quality, assurance, impact, and eligibility are configuration governance state spaces defined over $V_C$, whereas runtime is an execution state space defined over $V_R$.

\subsection{Lifecycle}

The lifecycle state space is defined as

\begin{equation}
	\mathcal{L}=
	\{
	Draft,
	Validated,
	Approved,
	Released,
	Deprecated,
	Archived
	\}.
	\label{eq:lifecycle_space}
\end{equation}

The ordering relation is the monotonic lifecycle order introduced in Equation~(\ref{eq:lifecycle_order}).

---

\subsection{Quality}

The quality state space is defined as

\begin{equation}
	\mathcal{Q}=
	\{
	Undefined,
	Valid,
	Warning,
	Invalid
	\}.
	\label{eq:quality_space}
\end{equation}

The quality state reflects the intrinsic validation status of an individual revision independently of its dependencies.

No ordering relation is imposed on $\mathcal{Q}$ by the reference semantics; these values represent distinct local evaluation outcomes.

\subsection{Assurance}

The assurance state space is defined as

\begin{equation}
	\mathcal{A}=
	\{
	Unassessed,
	Partial,
	Satisfied
	\}.
	\label{eq:assurance_space}
\end{equation}

The assurance state represents the availability of governance evidence associated with a revision.

The corresponding ordering relation is

\begin{equation}
	Unassessed
	<
	Partial
	<
	Satisfied.
	\label{eq:assurance_order}
\end{equation}

---

\subsection{Impact}

The impact state space is defined as

\begin{equation}
	\mathcal{I}=
	\{
	None,
	Local,
	Propagated
	\}.
	\label{eq:impact_space}
\end{equation}

The impact space is equipped with the total order

\begin{equation}
	None
	\prec_{\mathcal I}
	Local
	\prec_{\mathcal I}
	Propagated.
	\label{eq:impact_order}
\end{equation}

Its least element is

\begin{equation}
	\bot_{\mathcal{I}}
	=
	None.
	\label{eq:impact_bottom}
\end{equation}

The least upper bound of two impact states is defined by

\begin{equation}
	i_1
	\sqcup_{\mathcal{I}}
	i_2
	=
	\max\nolimits_{\prec_{\mathcal{I}}}
	\left(i_1,i_2\right).
	\label{eq:impact_join}
\end{equation}

Consequently,
$\left(\mathcal{I},\preceq_{\mathcal{I}}\right)$ is a finite lattice.

The impact state identifies whether a revision has been affected by a propagated governance change.
---

\subsection{Eligibility}

The eligibility state space is defined as

\begin{equation}
	\mathcal{E}=
	\{
	Blocked,
	Restricted,
	Eligible
	\}.
	\label{eq:eligibility_space}
\end{equation}

The corresponding ordering relation is

\begin{equation}
	Blocked
	<
	Restricted
	<
	Eligible.
	\label{eq:eligibility_order}
\end{equation}

The eligibility state summarizes whether a revision satisfies the governance conditions required for operational use.

---

\subsection{Runtime}

The runtime state space is defined as

\begin{equation}
	\mathcal{R}=
	\{
	Created,
	Ready,
	Running,
	Waiting,
	Completed,
	Failed,
	Cancelled,
	Terminated
	\}.
	\label{eq:runtime_space}
\end{equation}

The runtime state represents the execution status reconstructed from runtime observations independently of the configuration lifecycle.

Unlike the configuration governance state spaces, runtime states are reconstructed from execution events and characterize runtime entities rather than configuration revisions.

Accordingly, runtime state is assigned to runtime entities rather than
configuration revisions:

\begin{equation}
	Rt:V_R\rightarrow\mathcal{R}.
	\label{eq:runtime_state_assignment}
\end{equation}

\section{Propagation Policies}
\label{appendix:propagation_policies}

This appendix formally defines the propagation policies introduced in Section~\ref{sec:propagation_policies}. Propagation policies characterize how governance information traverses configuration relationships independently of the governance state evaluation.

\subsection{Policy Space}

Let $T_R$ denote the finite set of relationship types defined by the ACM reference model.

Each configuration relationship is associated with exactly one propagation policy through the resolved policy view

\begin{equation}
	\Pi:E_C\rightarrow\mathcal P,
	\label{eq:policy_mapping_formal}
\end{equation}

where the policy space is defined as

\begin{equation}
	\mathcal{P}=
	\left\{
	\textit{Blocking},
	\textit{Warning},
	\textit{Informational},
	\textit{None}
	\right\}.
	\label{eq:policy_set}
\end{equation}

A propagation policy is declarative: it does not itself modify an impact state. Instead, it selects the impact-transfer semantics applied when propagation traverses a governed configuration relationship.
The mapping $\Pi$ is the policy view resolved from the propagation policies represented in the Assurance Graph $G_A$. For a fixed governance evaluation, it is static and determined by relationship type. Consequently, identical relationship types induce identical propagation behavior independently of the connected revisions. The mapping satisfies the type-consistency constraint defined in Equation~(\ref{eq:policy_type_consistency}).

\subsection{Policy Precedence}

When several propagation paths converge on the same revision, multiple propagation policies may simultaneously apply. Their effective behavior is determined by a total precedence relation over $\mathcal{P}$.

The precedence relation is defined as

\begin{equation}
	\textit{Blocking}
	>
	\textit{Warning}
	>
	\textit{Informational}
	>
	\textit{None}.
	\label{eq:policy_order}
\end{equation}

This ordering defines policy precedence for deterministic composition; it does not constitute a governance state ordering. In particular, $\mathcal P$ is not a component of the configuration governance descriptor $\Gamma_C$.

\subsection{Policy Composition}

When multiple propagation paths converge on the same revision, the effective
propagation policy is obtained by selecting the highest-precedence applicable
policy.

The composition operator

\begin{equation}
	\otimes :
	\mathcal{P}\times\mathcal{P}
	\rightarrow
	\mathcal{P},
	\label{eq:policy_composition}
\end{equation}

is therefore defined as

\begin{equation}
	p_1\otimes p_2
	=
	\max\nolimits_{(\ref{eq:policy_order})}
	\left(p_1,p_2\right),
	\label{eq:policy_max}
\end{equation}

where the maximum is evaluated according to the precedence relation defined in Equation~(\ref{eq:policy_order}).

By construction, the composition operator satisfies

\begin{itemize}
	\item commutativity,
	\item associativity,
	\item idempotence.
\end{itemize}

Consequently, policy composition is independent of propagation order, parent ordering, and graph traversal strategy.

These properties ensure that policy composition introduces no order-dependent behavior into the propagation semantics defined in Appendix~\ref{appendix:propagation_semantics}.

Policy precedence determines how concurrent policy contributions are combined.
The effect of an individual policy on an impact state is specified separately
by the local transfer semantics introduced in
Appendix~\ref{appendix:propagation_semantics}. Policy composition and impact
transfer therefore remain distinct operations: the former selects the effective
policy, whereas the latter determines its contribution to impact propagation.

\section{Formal Evaluation Functions}
\label{appendix:evaluation_functions}

This appendix formally defines the governance evaluation functions introduced in Section~\ref{sec:evaluation_functions}. These functions assign governance states independently of propagation semantics.

\subsection{Quality Evaluation}

The quality evaluation function

\begin{equation}
	f_{\mathrm{quality}} : V_C \rightarrow \mathcal Q
	\label{eq:quality_function}
\end{equation}

assigns exactly one quality state to every ACI revision.

Consequently,

\begin{equation}
	\forall v\in V_C,\;
	\exists!\;
	Q(v)\in\mathcal{Q}.
	\label{eq:quality_total}
\end{equation}

\subsection{Assurance Evaluation}

Similarly,

\begin{equation}
	f_{\mathrm{assurance}} : V_C \rightarrow \mathcal A,
	\label{eq:assurance_function}
\end{equation}

assigns a unique assurance state to every revision.

Therefore,

\begin{equation}
	\forall v\in V_C,\;
	\exists!\;
	A(v)\in\mathcal{A}.
	\label{eq:assurance_total}
\end{equation}

\subsection{Eligibility Evaluation}

Eligibility is evaluated locally from the lifecycle, quality, assurance, and stabilized impact states associated with a revision. Propagation policies do not constitute an argument of the eligibility function; their effects are already reflected in the propagated impact state.

Formally,

\begin{equation}
	f_{\mathrm{elig}}
	:
	\mathcal{L}
	\times
	\mathcal{Q}
	\times
	\mathcal{A}
	\times
	\mathcal{I}
	\rightarrow
	\mathcal{E}
	\label{eq:eligibility_function_formal}
\end{equation}

The function is deterministic.

\begin{equation}
	\forall
	(l,q,a,i)
	\in
	\mathcal{L}
	\times
	\mathcal{Q}
	\times
	\mathcal{A}
	\times
	\mathcal{I},
	\;
	\exists!\;
	e
	\in
	\mathcal{E}
	\text{ such that }
	e
	=
	f_{\mathrm{elig}}(l,q,a,i).
	\label{eq:eligibility_deterministic}
\end{equation}
For each revision $v\in V_C$, eligibility is therefore obtained from the final
impact valuation $\iota^{*}$ computed by the propagation semantics:

\begin{equation}
	El(v)
	=
	f_{\mathrm{elig}}
	\left(
	L(v),
	Q(v),
	A(v),
	\iota^{*}(v)
	\right).
	\label{eq:eligibility_from_fixed_impact}
\end{equation}

Eligibility evaluation is not part of the fixed-point iteration and does not modify the propagated impact valuation.

The evaluation functions are defined as total deterministic mappings over their respective finite domains. They introduce no dependency traversal and do not modify the configuration graph. In particular, eligibility evaluation consumes the stabilized impact valuation produced by propagation but does not participate in the propagation fixed-point computation.

\section{Propagation Semantics}
\label{appendix:propagation_semantics}

This appendix formalizes the propagation semantics introduced in
Section~\ref{subsec:propagation_semantics}. It specifies the propagation operator, the fixed-point semantics, and the reference worklist algorithm.
The mathematical properties of the propagation semantics, including monotonicity, termination, convergence, and correctness, are established separately in Appendix~\ref{appendix:formal_properties}.

\subsection{Propagation Operator}

Propagation operates over the fixed Configuration Graph
$G_C=(V_C,E_C)$ defined in Appendix~\ref{appendix:formal_foundations}.
Its structure remains unchanged throughout one propagation computation; only the impact valuation evolves.

As defined in Equation~(\ref{eq:impact_valuation}), the current impact valuation $\iota\in\mathcal D_{G_C}$ assigns one impact state to each configuration revision.

The propagation process starts from an initial valuation

\begin{equation}
	\iota^{(0)} : V_C \rightarrow \mathcal{I},
	\label{eq:initial_impact_valuation}
\end{equation}

which identifies the revisions directly affected by the configuration change or
governance condition initiating the propagation computation. Revisions not
directly affected are assigned the least impact state

\[
\bot_{\mathcal I}=None.
\]

The graph structure and the propagation-policy mapping remain fixed throughout the computation. Consequently, propagation modifies only the impact valuation.


Each relationship

\[
e=(u,v)\in E_C
\]

is associated with a propagation policy

\[
\Pi(e)\in\mathcal P,
\]

defined in Appendix~\ref{appendix:propagation_policies}.

For formal reasoning, each propagation policy induces a local transfer function

\begin{equation}
	\tau_p :
	\mathcal I
	\rightarrow
	\mathcal I,
	\qquad
	p\in\mathcal P,
	\label{eq:transfer_function}
\end{equation}

which specifies how the propagated impact state is transmitted across a single relationship governed by policy $p$.

The concrete mapping associated with each propagation policy is intentionally left parameterized by the governance profile. The formal propagation results require every transfer function $\tau_p$ to be deterministic, monotone with respect to $\preceq_{\mathcal I}$, and bottom-preserving. Under these assumptions, the propagation semantics and the corresponding fixed-point properties hold independently of the concrete policy-specific transfer mapping.


For every revision $v\in V_C$, propagation is defined by the local evaluation function

\begin{equation}
	\label{eq:local_transfer}
	F_v(\iota)
	=
	\iota^{(0)}(v)
	\sqcup_{\mathcal I}
	\bigvee_{(u,v)\in E_C}
	\tau_{\Pi(u,v)}
	\!\left(
	\iota(u)
	\right),
\end{equation}

where $\sqcup_{\mathcal I}$ denotes the least upper bound on the impact lattice.

For a revision with no incoming relationship, the empty join is defined as the least impact state:

\begin{equation}
	\bigvee_{\varnothing}
	=
	\bot_{\mathcal I}.
	\label{eq:empty_impact_join}
\end{equation}

The first term preserves the initial impact assignment, whereas the second term
aggregates the propagated contributions received from predecessor revisions.

Here, the orientation $(u,v)\in E_C$ follows the propagation direction encoded
by the Configuration Graph: an impact contribution is transferred from $u$ to
the dependent revision $v$.


The global propagation operator is therefore defined as

\begin{equation}
	\label{eq:impact_operator}
	\widehat{\mathrm{Prop}}_{G_C,\Pi}
	:
	\mathcal D_{G_C}
	\rightarrow
	\mathcal D_{G_C},
\end{equation}
with

\begin{equation}
	\label{eq:impact_operator_definition}
	\widehat{\mathrm{Prop}}_{G_C,\Pi}(\iota)(v)
	=
	F_v(\iota),
	\qquad
	\forall v\in V_C.
\end{equation}

The operator acts exclusively on impact valuations.
Lifecycle, quality, assurance, and eligibility states are not modified during propagation. Once the impact valuation has stabilized, eligibility is evaluated locally using the function
$f_{\mathrm{elig}}$ defined in Appendix~\ref{appendix:evaluation_functions}.

For readability, the remainder of this appendix denotes
$\widehat{\mathrm{Prop}}_{G_C,\Pi}$ simply by $\widehat{\mathrm{Prop}}$ whenever the Configuration Graph and
propagation-policy mapping are fixed.

\subsection{Least Fixed-Point Semantics}

Starting from the initial impact valuation $\iota^{(0)}$, successive applications of the propagation operator generate the sequence

\begin{equation}
	\label{eq:impact_iteration}
	\iota^{(k+1)}
	=
	\widehat{\mathrm{Prop}}
	\!\left(
	\iota^{(k)}
	\right),
\end{equation}

where $k$ denotes the propagation iteration.

Propagation continues until the impact valuation becomes stable, that is, until an additional application of the propagation operator produces no further modification.

The stabilized valuation therefore satisfies

\begin{equation}
	\label{eq:impact_fixed_point}
	\iota^{*}
	=
	\widehat{\mathrm{Prop}}
	\!\left(
	\iota^{*}
	\right).
\end{equation}

The valuation $\iota^{*}$ assigns the final stabilized impact state to every
revision of the Configuration Graph.

Once the impact valuation has stabilized, eligibility is evaluated
independently for every revision using the local evaluation function defined in Appendix~\ref{appendix:evaluation_functions}. Formally,

\begin{equation}
	\label{eq:eligibility_after_propagation}
	El(v)
	=
	f_{\mathrm{elig}}
	\!\left(
	L(v),
	Q(v),
	A(v),
	\iota^{*}(v)
	\right),
	\qquad
	\forall v\in V_C.
\end{equation}

Consequently, the propagation phase computes only the propagated impact valuation, whereas eligibility evaluation is performed afterwards as an independent local computation. This separation eliminates any circular dependency between propagation and eligibility evaluation while preserving the normative semantics introduced in Section~\ref{subsec:propagation_semantics}.

The existence, uniqueness, and convergence properties of the stabilized impact valuation, together with the mathematical justification of the fixed-point construction, are established in Appendix~\ref{appendix:formal_properties}.

\subsection{Reference Worklist Algorithm}
\label{appendix:ref_worklist_algo}
The propagation semantics specify the stable impact valuation independently of
the evaluation strategy used to compute it. An implementation is compliant with
the reference semantics if it computes the least fixed point above the initial
valuation $\iota^{(0)}$ characterized in
Appendix~\ref{appendix:formal_properties}.

The reference implementation adopts a worklist-based evaluation strategy that incrementally updates only the revisions whose propagated impact state may change.

Initially, the current valuation is set to the initial impact valuation

\[
\iota
=
\iota^{(0)},
\]

and the worklist

\[
W
\subseteq
V_C
\]

contains every revision directly affected by the triggering governance event, together with their immediate successors.

At each iteration, one revision

\[
v\in W
\]

is removed from the worklist and its local propagation equation
(Equation~(\ref{eq:local_transfer})) is evaluated using the current impact valuation.

If the evaluation produces a strictly greater impact state,

\begin{equation}
	\label{eq:worklist_update}
	F_v(\iota)
	\succ_{\mathcal I}
	\iota(v),
\end{equation}

the valuation is updated according to

\begin{equation}
	\label{eq:worklist_assignment}
	\iota(v)
	:=
	F_v(\iota),
\end{equation}

and every successor

\[
w
\quad
\text{such that}
\quad
(v,w)\in E_C
\]

is inserted into the worklist if not already present.

Otherwise, no successor is scheduled and propagation continues with the next revision contained in the worklist.

The evaluation terminates when the worklist becomes empty,

\begin{equation}
	\label{eq:worklist_stop}
	W
	=
	\varnothing.
\end{equation}

Under the worklist initialization, successor rescheduling, bottom-preserving transfer semantics, and fairness conditions defined above, an empty worklist corresponds to a valuation satisfying every local propagation equation. The formal implication

\begin{equation}
	W
	=
	\varnothing
	\Longrightarrow
	\forall v\in V_C,\;
	F_v(\iota)
	=
	\iota(v)
	\label{eq:worklist_local_equilibrium}
\end{equation}

and the equality of the resulting valuation with $\iota^{*}$ are established in Appendix~\ref{appendix:formal_properties}.

The reference worklist algorithm assumes a fair evaluation strategy, meaning that every revision inserted into the worklist is eventually processed.
Different scheduling strategies, including FIFO, LIFO, or priority-based queues, remain compliant provided that they satisfy this fairness condition and compute the same stabilized valuation.

The correctness, termination, convergence, and complexity of the reference worklist algorithm are established in Appendix~\ref{appendix:formal_properties}.

\section{Formal Properties of the Propagation Semantics}
\label{appendix:formal_properties}

This appendix establishes the mathematical properties of the propagation
semantics introduced in Appendix~\ref{appendix:propagation_semantics}. It proves
the monotonicity of the propagation operator, the existence of a least fixed
point above the initial valuation, the convergence of the iterative evaluation
toward this fixed point, and the correctness of the reference worklist
algorithm. Throughout this appendix, the Configuration Graph and the
propagation-policy mapping are assumed to be fixed.

\subsection{Formal Properties of the Propagation Domain}
\label{appendix:finite_lattice}
The propagation operator acts on the impact valuation domain

\[
\mathcal D_{G_C}
=
\mathcal I^{V_C},
\]

introduced in Appendix~\ref{appendix:formal_foundations}, where each valuation associates one impact state with every revision of the fixed Configuration
Graph.

The ordering relation on $\mathcal D_{G_C}$ is defined componentwise.

\begin{equation}
	\label{eq:product_order}
	\iota_1
	\sqsubseteq
	\iota_2
	\iff
	\forall v\in V_C,\;
	\iota_1(v)
	\preceq_{\mathcal I}
	\iota_2(v).
\end{equation}

Since both the configuration graph and the impact state space are finite (Appendices~\ref{appendix:formal_foundations} and
\ref{appendix:state_spaces}), the valuation domain
$\mathcal D_{G_C}$ is finite.

Moreover, because
$\left(\mathcal I,\preceq_{\mathcal I}\right)$
is a finite lattice, the product domain
$\left(\mathcal D_{G_C},\sqsubseteq\right)$
is itself a finite lattice under the componentwise ordering, whose least element is the valuation

\begin{equation}
	\label{eq:bottom_valuation}
	\bot_{\mathcal D_{G_C}}(v)
	=
	\bot_{\mathcal I},
	\qquad
	\forall v\in V_C.
\end{equation}

The least upper bound of two valuations is defined componentwise by

\begin{equation}
	\label{eq:valuation_join}
	(\iota_1
	\sqcup
	\iota_2)(v)
	=
	\iota_1(v)
	\sqcup_{\mathcal I}
	\iota_2(v),
	\qquad
	\forall v\in V_C.
\end{equation}

The finite lattice
$\left(\mathcal D_{G_C},\sqsubseteq\right)$
constitutes the mathematical domain on which the propagation operator
$\widehat{\mathrm{Prop}}$ is defined.

\subsection{Monotonicity of Local Transfer Functions}
\label{appendix:local_monotonicity}
The propagation operator introduced in
Appendix~\ref{appendix:propagation_semantics}
is constructed from the local transfer functions

\[
\tau_p :
\mathcal I
\rightarrow
\mathcal I,
\qquad
p\in\mathcal P,
\]

associated with the propagation policies.

Under the transfer-function assumptions stated in Appendix~\ref{appendix:propagation_semantics}, each policy-specific transfer function preserves the ordering relation defined on the impact lattice.

Formally,

\begin{equation}
	\label{eq:transfer_monotone}
	\forall
	x,y
	\in
	\mathcal I,
	\qquad
	x
	\preceq_{\mathcal I}
	y
	\Longrightarrow
	\tau_p(x)
	\preceq_{\mathcal I}
	\tau_p(y).
\end{equation}

In addition, each transfer function is assumed to preserve the bottom element:
\begin{equation}
	\tau_p
	\left(
	\bot_{\mathcal I}
	\right)
	=
	\bot_{\mathcal I},
	\qquad
	\forall p\in\mathcal P.
	\label{eq:transfer_bottom_preservation}
\end{equation}

Consequently, a relationship cannot generate a propagated impact in the absence of an impacted predecessor. The transfer functions may transmit or suppress an existing impact according to the associated propagation policy, but they do not create impact independently of the initial valuation.
The local propagation equation associated with a revision $v\in V_C$ is
given by Equation~(\ref{eq:local_transfer}).

\paragraph{Monotonicity of the local propagation equation.}

Let $\iota_1\sqsubseteq\iota_2$.

By definition of the product ordering,
$\iota_1(u)\preceq_{\mathcal I}\iota_2(u)$ for all $u\in V_C$.

Applying Equation~(\ref{eq:transfer_monotone}) to every incoming relationship
yields

\begin{equation}
\label{eq:tau_pi_def1}
\tau_{\Pi(u,v)}
(
\iota_1(u)
)
\preceq_{\mathcal I}
\tau_{\Pi(u,v)}
(
\iota_2(u)
).
\end{equation}

Since the least upper bound operator $\sqcup_{\mathcal I}$
is monotone on the finite lattice $\mathcal I$, the aggregation of propagated contributions also preserves the ordering.

Therefore,

\begin{equation}
	\label{eq:local_monotone}
	F_v(\iota_1)
	\preceq_{\mathcal I}
	F_v(\iota_2),
	\qquad
	\forall v\in V_C.
\end{equation}

Consequently, every local propagation equation is monotone with respect to the componentwise ordering defined on the impact valuation domain.

\subsection{Global Monotonicity of the Propagation Operator}
\label{appendix:global_monotonicity}
The global propagation operator is obtained by assembling the local propagation equations associated with each revision of the fixed configuration graph.

Formally,

\begin{equation}
	\label{eq:global_operator}
	\widehat{\mathrm{Prop}}(\iota)
	=
	\bigl(
	F_v(\iota)
	\bigr)_{v\in V_C},
\end{equation}

where each local equation $F_v$ is defined as in
Appendix~\ref{appendix:local_monotonicity}.

\paragraph{Proposition.}
The global propagation operator

\[
\widehat{\mathrm{Prop}}
:
\mathcal D_{G_C}
\rightarrow
\mathcal D_{G_C}
\]

is monotone with respect to the product ordering
$\sqsubseteq$.

\paragraph{Proof.}

Let

\[
\iota_1
\sqsubseteq
\iota_2.
\]

By Appendix~\ref{appendix:local_monotonicity}, every local propagation equation is monotone, that is,

\[
F_v(\iota_1)
\preceq_{\mathcal I}
F_v(\iota_2),
\qquad
\forall v\in V_C.
\]

Since the product ordering on

\[
\mathcal D_{G_C}
=
\mathcal I^{V_C}
\]

is defined componentwise, the monotonicity of every component immediately implies

\[
\widehat{\mathrm{Prop}}(\iota_1)
\sqsubseteq
\widehat{\mathrm{Prop}}(\iota_2).
\]

Therefore,

\begin{equation}
	\label{eq:global_monotonicity}
	\iota_1
	\sqsubseteq
	\iota_2
	\Longrightarrow
	\widehat{\mathrm{Prop}}(\iota_1)
	\sqsubseteq
	\widehat{\mathrm{Prop}}(\iota_2).
\end{equation}

Hence, the propagation operator is monotone over the finite lattice
$\mathcal D_{G_C}$.

\subsection{Existence of the Least Fixed Point}

The previous results establish that the propagation domain

\[
\mathcal D_{G_C}
=
\mathcal I^{V_C}
\]

is a finite lattice (Appendix~\ref{appendix:finite_lattice}) and that the global propagation operator

\[
\widehat{\mathrm{Prop}}
:
\mathcal D_{G_C}
\rightarrow
\mathcal D_{G_C}
\]

is monotone (Appendix~\ref{appendix:global_monotonicity}).

The classical Knaster--Tarski Fixed Point Theorem therefore applies directly.

\paragraph{Theorem (Knaster--Tarski).}
For the propagation computation initialized by $\iota^{(0)}$, consider the
principal upper set

\begin{equation}
	\mathcal D_{G_C}^{\iota^{(0)}}
	=
	\left\{
	\iota\in\mathcal D_{G_C}
	\;\middle|\;
	\iota^{(0)}
	\sqsubseteq
	\iota
	\right\}.
	\label{eq:initial_upper_domain}
\end{equation}

Because $\mathcal D_{G_C}$ is a finite lattice,
$\mathcal D_{G_C}^{\iota^{(0)}}$ is itself a finite complete lattice whose least element is $\iota^{(0)}$.

Moreover, the preservation of the initial valuation in every local propagation equation implies

\begin{equation}
	\iota^{(0)}
	\sqsubseteq
	\widehat{\mathrm{Prop}}(\iota),
	\qquad
	\forall
	\iota
	\in
	\mathcal D_{G_C}^{\iota^{(0)}}.
	\label{eq:operator_preserves_initial_domain}
\end{equation}

Hence,
$\widehat{\mathrm{Prop}}$ maps
$\mathcal D_{G_C}^{\iota^{(0)}}$ into itself. Since the operator is monotone, the Knaster--Tarski theorem guarantees the existence of a least fixed point in this restricted domain.

This fixed point is denoted by $\iota^{*}$ and satisfies

\begin{equation}
	\widehat{\mathrm{Prop}}
	\left(
	\iota^{*}
	\right)
	=
	\iota^{*}.
	\label{eq:least_fixed_point_above_initial}
\end{equation}

It is characterized by

\begin{equation}
	\iota^{*}
	=
	\bigwedge
	\left\{
	\iota
	\in
	\mathcal D_{G_C}^{\iota^{(0)}}
	\;\middle|\;
	\widehat{\mathrm{Prop}}(\iota)
	=
	\iota
	\right\}.
	\label{eq:least_fixed_point_characterization}
\end{equation}

Therefore, $\iota^{*}$ is the unique least fixed point above $\iota^{(0)}$, rather than necessarily the least fixed point of the operator over the entire valuation domain.

\subsection{Convergence by Increasing Iteration}

The existence of a least fixed point does not by itself establish that the iterative evaluation defined in Appendix~\ref{appendix:propagation_semantics}
reaches that fixed point. This subsection establishes the increasing character, finite stabilization, and minimality of the propagation sequence.

Starting from the initial impact valuation $\iota^{(0)}$, the sequence is defined recursively by

\begin{equation}
	\iota^{(k+1)}
	=
	\widehat{\mathrm{Prop}}
	\!\left(
	\iota^{(k)}
	\right),
	\qquad
	k\geq 0.
	\label{eq:kleene_iteration}
\end{equation}

\paragraph{Increasing character of the propagation sequence.}

For every revision $v\in V_C$, the local propagation equation gives

\begin{equation}
	F_v\!\left(\iota^{(0)}\right)
	=
	\iota^{(0)}(v)
	\sqcup_{\mathcal I}
	\bigvee_{(u,v)\in E_C}
	\tau_{\Pi(u,v)}
	\!\left(
	\iota^{(0)}(u)
	\right).
	\label{eq:initial_local_expansion}
\end{equation}

Since each operand of a join is lower than or equal to the resulting least upper bound,

\begin{equation}
	\iota^{(0)}(v)
	\preceq_{\mathcal I}
	F_v\!\left(\iota^{(0)}\right),
	\qquad
	\forall v\in V_C.
	\label{eq:initial_component_growth}
\end{equation}

By the componentwise ordering on $\mathcal D_{G_C}$, it follows that

\begin{equation}
	\iota^{(0)}
	\sqsubseteq
	\widehat{\mathrm{Prop}}
	\!\left(
	\iota^{(0)}
	\right)
	=
	\iota^{(1)}.
	\label{eq:initial_iteration_growth}
\end{equation}

Assume now that, for some $k\geq 0$,

\begin{equation}
	\iota^{(k)}
	\sqsubseteq
	\iota^{(k+1)}.
	\label{eq:iteration_induction_hypothesis}
\end{equation}

By the global monotonicity of $\widehat{\mathrm{Prop}}$ established in the
previous subsection,

\begin{equation}
	\widehat{\mathrm{Prop}}
	\!\left(
	\iota^{(k)}
	\right)
	\sqsubseteq
	\widehat{\mathrm{Prop}}
	\!\left(
	\iota^{(k+1)}
	\right).
	\label{eq:iteration_monotone_step}
\end{equation}

Using Equation~(\ref{eq:kleene_iteration}), this is equivalent to

\begin{equation}
	\iota^{(k+1)}
	\sqsubseteq
	\iota^{(k+2)}.
	\label{eq:iteration_induction_step}
\end{equation}

Therefore, by induction, the propagation sequence forms an increasing chain:

\begin{equation}
	\iota^{(0)}
	\sqsubseteq
	\iota^{(1)}
	\sqsubseteq
	\cdots
	\sqsubseteq
	\iota^{(k)}
	\sqsubseteq
	\iota^{(k+1)}
	\sqsubseteq
	\cdots.
	\label{eq:increasing_chain}
\end{equation}

\paragraph{Finite stabilization.}

The valuation domain $\mathcal D_{G_C}$ is finite. Every increasing chain in $\mathcal D_{G_C}$ must therefore become stationary after a finite number of iterations. Hence, there exists an integer $N\geq 0$ such that

\begin{equation}
	\iota^{(N)}
	=
	\iota^{(N+1)}.
	\label{eq:stabilization}
\end{equation}

By Equation~(\ref{eq:kleene_iteration}),

\begin{equation}
	\widehat{\mathrm{Prop}}
	\!\left(
	\iota^{(N)}
	\right)
	=
	\iota^{(N)}.
	\label{eq:iteration_fixed_point}
\end{equation}

Thus, the stabilized valuation is a fixed point of the propagation operator.

\paragraph{Minimality of the stabilized valuation.}

Let $\mu\in\mathcal D_{G_C}$ be any fixed point of
$\widehat{\mathrm{Prop}}$ satisfying

\begin{equation}
	\iota^{(0)}
	\sqsubseteq
	\mu.
	\label{eq:fixed_point_above_initial}
\end{equation}

It follows by induction that

\begin{equation}
	\iota^{(k)}
	\sqsubseteq
	\mu,
	\qquad
	\forall k\geq 0.
	\label{eq:iteration_below_fixed_point}
\end{equation}

The base case follows from
Equation~(\ref{eq:fixed_point_above_initial}). For the induction step, assume
$\iota^{(k)}\sqsubseteq\mu$. Global monotonicity and the fixed-point property of
$\mu$ yield

\begin{equation}
	\iota^{(k+1)}
	=
	\widehat{\mathrm{Prop}}
	\!\left(
	\iota^{(k)}
	\right)
	\sqsubseteq
	\widehat{\mathrm{Prop}}(\mu)
	=
	\mu.
	\label{eq:minimality_induction_step}
\end{equation}

In particular,

\begin{equation}
	\iota^{(N)}
	\sqsubseteq
	\mu.
	\label{eq:stabilized_minimality}
\end{equation}

Consequently, $\iota^{(N)}$ is the least fixed point of the propagation
operator above the initial valuation. Therefore,

\begin{equation}
	\iota^{(N)}
	=
	\iota^{*}.
	\label{eq:iteration_lfp}
\end{equation}

The propagation iteration thus converges after finitely many steps to the
unique least stable impact valuation consistent with the initial impact
assignment and the propagation policies.

\subsection{Correctness of the Reference Worklist Algorithm}

The previous subsections establish that the propagation operator admits a
unique least fixed point above the initial impact valuation and that the
iterative propagation semantics converge toward this valuation. This subsection
shows that the reference worklist algorithm computes exactly the same result.

The worklist algorithm performs asynchronous evaluations of the local
propagation equations defined in
Equation~(\ref{eq:local_transfer}).
Only revisions whose propagated impact state may change are scheduled for
re-evaluation.

\paragraph{Soundness.}

Assume that the worklist algorithm terminates with the valuation

\[
\iota.
\]

By definition of the algorithm, the worklist is empty,

\[
W=\varnothing.
\]

Furthermore, every revision whose local propagation equation changes is
immediately reinserted into the worklist. Consequently, an empty worklist
implies that no local propagation equation can further increase the current
valuation.

Therefore,

\[
F_v(\iota)
=
\iota(v),
\qquad
\forall v\in V_C.
\]

Using the definition of the global propagation operator,

\[
\widehat{\mathrm{Prop}}(\iota)
=
\bigl(
F_v(\iota)
\bigr)_{v\in V_C},
\]

it follows immediately that

\begin{equation}
	\label{eq:worklist_soundness}
	\widehat{\mathrm{Prop}}(\iota)
	=
	\iota.
\end{equation}

Hence, every valuation returned by the worklist algorithm is a fixed point of the propagation operator.

\paragraph{Completeness.}

The worklist algorithm starts from the initial impact valuation
$\iota^{(0)}$ and applies exactly the same local propagation equations as the global iterative evaluation.

Each update replaces the current valuation of one revision by the value prescribed by its local propagation equation while leaving every other component unchanged.

Since every scheduled update is monotone
(Appendix~\ref{appendix:local_monotonicity}), the sequence of valuations generated by the worklist algorithm remains increasing.

Moreover, the fairness assumption introduced in
Appendix~\ref{appendix:propagation_semantics} guarantees that every revision whose propagated impact state may still change
is eventually processed.

Consequently, no local propagation equation capable of modifying the valuation can remain permanently unapplied.

The stabilized valuation computed by the worklist algorithm therefore satisfies all local propagation equations and is reachable from the initial valuation through a finite sequence of monotone updates.

Since the preceding fixed-point and minimality results establish a least fixed
point above $\iota^{(0)}$, the valuation produced by the worklist algorithm must
coincide with this least fixed point.

Therefore,

\begin{equation}
	\label{eq:worklist_correctness}
	\iota
	=
	\iota^{*}.
\end{equation}

The reference worklist algorithm is therefore correct: it computes exactly the least fixed point above the initial impact valuation defined by the propagation semantics while avoiding the unnecessary global re-evaluation of unaffected revisions.

\subsection{Termination and Complexity}

The previous subsections establish that the propagation sequence forms an increasing chain in the finite lattice $\mathcal D_{G_C}$
and that the reference worklist algorithm computes the same least fixed point above the initial impact valuation as the global propagation operator.

\paragraph{Termination.}

Each update performed by the worklist algorithm replaces the current impact state of a revision only when the corresponding local propagation equation produces a strictly greater value with respect to the ordering $\preceq_{\mathcal I}$.

Consequently, every successful update strictly increases one component of the current impact valuation while leaving all other components unchanged. 

Since the impact lattice $\mathcal I$
is finite, every revision can undergo only a finite number of strictly increasing updates.

Furthermore, the configuration graph contains a finite number of revisions.

Therefore, only finitely many successful updates can occur during one propagation computation.

Because every unsuccessful evaluation removes one revision from the worklist without modifying the current valuation, the worklist eventually becomes empty.

Hence, the reference worklist algorithm always terminates after a finite number of iterations.

\paragraph{Complexity.}

Let

\[
n
=
|V_C|,
\qquad
m
=
|E_C|,
\]

and let

\[
h_{\mathcal I}
\]

denote the height of the impact lattice.

Since each revision can experience at most $h_{\mathcal I}-1$ strictly increasing updates, each outgoing relationship is evaluated at most $h_{\mathcal I}$
times.

The overall propagation complexity is therefore bounded by

\begin{equation}
	\label{eq:worklist_complexity_final}
	O
	\!\left(
	h_{\mathcal I}
	\,m
	\right).
\end{equation}

For the three-level impact lattice introduced in
Appendix~\ref{appendix:state_spaces},

\[
\mathcal I
=
\{
None,
Local,
Propagated
\},
\]

the lattice height is

\[
h_{\mathcal I}=3.
\]

Consequently, each revision can undergo at most two strictly increasing impact updates, and the propagation complexity becomes

\begin{equation}
	\label{eq:worklist_complexity_acm}
	O(3m)
	=
	O(m).
\end{equation}

Thus, propagation is linear in the number of configuration relationships for the reference ACM impact model.

\section{Runtime Semantics}
\label{appendix:runtime_semantics}

This appendix formalizes the runtime semantics introduced in
Section~\ref{subsec:runtime_semantics}. Runtime semantics are intentionally defined independently from the propagation operator. Whereas propagation computes the stabilized impact valuation over immutable configuration revisions, runtime semantics reconstruct execution states from normalized runtime events. Throughout this appendix, the released configuration baseline remains fixed and immutable.

\subsection{Runtime Event Model}
\label{appendix:runtime_event_model}

Let

\[
B=(V_B,E_B)
\]

denote a released configuration baseline.
The baseline vertices are immutable configuration revisions represented in the
Configuration Graph; therefore $V_B\subseteq V_C$.

A runtime execution is represented by a finite ordered sequence of normalized runtime events

\begin{equation}
\Sigma
=
\langle
e_1,e_2,\ldots,e_n
\rangle.
\label{eq:runtime_e1_e2}
\end{equation}

Each event is represented by the tuple

\begin{equation}
	e
	=
	(
	id,
	t,
	x,
	r,
	\alpha,
	\mu
	),
	\label{eq:runtime_event}
\end{equation}

where

\begin{itemize}
	\item $id$ uniquely identifies the event;
	\item $t$ is its logical timestamp;
	\item $x$ identifies the runtime entity affected by the event;
	\item $r$ identifies the immutable ACI revision providing the configuration
	provenance of $x$;
	\item $\alpha$ denotes the event type;
	\item $\mu$ contains optional execution metadata.
\end{itemize}

The event sequence is assumed to be totally ordered by timestamp.

Each runtime entity is associated through provenance with exactly one immutable
configuration revision governing its instantiation. Runtime events therefore
record execution observations without modifying the governed configuration.

\subsection{Replay Function}
\label{appendix:replay_function}

Let $R_0$ denote the initial Runtime Graph associated with the released
baseline $B$. The baseline provides the immutable configuration reference from
which runtime entities may be instantiated, but its configuration revisions are
not themselves runtime entities. Accordingly, $B$ remains distinct from
$R_0$ throughout replay.

Let $\mathfrak R$ denote the set of Runtime Graph states and
$\mathcal E_r$ the set of normalized runtime events. A replay step is defined by

\begin{equation}
	Replay
	:
	\mathfrak R
	\times
	\mathcal E_r
	\rightarrow
	\mathfrak R.
	\label{eq:runtime_replay}
\end{equation}

Applying an event updates only the Runtime Graph, including the runtime entity
identified by the event and its execution relationships when applicable.

The complete runtime reconstruction is therefore defined recursively as

\begin{equation}
R_i
=
Replay
(
R_{i-1},
e_i
),
\qquad
1\le i\le n.
\label{eq:runtime_replay_recursion}
\end{equation}

The reconstructed Runtime Graph after processing the complete execution trace is

\[
R_n.
\]

The replay function never modifies immutable configuration revisions. Runtime
events update only the Runtime Graph while preserving their relationship to the
released configuration baseline.

\subsection{Incremental Runtime Reconstruction}
\label{appendix:incremental_runtime_reconstruction}

Because replay is defined event by event, runtime reconstruction is naturally incremental.

Let

\begin{equation}
\Sigma_k
=
\langle
e_1,\ldots,e_k
\rangle
\label{eq:runtime_e1_ek}
\end{equation}

denote the prefix of length $k$.

The Runtime Graph reconstructed after observing this prefix is

\begin{equation}
	R_k
	=
	Replay^*
	(
	R_0,
	\Sigma_k
	),
	\label{eq:runtime_prefix_replay}
\end{equation}

where $Replay^*$ denotes repeated application of the replay function from the
initial Runtime Graph $R_0$ associated with baseline $B$.

Consequently, extending an execution trace requires replaying only newly observed events,

\begin{equation}
R_{k+1}
=
Replay(R_k,e_{k+1}),
\label{eq:runtime_incremental_replay}
\end{equation}

without rebuilding the Runtime Graph from scratch.

This property enables efficient online reconstruction while preserving exactly the same semantics as complete replay.

\subsection{Provenance Preservation}
\label{appendix:runtime_provenance}

Every runtime entity remains associated with the immutable configuration
revision governing its provenance.

Formally, let

\begin{equation}
	prov
	:
	V_R
	\rightarrow
	V_C
	\label{eq:runtime_provenance}
\end{equation}

associate each runtime node with its originating configuration revision.

For every runtime node

\[
x\in V_R,
\]

the provenance mapping satisfies

\begin{equation}
	prov(x)
	\in
	V_C.
	\label{eq:runtime_baseline_provenance}
\end{equation}

Consequently,

\begin{itemize}
	\item immutable revision identities remain unchanged;
	\item runtime entities remain distinct from the governed revisions to which
	they refer;
	\item every runtime state can be traced back to a unique immutable
	configuration revision.
\end{itemize}

Runtime replay therefore preserves the complete provenance chain established by the released baseline.

\subsection{Determinism of Runtime Replay}
\label{appendix:runtime_determinism}

Runtime replay is deterministic under the following assumptions:

\begin{enumerate}
	\item the released baseline and its associated initial Runtime Graph $R_0$
	are fixed;
	\item the normalized runtime event sequence is identical;
	\item events are replayed in the same total order;
	\item $Replay$ is deterministic for a given Runtime Graph and normalized
	event.
\end{enumerate}

Under these assumptions, the complete replay result

\begin{equation}
	R_n
	=
	Replay^*(R_0,\Sigma)
	\label{eq:runtime_complete_replay}
\end{equation}

is uniquely determined.

\paragraph*{Proof.}

The proof proceeds by induction on the length of the event sequence.

The base case is immediate since replay over the empty sequence returns the
fixed initial Runtime Graph $R_0$ associated with the released baseline.
Assume the property holds for the first $k$ events.

Since the replay function is deterministic and event ordering is fixed, processing event $e_{k+1}$ produces a uniquely determined successor Runtime Graph.

Therefore the reconstructed Runtime Graph is uniquely determined after every replay step.

By induction, replay deterministically reconstructs the Runtime Graph for the complete event sequence.

\subsection{Reconstruction Properties}
\label{appendix:runtime_properties}

The runtime reconstruction defined above satisfies the following properties.

\paragraph*{Baseline preservation}

Replay never modifies immutable configuration revisions.

\paragraph*{Provenance preservation}

Every runtime entity remains associated through $prov$ with exactly one
immutable configuration revision; multiple runtime entities may share the same
configuration provenance.

\paragraph*{Incrementality}

Replaying an additional event requires updating only the previously
reconstructed Runtime Graph.

\paragraph*{Determinism}

Identical released baselines and identical normalized event sequences always produce identical Runtime Graphs.

\paragraph*{Framework independence}

Because replay operates exclusively on the canonical Runtime Graph and
normalized runtime events, the reconstruction semantics do not depend on native
framework objects once configuration projection and runtime-event normalization
have been completed.

\subsection{Reproducibility of Governance Evaluation}
\label{appendix:governance_reproducibility}

The governance semantics defined throughout this paper are fully deterministic once the governed configuration, propagation policies, and runtime observations are fixed. Consequently, governance evaluation is reproducible across repeated
executions.

More precisely, let

\[
B
\]

be a released baseline,

\[
\Pi
\]

the propagation policy assignment,

and

\[
\Sigma
\]

a normalized runtime event sequence.

For fixed inputs $(B,\Pi,\Sigma)$ and a fixed governance evaluation context, the
complete governance result is uniquely determined by the deterministic
composition of impact propagation, local governance evaluation, and runtime
replay.

\paragraph*{Proposition.}

For identical inputs

\[
(B,\Pi,\Sigma)
\]

and an identical governance evaluation context, ACM always produces the same
governance evaluation.

\paragraph*{Proof.}

The result follows directly from the previous appendices.

Impact propagation converges toward the unique least fixed point above the initial impact valuation (Appendix~\ref{appendix:formal_properties}).

The evaluation functions
$f_{\mathrm{quality}}$,
$f_{\mathrm{assurance}}$,
and
$f_{\mathrm{elig}}$
are deterministic local functions.

Runtime replay is deterministic for a fixed released baseline and an identical ordered event sequence (Appendix~\ref{appendix:runtime_determinism}).

Since each stage of the evaluation pipeline is deterministic, their composition is also deterministic.

Therefore identical inputs always produce identical governance states.

\subsection{Framework-Independent Consistency}
\label{appendix:framework_independent_consistency}

The ACM governance semantics are intentionally defined over the normalized reference model rather than over framework-specific execution representations.

Assume two heterogeneous execution frameworks

\[
F_1
\quad\text{and}\quad
F_2,
\]

with native configuration models $M_{F_1}$ and $M_{F_2}$ projected through
their respective semantic adapters.

Let the resulting Configuration Graphs be

\[
G_C^{(1)}
\quad\text{and}\quad
G_C^{(2)}.
\]

Assume that these projected graphs preserve equivalent governance-relevant
configuration semantics and that they are subsequently governed under
equivalent release conditions.

Similarly, assume that the two executions produce equivalent normalized runtime
event sequences

\[
\Sigma_1^{\mathrm{norm}}
\quad\text{and}\quad
\Sigma_2^{\mathrm{norm}}.
\]

\paragraph*{Proposition.}

Equivalent governance-relevant projected configurations, equivalent propagation
policies, identical governance evaluation contexts, and equivalent normalized
runtime event sequences produce equivalent ACM governance results independently
of the original execution framework.

\paragraph*{Proof.}

After semantic projection and release governance, evaluation depends exclusively
on the resulting ACM configuration representation, the resolved propagation
policies, the fixed governance evaluation context, and the normalized runtime
event sequence.

Framework-specific implementation details that are outside the governance scope
do not participate in the governance semantics, while source provenance is
retained explicitly for traceability.

Since propagation, evaluation, and runtime replay depend only on the
governance-relevant normalized representation and the fixed governance context,
equivalent normalized inputs produce equivalent ACM governance results.

Consequently, governance evaluation is independent of the original execution framework, provided that semantic projection preserves the governance-relevant information represented by the ACM reference model.

\subsection{Determinism of the Complete Governance Semantics}
\label{appendix:global_determinism}

The previous appendices establish the determinism of each individual stage of the ACM governance semantics.

Impact propagation converges toward a unique stabilized impact valuation (Appendix~\ref{appendix:propagation_policies}).

The governance evaluation functions are deterministic by construction, since they operate exclusively on local governance states.

Runtime replay deterministically reconstructs the Runtime Graph from an immutable released baseline and an ordered sequence of normalized runtime events
(Appendix~\ref{appendix:runtime_determinism}).

These results immediately imply the following property.

\paragraph*{Theorem.}

Let $B$ be a released baseline, $\Pi$ a fixed propagation-policy assignment,
$\Sigma$ a normalized runtime event sequence, and let the governance evaluation
context be fixed.

Under these conditions, the ACM governance semantics produces a unique
governance result.

\paragraph*{Proof.}

The complete governance result combines three deterministic computations:
impact propagation, local governance evaluation, and runtime replay. Impact
propagation and local governance evaluation operate on the governed
configuration, whereas runtime replay reconstructs the execution view from the
fixed released baseline and normalized runtime events.

Since each component is deterministic under the hypotheses established in the previous appendices, their composition is deterministic.

Therefore the complete governance evaluation is uniquely determined.

\subsection{Complexity Analysis}
\label{appendix:complexity_analysis}

Let

\[
|V_C|
\]

denote the number of immutable revisions in the Configuration Graph,

\[
|E_C|
\]

the number of configuration relationships participating in propagation,

and

\[
h_{\mathcal I}
\]

the height of the impact lattice.

As established in Appendix~\ref{appendix:formal_properties},
each revision may undergo at most

\[
h_{\mathcal I}-1
\]

strict impact increases.

Consequently, the reference worklist algorithm performs at most

\[
O(h_{\mathcal I}|E_C|)
\]

successful propagation updates.

Since the impact lattice adopted by ACM is finite and consists of the ordered
states

\[
None
<
Local
<
Propagated,
\]

its height is

\[
h_{\mathcal I}=3.
\]

Therefore the propagation complexity becomes

\[
O(|E_C|).
\]

Runtime replay processes each normalized event exactly once.

For an execution trace containing

\[
|\Sigma|
\]

events, replay therefore requires

\[
O(|\Sigma|)
\]

time.

The overall governance evaluation consequently remains linear with respect to
the size of the dependency graph and the observed runtime trace.

\subsection{Height of the Product Lattice}
\label{appendix:lattice_height}

The convergence proof presented in Appendix~\ref{appendix:formal_properties}
relies on the finiteness of the impact valuation domain.

Let

\[
\mathcal D_{G_C}
=
\mathcal I^{V_C}
\]

denote the product lattice associated with the configuration graph.

Because the impact lattice has height

\[
h_{\mathcal I},
\]

the height of the product lattice satisfies

\begin{equation}
	\label{eq:product_lattice_height}
	H_{G_C}
	=
	1
	+
	|V_C|
	\left(
	h_{\mathcal I}-1
	\right).
\end{equation}

For the impact lattice adopted by ACM,

\[
h_{\mathcal I}=3,
\]

yielding

\begin{equation}
H_{G_C}
=
1+2|V_C|.
\label{eq:h_equality}
\end{equation}

This finite upper bound guarantees that every strictly increasing propagation sequence eventually stabilizes, providing the theoretical foundation for the termination of both the increasing iteration and the reference worklist algorithm established in Appendix~\ref{appendix:complexity_analysis}.

\section{Operationalization: Projection Formalism}
\label{appendix:projection_formalism}
\subsection{Native Structure Extraction}

Let

\[
F
\]

denote an agentic framework and let

\[
M_F
\]

denote a native framework model extracted from a particular agentic system.

The extracted model is represented as

\begin{equation}
M_F=(N_F,R_F,P_F),
\label{eq:M_equality}
\end{equation}

where

\begin{itemize}
	\item $N_F$ is the set of native framework entities;
	\item $R_F$ is the set of native relationships between these entities;
	\item $P_F$ denotes their framework-specific properties and metadata.
\end{itemize}

Depending on the framework, native entities may include agents, workflow nodes, tasks, tools, prompts, language-model configurations, routing conditions, memory components, or execution policies.

Extraction is performed by a framework-specific projection adapter. The adapter does not assign governance meaning directly. Its responsibility is limited to identifying native objects and recovering the structural and configuration information exposed by the framework.

This distinction prevents framework-specific implementation concepts from entering the ACM governance kernel.

\subsection{Semantic Classification}

The extracted native entities are subsequently classified according to the ACI taxonomy defined in Section~\ref{sec:reference_model}.

Let

\[
\tau_F :
N_F
\rightarrow
T_{\mathrm{ACI}}
\cup
\{\bot\}
\]

be the semantic classification function, where

\[
T_{\mathrm{ACI}}
\]

is the set of ACM configuration-item types and

\[
\tau_F(n)=\bot
\]

indicates that the native entity $n$ has no direct canonical representation in
the current ACM model.

A native entity is classified as an ACI only when it satisfies at least one of
the following conditions:

\begin{itemize}
	\item it contributes to the declared structure of the agentic system;
	\item it affects execution behaviour;
	\item it carries policy, permission, assurance, or deployment information;
	\item it must be versioned to support reproducibility or auditability.
\end{itemize}

Native implementation artifacts that do not carry configuration significance
may be deliberately excluded. Such exclusions must nevertheless remain
observable through the coverage analysis introduced below.

Native relationships are classified independently through

\[
\rho_F :
R_F
\rightarrow
T_R
\cup
\{\bot\},
\]

where $T_R$ is the set of canonical ACM relationship types.

This independent treatment is necessary because successful projection of two
entities does not imply that the semantic relationship connecting them has also
been preserved.

\subsection{Canonical Normalization}

Once classified, native entities and relationships are normalized into the ACM
canonical representation.

The entity projection function is defined as

\[
	\phi_F :
	N_F
	\rightharpoonup
	V_C,
\]

where $V_C$ is the vertex set of the resulting Configuration Graph
$G_C=(V_C,E_C)$ defined by the ACM reference model.

The partial nature of $\phi_F$ reflects that not every native entity is
necessarily representable by the current ACM reference model.

For every entity $n\in N_F$ such that

\[
\tau_F(n)\neq\bot,
\]

the projection produces an immutable ACI revision containing:

\begin{itemize}
	\item a canonical ACI type;
	\item a stable logical identifier;
	\item a revision identifier;
	\item a content digest;
	\item normalized configuration attributes;
	\item provenance linking the ACI revision to the native object.
\end{itemize}

Similarly, projected relationships are obtained through

\[
	\psi_F :
	R_F
	\rightharpoonup
	E_C.
\]

Each canonical relationship records its type, source and target ACIs, its
cardinality constraints, and the governance metadata required by the
propagation semantics.

The complete semantic projection induced by these mappings is denoted

\begin{equation}
	\mathrm{Proj}_F(M_F)
	=
	G_C,
	\label{eq:framework_projection}
\end{equation}

where $G_C$ contains the projected configuration revisions and relationships
obtained from the governance-relevant constructs represented in $M_F$.

Normalization therefore produces a framework-independent graph over which the
governance kernel can apply lifecycle validation, impact propagation, assurance
evaluation, and eligibility computation.

\subsection{Projection Traceability}

Every projected ACI retains explicit provenance to its source framework object.

Let

\begin{equation}
	src_F :
	V_C^{F}
	\rightarrow
	N_F
	\label{eq:projection_source}
\end{equation}

associate each projected ACI revision with the native entity from which it was
derived, where $V_C^{F}\subseteq V_C$ denotes the set of revisions produced by
the projection of $M_F$.

For every successfully projected entity,

\begin{equation}
	src_F(\phi_F(n))=n.
	\label{eq:projection_traceability}
\end{equation}

A corresponding mapping is maintained for relationships.

This bidirectional traceability serves three purposes. First, it allows the
projection result to be audited against the original framework definition.
Second, it identifies unsupported or partially represented constructs. Third,
it allows a governance finding produced by ACM to be related back to the native
framework object affected by that finding.

\subsubsection{Projection Coverage}

Projection expressiveness cannot be assessed solely by demonstrating that a
framework can be translated into ACM. It is also necessary to quantify how much
of the source framework model is represented by the projection.

Let

\[
N_F^{\mathrm{rel}}
\subseteq
N_F
\]

denote the set of governance-relevant native entities considered within the
scope of the adapter evaluation.

Entity Projection Coverage is defined as

\begin{equation}
	\label{eq:entity_projection_cover}
C_N(F)
=
\frac{
	\left|
	\left\{
	n\in N_F^{\mathrm{rel}}
	\mid
	\phi_F(n)\ \text{is defined}
	\right\}
	\right|
}{
	|N_F^{\mathrm{rel}}|
}.
\end{equation}

Similarly, let

\[
R_F^{\mathrm{rel}}
\subseteq
R_F
\]

denote the governance-relevant native relationships. Relationship Projection
Coverage is defined as

\begin{equation}
C_R(F)
=
\frac{
	\left|
	\left\{
	r\in R_F^{\mathrm{rel}}
	\mid
	\psi_F(r)\ \text{is defined}
	\right\}
	\right|
}{
	|R_F^{\mathrm{rel}}|
}.
\label{eq:relationship_projection_cover}
\end{equation}

The two measures are reported separately because a projection may preserve all
major entities while losing routing, dependency, delegation, or containment
semantics between them.

An aggregate Projection Coverage score may be computed as

\begin{equation}
C_P(F)
=
\frac{
	|N_F^{\mathrm{proj}}|
	+
	|R_F^{\mathrm{proj}}|
}{
	|N_F^{\mathrm{rel}}|
	+
	|R_F^{\mathrm{rel}}|
},
\label{eq:agg_projection_converage}
\end{equation}

where

\[
N_F^{\mathrm{proj}}
=
\left\{
n\in N_F^{\mathrm{rel}}
\mid
\phi_F(n)\ \text{is defined}
\right\}
\]

and

\[
R_F^{\mathrm{proj}}
=
\left\{
r\in R_F^{\mathrm{rel}}
\mid
\psi_F(r)\ \text{is defined}
\right\}.
\]

The aggregate score provides a concise summary, whereas $C_N(F)$ and $C_R(F)$
retain the distinction between structural and relational expressiveness.

The denominator deliberately includes only governance-relevant constructs
within the declared evaluation scope. Counting all internal framework objects
would introduce implementation details that ACM does not intend to model and
would make coverage scores incomparable across frameworks.

\subsubsection{Projection Outcomes}

Each native construct is assigned one of the following projection outcomes:

\begin{itemize}
	\item \emph{complete}: the construct and its governance-relevant semantics
	are preserved in the canonical model;
	\item \emph{partial}: the construct is represented, but one or more
	governance-relevant attributes or relationships are not preserved;
	\item \emph{unsupported}: no canonical representation exists in the current
	ACM model;
	\item \emph{out of scope}: the construct is intentionally excluded because
	it carries no configuration or governance significance under the evaluation
	protocol.
\end{itemize}

A projected entity contributes to $C_N(F)$ only when a canonical representation
is produced. However, partial projections are reported separately and are not
treated as semantically equivalent to complete projections.

Let

\begin{equation}
N_F^{\mathrm{complete}},
\quad
N_F^{\mathrm{partial}},
\quad
N_F^{\mathrm{unsupported}}
\label{eq:proj_entity_state_repr}
\end{equation}

denote the corresponding subsets. A stricter semantic coverage measure may
therefore be defined as

\[
C_N^{\mathrm{strict}}(F)
=
\frac{
	|N_F^{\mathrm{complete}}|
}{
	|N_F^{\mathrm{rel}}|
}.
\]

The difference

\[
C_N(F)-C_N^{\mathrm{strict}}(F)
\]
quantifies the proportion of native entities that are structurally projected
but only partially preserved semantically.

\subsubsection{Projection Validation}

The resulting canonical graph is validated before it is accepted by the
governance kernel.

Validation verifies:

\begin{itemize}
	\item conformance of each projected entity to its ACI schema;
	\item uniqueness and stability of logical and revision identifiers;
	\item integrity of content digests;
	\item validity of relationship endpoints and cardinalities;
	\item preservation of source provenance;
	\item explicit reporting of partial and unsupported constructs.
\end{itemize}

The adapter therefore produces both a canonical Configuration Graph and a
Projection Report:

\[
	Adapter_F(M_F)
	=
	(G_C,\mathcal{R}_F),
\]

where $\mathcal{R}_F$ contains the projection outcome of every in-scope native
entity and relationship, together with the coverage measures

\[
C_N(F),\quad
C_R(F),\quad
C_P(F),\quad
C_N^{\mathrm{strict}}(F).
\]

This report makes projection loss explicit rather than allowing unsupported
framework semantics to disappear silently during normalization.

\section{Consolidated Experimental Reports}
\label{app:experimental-reports}

This appendix reproduces the automatically generated reports underlying
the evaluation of Section~\ref{sec:evaluation}. They are included in full so that the reported figures can be traced back to their source
artifacts rather than only to the summary tables of the main text. Four
reports are consolidated: (i)~the consolidated scenario evaluation, which records the pass status of the twenty-seven normative governance
scenarios and the propagation metrics of the fixture harness;
(ii)~the information-preservation report, which quantifies semantic
projection fidelity across the three frameworks and their distinct
introspection regimes; (iii)~the comparative impact study, which opposes
ACM propagation to manual investigation on a shared configuration; and
(iv)~the quantitative impact experiment, which reports the nine
cross-framework impact cases used in Campaign~B.

All reports are generated directly from the execution environment; every
status and metric is derived from the executed tests and propagation
runs rather than hard-coded. The reports were produced on Python~3.13.14
(Linux~x86\_64) with the three evaluated frameworks available
(\texttt{langgraph}, \texttt{crewai}, \texttt{openai\_agents}). One methodological caveat is retained here, since it bounds the strength of the reported evidence: three scenarios pass with a documented, deliberate deviation from the original plan (\texttt{pass\_with\_deviation}). Oracle integrity in the impact run is attested by content digest and version-control provenance (\texttt{digest\_verified: true}). Both are discussed in
Section~\ref{sec:threats_validity}.

\subsection{Consolidated Scenario Evaluation}
\label{app:report-evaluation}

The scenario evaluation covers all twenty-seven governance scenarios.
Each scenario status is derived from the underlying tests. Across the
campaign, $379$ tests passed with no failures and no skipped tests;
$24$ scenarios pass without deviation and $3$ pass with a documented
deviation.

\begin{table}[htbp]
	\centering
	\caption{Scenario evaluation summary.}
	\label{tab:app-eval-summary}
	\small
	\begin{tabular}{@{}lc@{}}
		\toprule
		\textbf{Indicator} & \textbf{Value} \\
		\midrule
		Scenarios covered by at least one test & 27 / 27 \\
		\quad of which \texttt{pass} & 24 \\
		\quad of which \texttt{pass\_with\_deviation} & 3 \\
		\quad of which \texttt{skipped} & 0 \\
		\quad of which \texttt{fail} & 0 \\
		\quad of which \texttt{not\_executed} & 0 \\
		Tests passed (cumulative ) & 379 \\
		Tests failed & 0 \\
		Tests skipped & 0 \\
		YAML fixtures measured & 11 \\
		\bottomrule
	\end{tabular}
\end{table}

\begin{table}[htbp]
	\centering
	\caption{Per-scenario results. Priority is the internal test priority
		(P0--P2); the triple $P/F/S$ reports passed, failed, and skipped tests.}
	\label{tab:app-eval-scenarios}
	\small
	\begin{tabularx}{\columnwidth}{@{}llXcl@{}}
		\toprule
		\textbf{ID} & \textbf{Pri.} & \textbf{Object} & \textbf{P/F/S} & \textbf{Status} \\
		\midrule
		\multicolumn{5}{@{}l}{\emph{Group A --- Configuration and baseline}}\\
		S01 & P0 & Nominal configuration, promotion & 5/0/0 & \texttt{pass} \\
		S02 & P0 & Missing mandatory reference & 6/0/0 & \texttt{pass} \\
		S03 & P0 & Exact evidence identity (digest) & 2/0/0 & \texttt{pass} \\
		S04 & P0 & Immutability of a released baseline & 13/0/0 & \texttt{pass} \\
		\midrule
		\multicolumn{5}{@{}l}{\emph{Group B --- Propagation and assurance}}\\
		S05 & P0 & Blocking dependency NOK & 9/0/0 & \texttt{pass} \\
		S06 & P1 & Non-blocking dependency (warning) & 2/0/0 & \texttt{pass} \\
		S07 & P0 & New prompt revision, invalidation & 9/0/0 & \texttt{pass} \\
		S08 & P0 & Distributed assurance coverage & 6/0/0 & \texttt{pass} \\
		S09 & P0 & Complete evidence, failing result & 2/0/0 & \texttt{pass} \\
		S10 & P0 & Direct/aggregate/hybrid modes & 7/0/0 & \texttt{pass} \\
		S11 & P0 & Absent vs.\ empty policy & 4/0/0 & \texttt{pass} \\
		\midrule
		\multicolumn{5}{@{}l}{\emph{Group C --- Runtime, replay and drift}}\\
		S12 & P0 & Deterministic nominal replay & 4/0/0 & \texttt{pass} \\
		S13 & P0 & Invalid runtime sequence & 2/0/0 & \texttt{pass} \\
		S14 & P0 & Authorized runtime mutation & 3/0/0 & \texttt{pass\_w/dev.} \\
		S15 & P0 & Undeclared runtime mutation & 4/0/0 & \texttt{pass\_w/dev.} \\
		S16 & P1 & Configuration drift of a prompt & 4/0/0 & \texttt{pass\_w/dev.} \\
		S17 & P1 & Baseline withdrawn after run & 7/0/0 & \texttt{pass} \\
		\midrule
		\multicolumn{5}{@{}l}{\emph{Group D --- Dynamic agents and permissions}}\\
		S18 & P0 & Conformant dynamic agent & 14/0/0 & \texttt{pass} \\
		S19 & P0 & Permission escalation refused & 13/0/0 & \texttt{pass} \\
		S20 & P1 & Forbidden behavioral override & 5/0/0 & \texttt{pass} \\
		S21 & P1 & Promotion of a runtime agent & 5/0/0 & \texttt{pass} \\
		\midrule
		\multicolumn{5}{@{}l}{\emph{Group E --- Portability and robustness}}\\
		S22 & P0 & Same system in LangGraph/CrewAI & 2/0/0 & \texttt{pass} \\
		S23 & P1 & Explicit vs.\ distributed topology & 1/0/0 & \texttt{pass} \\
		S24 & P1 & Runtime event equivalence & 2/0/0 & \texttt{pass} \\
		S25 & P0 & Invariance to input ordering & 3/0/0 & \texttt{pass} \\
		S26 & P1 & Dependency cycles & 5/0/0 & \texttt{pass} \\
		S27 & P2 & Local volume and scaling & 3/0/0 & \texttt{pass} \\
		\bottomrule
	\end{tabularx}
\end{table}

\begin{table}[htbp]
	\centering
	\caption{Documented deviations (\texttt{pass\_with\_deviation}). All
		tests pass; the behavior departs from the original plan as a deliberate
		design choice rather than a defect.}
	\label{tab:app-eval-deviations}
	\small
	\begin{tabularx}{\columnwidth}{@{}lX@{}}
		\toprule
		\textbf{Scenario} & \textbf{Justification} \\
		\midrule
		S14 & Drift classification detail (\texttt{declared\_extension})
		derived in a conformance layer, not carried natively by the engine
		enumeration. \\
		S15 & \texttt{untraceable\_instance} classification derived; the engine
		carries the discrete \texttt{drift\_state} (\texttt{undeclared\_instance})
		without the explanatory detail. \\
		S16 & Configuration conformance (mismatch) evaluated as a separate
		result, orthogonal to \texttt{drift\_state}; not promoted to a
		first-order judgment. \\
		\bottomrule
	\end{tabularx}
\end{table}

\begin{table}[htbp]
	\centering
	\caption{Propagation-fixture metrics (harness). Iterations denote
		fixed-point iterations; convergence was reached in every case. Timings are indicative, hardware-dependent.}
	\label{tab:app-eval-fixtures}
	\small
	\begin{tabular}{@{}llccc@{}}
		\toprule
		\textbf{Scenario} & \textbf{Pri.} & \textbf{Iter.} & \textbf{Time (ms)} & \textbf{Conv.} \\
		\midrule
		ACM-S01 & P0 & 3 & 0.73 & yes \\
		ACM-S02 & P0 & 2 & 0.22 & yes \\
		ACM-S03 & P0 & 2 & 0.22 & yes \\
		ACM-S05 & P0 & 3 & 0.28 & yes \\
		ACM-S06 & P1 & 3 & 0.20 & yes \\
		ACM-S07 & P0 & 2 & 0.16 & yes \\
		ACM-S08 & P0 & 2 & 0.12 & yes \\
		ACM-S09 & P0 & 2 & 0.10 & yes \\
		ACM-S10 & P0 & 2 & 0.16 & yes \\
		ACM-S11 & P0 & 2 & 0.09 & yes \\
		ACM-S13 & P0 & 1 & 0.04 & yes \\
		\bottomrule
	\end{tabular}
\end{table}

\subsection{Information-Preservation Report}
\label{app:report-preservation}

For each native workflow $F$, the extraction $E(F)$ is compared against a
manually established golden representation over the normative ACM
perimeter. Each property is classified as \texttt{preserved} (exactly
reconstructed), \texttt{approximated} (ACM abstraction retained, e.g.\ an
opaque condition), or \texttt{unsupported} (no corresponding ACM concept).
The distinction between nodes that are \emph{extracted} directly and
nodes that are \emph{declared by the adapter} through metadata is
reported explicitly, since it characterizes the introspection regime of
each framework rather than a loss of information.

\begin{table}[htbp]
	\centering
	\caption{Information preservation across frameworks and abstraction
		levels. Coverage figures are node/relationship/branch coverage over the
		normative perimeter; $E$/$A$/$U$ report the counts of
		\texttt{preserved}/\texttt{approximated}/\texttt{unsupported}
		properties; \emph{Extr.}/\emph{Decl.} report nodes obtained by direct
		extraction vs.\ adapter declaration; \emph{Unres.} reports unresolved
		topology elements.}
	\label{tab:app-preservation}
	\small
	\begin{tabularx}{\columnwidth}{@{}Xccccc@{}}
		\toprule
		\textbf{Case} & \textbf{Cov.} & \textbf{$E/A/U$} & \textbf{Extr.} & \textbf{Decl.} & \textbf{Unres.} \\
		\midrule
		\multicolumn{6}{@{}l}{\emph{LangGraph}}\\
		Conditional branch & 100\% & 10/1/0 & 3 & 4 & 0 \\
		\midrule
		\multicolumn{6}{@{}l}{\emph{CrewAI --- three abstraction levels}}\\
		Crew-only & 100\% & 10/1/0 & 0 & 3 & 0 \\
		Flow-only & 100\% & 9/1/1 & 4 & 2 & 0 \\
		Flow+Crew & 100\% & 9/1/1 & 4 & 5 & 0 \\
		\midrule
		\multicolumn{6}{@{}l}{\emph{OpenAI Agents SDK --- two abstraction levels}}\\
		Agent only & 100\% & 10/1/0 & 0 & 1 & 0 \\
		Agent + handoffs & 100\% & 10/1/0 & 0 & 5 & 0 \\
		\bottomrule
	\end{tabularx}
\end{table}

\begin{table}[htbp]
	\centering
	\caption{Properties not fully preserved after semantic projection, with
		their status and the reason recorded by the extractor.}
	\label{tab:app-preservation-losses}
	\small
	\begin{tabularx}{\columnwidth}{@{}llX@{}}
		\toprule
		\textbf{Property} & \textbf{Status} & \textbf{Detail} \\
		\midrule
		\texttt{conditional\_branches} & \texttt{approximated} &
		Topological abstraction retained, opaque semantics (all frameworks). \\
		\texttt{state\_schema} & \texttt{unsupported} &
		No element extracted for a property present in the golden
		(CrewAI Flow-only and Flow+Crew cases). \\
		\bottomrule
	\end{tabularx}
\end{table}

The three frameworks exercise three distinct introspection regimes at a
constant ACM normative perimeter. LangGraph exposes its topology
explicitly and extracts it, signalling opaque conditions. For LangGraph, the two opaque conditional-branch predicates are retained as topological abstractions (status \texttt{approximated}). The execution topology itself is fully extracted; no governance-relevant relationship is left unrecovered. The CrewAI Flow topology is not statically introspectable in recent versions (edges are resolved at execution time) and is therefore partly supplied by adapter metadata; the state schema is reported \texttt{unsupported} rather than silently dropped. The OpenAI Agents SDK exposes the handoff delegation topology directly on the objects, so edges are extracted without adapter-declared topology, while the mapping of nodes to ACM references (agent/prompt/model/tool) remains adapter-declared, exactly as for LangGraph. This documented-loss behavior is the property that distinguishes semantic projection from ad hoc per-framework extraction.

\subsection{Comparative Impact Study}
\label{app:report-impact-comparative}

This study opposes two ways of answering the question ``if this ACI is
modified, what is affected?'' on a shared configuration of $13$ ACIs and
$18$ relations, changing the shared component \texttt{aci:model:shared-llm}.
It contrasts ACM fixed-point propagation with manual investigation, and
in particular with a na\"ive one-level investigation.

\begin{table}[htbp]
	\centering
	\caption{Comparative impact study on a 13-ACI, 18-relation configuration.}
	\label{tab:app-impact-comparative}
	\small
	\begin{tabularx}{\columnwidth}{@{}Xcc@{}}
		\toprule
		\textbf{Metric} & \textbf{Manual} & \textbf{ACM} \\
		\midrule
		Affected items identified & 9 (exhaustive) & 9 \\
		Cost & 10 inspections & 1 query \\
		Propagation depth & 2 levels & 3 iterations (fixed point) \\
		\bottomrule
	\end{tabularx}
\end{table}

A manual investigation that stops at direct dependents (one level)
identifies only $6$ of the $9$ genuinely affected items, missing three
items affected \emph{indirectly} through an intermediate ACI:
\texttt{aci:workflow:w0}, \texttt{aci:workflow:w1}, and
\texttt{aci:workflow:w2}. The affected set computed by ACM coincides
exactly with the exhaustive manual investigation: no false positives and
no false negatives. The difference lies in cost and reliability---ACM
guarantees exhaustiveness in a single propagation, whereas the manual
investigation requires perfect transitive discipline.

\subsection{Quantitative Impact Experiment (Campaign~B)}
\label{app:report-impact-quantitative}

Nine cross-framework cases combine the three frameworks with three change
classes (local, intermediate, global). The chain is: native-equivalent
system $\rightarrow$ ACM fixed-point engine $\rightarrow$ $P_f(c)$
$\rightarrow$ comparison against a frozen oracle. Each case was executed
five times; results were deterministic across repetitions, and the static
containment $P_f(c) \subseteq \mathrm{reach}(\mathit{root})$ held
throughout.
Each case records a SHA-256 digest of its frozen oracle, verified against a manifest generated prior to the run (\texttt{digest\_verified: true} across all nine cases). The oracles and their manifest are committed together under version control, and each result records the originating commit (\texttt{oracle\_git\_commit}), establishing that every oracle predates the run that consumes it. Oracle integrity is therefore attested both by content digest and by version-control provenance rather than by construction of the harness alone. 

\begin{table}[htbp]
	\centering
	\caption{Quantitative impact experiment. Impact size $|P_f(c)|$ and
		ratio $|P_f(c)|/|V|$ are identical across frameworks by construction of
		the governance-equivalent projected dependency structures; iterations
		denote fixed-point iterations; \emph{Red.} is the strict inspection-scope
		reduction (assisted vs.\ manual inspections). All cases are deterministic
		over five repetitions. Timings are indicative, hardware-dependent.}
	\label{tab:app-impact-quantitative}
	\small
	\begin{tabular}{@{}llccccc@{}}
		\toprule
		\textbf{Framework} & \textbf{Class} & \textbf{$|P_f|$} & \textbf{Ratio} & \textbf{Iter.} & \textbf{ms} & \textbf{Red.} \\
		\midrule
		LangGraph      & local        & 2 & 0.182 & 3 & 2.20 & 0.846 \\
		LangGraph      & intermediate & 2 & 0.182 & 3 & 2.10 & 0.692 \\
		LangGraph      & global       & 5 & 0.455 & 3 & 2.09 & 0.615 \\
		\midrule
		CrewAI         & local        & 2 & 0.182 & 3 & 2.12 & 0.846 \\
		CrewAI         & intermediate & 2 & 0.182 & 3 & 2.09 & 0.692 \\
		CrewAI         & global       & 5 & 0.455 & 3 & 2.12 & 0.615 \\
		\midrule
		OpenAI Agents  & local        & 2 & 0.182 & 3 & 2.11 & 0.846 \\
		OpenAI Agents  & intermediate & 2 & 0.182 & 3 & 2.10 & 0.692 \\
		OpenAI Agents  & global       & 5 & 0.455 & 3 & 2.11 & 0.615 \\
		\bottomrule
	\end{tabular}
\end{table}

\begin{table}[htbp]
	\centering
	\caption{Baseline reassessment (derived signal, outside $P_f(c)$). A
		released baseline is immutable; a change to a required item triggers an
		external operational reassessment computed from $P_f(c)$. The baseline
		itself does not appear in the affected set.}
	\label{tab:app-impact-reassessment}
	\small
	\begin{tabularx}{\columnwidth}{@{}llX@{}}
		\toprule
		\textbf{Class} & \textbf{Reassess.} & \textbf{Triggering items (all frameworks)} \\
		\midrule
		local & required &
		\texttt{aci:agent:finalizer}, \texttt{aci:prompt:finalize},
		\texttt{aci:workflow:main} \\
		intermediate & required &
		\texttt{aci:agent:researcher}, \texttt{aci:tool:web-search},
		\texttt{aci:workflow:main} \\
		global & required &
		\texttt{aci:agent:direct}, \texttt{aci:agent:finalizer},
		\texttt{aci:agent:researcher}, \texttt{aci:agent:reviewer},
		\texttt{aci:model:shared-llm}, \texttt{aci:workflow:main} \\
		\bottomrule
	\end{tabularx}
\end{table}

Across the nine cases, the affected sets produced by the fixed-point
engine matched the frozen reference sets exactly. Because these reference sets were established from the native definitions but were not
independently adjudicated, the agreement should be read as internal
consistency of the propagation engine with a pre-specified oracle rather
than as accuracy against an externally established ground truth. The
comparative study of Appendix~\ref{app:report-impact-comparative}
provides the complementary evidence that this oracle is non-trivial,
since a na\"ive traversal of the same dependency structure omits
transitively affected items.

\bibliographystyle{IEEEtran}
\bibliography{references}
\end{document}